\documentclass[a4paper,10pt,oneside,onecolumn,preprint]{elsarticle}

\usepackage{url}
\usepackage{hyperref}
\usepackage{color}
\usepackage{enumerate}
\usepackage[cmex10]{amsmath}
\usepackage{amsfonts}
\usepackage{amssymb}
\usepackage{listings}
\usepackage{floatrow} 
\usepackage{comment} 
\usepackage{multirow}
\usepackage{graphicx, subfigure}
\usepackage[dvipsnames]{xcolor}
\usepackage{listing}
\usepackage{draftwatermark}
\usepackage[outdir=./draws/]{epstopdf}
\usepackage{hyperref}

\journal{Computer Communications}

\begin{document}

\begin{frontmatter}

\title{A novel JXTA-based architecture for implementing heterogenous Networks of Things}

\author[mymainaddress]{Filippo Battaglia}
\ead{fbattaglia@unict.it}

\author[mymainaddress]{Lucia Lo Bello\corref{mycorrespondingauthor}}
\cortext[mycorrespondingauthor]{Corresponding author}
\ead{lobello@unict.it}

\address[mymainaddress]{DIEEI - Department of Electrical, Electronic and Computer Engineering, Viale Andrea Doria, 6, 95125 Catania}

\begin{abstract}

\small This paper presents EmbJXTAChord, a novel peer-to-peer (P2P) architecture that integrates the good features of different sources, 
such as JXTA, EXI, CoAP, combining and augmenting them to provide a framework that is specifically devised for developing 
IoT applications over heterogeneous networks. 

EmbJXTAChord provides for several interesting properties, such as,  distributed and fault-tolerant resource discovery, transparent routing over
subnetworks, application protocol independence from the transport protocol in narrowband Wireless Sensor Networks, 
thus eliminating the need for using dedicated software or configuring custom gateways to achieve these functionalities.

Moreover,  EmbJXTAChord offers native support not only for TCP/HTTP,
but also for Bluetooth RFCOMM and 6LoWPAN, thus opening to a broad range of IoT devices
in supernetworks composed of networks using different interconnection technologies, not necessarily IP-based.
In addition, EmbJXTAChord offers security over heterogeneous networks
providing support for secure peergroups (even nested) and for group encryption, thus allowing for unicast and multicast communication between groups of objects sharing
the same resources.
The users of the proposed architecture will benefit from an integrated solution and
the applications developed on the proposed framework will be able to reconfigure themselves, adapting automatically to the
network topology of the execution environment.

Finally, EmbJXTAChord provides jxCOAP-E, a new CoAP implementation that leverages on the transport mechanisms for heterogeneous 
networks offered by EmbJXTAChord. jxCOAP-E enables to realize a RESTful service architecture for peer-to-peer narrowband or broadband 
networks composed of devices connected via Ethernet, Wi-Fi, Bluetooth, BLE or IEEE 802.15.4. Differently from CoAP, jxCOAP-E provides 
a distributed and fault-tolerant service discovery mechanism and support for secure multicast communications.  
The paper presents EmbJXTAChord, discusses all the relevant design challenges and presents a comparative
experimental performance assessment with state-of-the-art solutions on
commercial-off-the-shelf devices.
\end{abstract}


\end{frontmatter}

\small
\textbf{Note:}\textit{This is a preprint version of an Elsevier accepted paper. The formal version of the work is
available at \href{https://doi.org/10.1016/j.comcom.2017.11.002}{https://doi.org/10.1016/j.comcom.2017.11.002}}

\textit{\copyright 2017. This manuscript version is made available under the CC-BY-NC-ND 4.0 license 
\href{http://creativecommons.org/licenses/by-nc-nd/4.0/}{http://creativecommons.org/licenses/by-nc-nd/4.0/}}
\normalsize 


\section{Introduction and motivation}\label{Sec:Introduction}

A typical domestic scenario of today includes several computers, smart phones 
and appliances connected through wired Ethernet, power lines or 
IEEE 802.11 (Wi-Fi) wireless links, which can, in turn, access 
sensors or actuators through multiple wireless protocols, such as Wi-Fi, Bluetooth or IEEE 802.15.4.
For instance, a palmtop can run an application that sets the start time for the washing 
machine that, in turn, can defer its work based on the constraints 
of the power meter and on the energy output currently provided by the solar panels.

This scenario is only a small-scale example of the Internet of Things (IoT), 
in which a number of interconnected cooperative smart objects
provide multi-agent services \cite{IoTSurvey05}. However, none of the frameworks proposed in the IoT field so far is able
to fully guarantee two fundamental properties, i.e., the application source code independence from the underlying network topology
and the suitability for any kind of subnetwork. 
Currently the IoT applications must be tailored to the middleware and the network configuration, as the available
solutions are mainly incompatible \cite{IoTSurvey06}. In fact, each middleware exploits its own API and data formats for requests
and responses.

\newcommand{\TABzb} 
{
	\begin{tabular}{|l l|l l|}%
		\hline 
		  3GPP 		&3rd Generation Partnership Project 	&NAT 	&Network Address Translation \\
		  6LoWPAN	&IPv6 over Low power Wireless 		&PAN	&P2P Peer-to-Peer \\
		  AES 		&Advanced Encryption Standard		&PAN	&Personal Area Network \\ 
		  ATT 		&Attribute Protocol (BLE)		&RaspPI	&Raspberry PI \\
		  BLE 		&Bluetooth Low Energy               	&REST 	&Representational State Transfer \\    
		  BNEP 		&Bluetooth Network Encapsulation Protocol&RPV 	&Rendezvous PeerView \\
		  CM 		&EmbJXTAChord Compressor Manager    	&RTT   	&Round Trip Time \\
		  CoAP 		&Constrained Application Protocol  	&SICS	&6LoWPAN \\
		  CSR		&Cambridge Silicon Radio 		&SOAP 	&Simple Object Access Protocol \\      
		  DHT 		&Distributed Hash Table 		&SRDI 	&Shared Resource Distributed Index \\ 
		  DTLS 		&Datagram Transport Layer Security	&TLS 	&Transport Layer Security \\
		  ETH		&Ethernet 				&URI 	&Uniform Resource Identifier \\
		  EXI 		&Efficient XML Interchange		&UDDI  	&Universal Description Discovery and Integration \\     
		  GATT 		&Generic Attribute Profile (BLE)	&UTF 	&Unicode Transformation Format\\
		  IETF 		&Internet Engineering Task Force    	&WAVE  	&Wireless Access in Vehicular Environment  \\
		  IMX 		&Intermessage compression		&WG	&Working Group \\
		  LRW 		&Limited Range Walker               	&WS	&Web Service \\           
		  LTE 		&Long Term Evolution 			&WSN  	&Wireless Sensor Network \\ 
		  MIME		&Multipurpose Internet Mail Extensions 	&XML 	&eXtensible Markup Language \\
		  MQTT 		&Message Queue Telemetry Transport  	&adv 	&Advertisement \\
		  MSA 		&Module Specification Advertisement	&rdv	&Rendezvous \\
		  MTB 		&Message Transport Binding		&uPNP   &Universal Plug and Play \\	
		\hline
	\end{tabular}
}
\ifdefined\ACM
\begin{table}
	\tbl{}
	{      
		\TABzb{List of the acronyms and abbreviations used in the paper.}
	}
	\label{AddedTable_II}
\end{table}
\else
\begin{table}[t]
	\setlength{\tabcolsep}{.16em}
	\scriptsize
	{
		\caption{\scriptsize {List of the acronyms and abbreviations used in the paper.} }
		\vspace*{-0.1in}
		\hspace*{-1.0in}
		\begin{center}
			\TABzb{}
		\end{center}\label{AddedTable_II}
	}
	\vspace*{-0.2in}
\end{table}
\fi

In the past, in an attempt of standardization, the use of web services for domestic or industrial applications based on SOAP
\footnote{For convenience of the reader, the acronyms most frequently used in this paper are re-
ported in Tab. \ref{AddedTable_II} and \ref{AddedTable_I}.}
(Simple Object Access Protocol) was proposed \cite{SOAPWSApp01}\cite{SOAPWSApp00}\cite{SOAPWSApp03}.
Other works proposed REST (Representational state transfer), a paradigm 
where all resources are accessible by a common HTTP client (as a browser) \cite{REST00}. 
Unfortunately, both SOAP and REST leverage on the transmission of large HTML or XML documents,
which can be too bandwidth-consuming for low-bandwidth wireless sensor networks (WSN)
protocols, such as the IEEE 802.15.4 \cite{IEEE802_15_4_00}\cite{SOAPWSApp04}\cite{SOAPWSApp05}.

As a consequence,  new lightweight and efficient protocols for IoT communication between 
smart devices, such as CoAP (Constrained Application Protocol) \cite{COAP00}\cite{COAP01}, MQTT
(Message Queue Telemetry Transport) \cite{IP00_Repl_00} or SMQ (Simple Message Queries) \cite{SMQ01}\cite{SMQ02}, 
were recently developed \cite{IP00_Repl_02}. 

Under CoAP, a new protocol recently proposed by the IETF CoRE Working Group, all the functionalities 
are seen as a set of resources, identified by a URI (Uniform Resource Identifier) and all the operations 
are performed through four methods (GET, PUT, POST and DELETE). Under MQTT and SMQ, instead, sensor
readings are seen as \textit{topics}, whose change can be notified by a \textit{central broker} to a
set of subscribers. 

The protocols cited above were designed minimizing the functionalities provided to the developer,
thus obtaining bandwidth-efficient and energy-efficient communications between resource-constrained
devices 
\cite{IP00_Repl_01}\cite{IP00_Repl_00}.

The solutions based on CoAP, MQTT, SOAP-WS or REST \cite{IoTSurvey00}\cite{IP00_Repl_02} 
need further components to provide the functionalities of distributed service discovery  \cite{UDDI03_Repl_01}
\cite{UDDI03_Repl_02}\cite{UDDI03_Repl_03} ,
fault tolerance  \cite{UDDI02_Repl_01}  
and resource protection \cite{IoTSecurity02}. 
Moreover, these IoT protocols  use the underlying IP network layer, 
assuming the availability of such a layer for each link-layer protocol.

The envisaged vision of all devices interconnected via IP 
determined the proliferation of middleware solutions for IoT that are designed to work over IPv4/IPv6.
However, in a heterogeneous network made up of devices connected through different kinds of protocols 
at the physical, network and transport levels, located at different places and separated by firewalls, gateways and NATs, the creation of an IPv6 network can be difficult.
\begin{figure}[t]
  \addtolength{\abovecaptionskip}{-2.0in}
  \vspace*{-0.23in}
  \subfigure
  {
    \includegraphics[width=5.00in, height=1.65in]{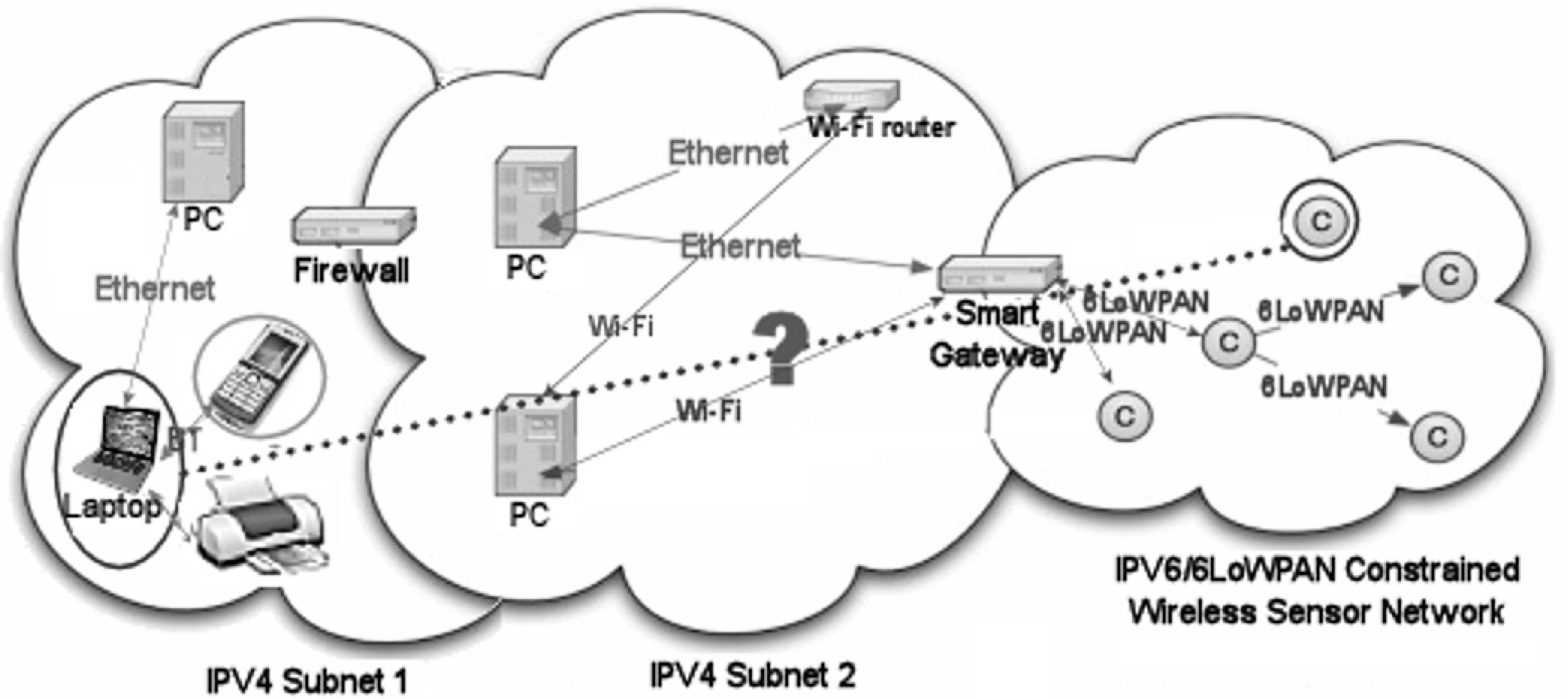} }
    \caption{A typical scenario for a heterogeneous network.}
    \vspace*{-0.12in}
\label{Scene01}
\end{figure}
For example, let us consider Fig. \ref{Scene01}.
The laptop is connected to a subnetwork consisting of a printer and a
mobile phone that is connected to the laptop 
via Bluetooth or Bluetooth Low Energy (BLE) , 
as this protocol is less power-consuming than Wi-Fi. 
A firewall operates between the first and the second subnetwork and
a 6LoWPAN smart gateway  between the second and the
third subnetwork.  The first two subnetworks are based on IPv4, while
the third one on IPv6/6LoWPAN.
Some issues can prevent the delivery of the CoAP messages
between the laptop or the mobile phone and the WSN 
in the third subnet, for instance

\begin{itemize}
\item \textbf{Bad firewall configuration}. The firewall can prevent the delivery
of the packets to the CoAP standard port;

\item \textbf{Different network protocols}. The three networks use different
network protocols, therefore there are different addressing schemes
(32b for IPv4, 128b for IPv6);

\item \textbf{Difficult use of IP in Bluetooth networks for some devices}.
The mobile phone firmware might not support the Ethernet emulation
for Bluetooth (BNEP, the Bluetooth Network Encapsulation Protocol), but
only connections via RFCOMM (serial port emulation). In such a case,
a custom gateway RFCOMM/IP should be implemented. 
Some issues may occur even if the mobile phone can use some of the
new BLE features, such as the IPSP (Internet Protocol
Support Profile) \cite{Bluetooth12} that allows to create
a 6LoWPAN network over BLE (6LoBLE). For instance, the BLE nodes that work as IPSP 
routers cannot be connected to a IPv4 subnetwork without an IPv6/IPv4
gateway. 
Moreover, some operating systems might also have limitations that 
hinder the use of BNEP or IPSP \cite{Bluetooth13}. 

\item \textbf{Complex security implementation}. 
IoT communications can be secured using IPSec \cite{IPSec00}, TLS (Transport 
Layer Security) \cite{TLS03} or DTLS (Datagram Transport Layer Security) \cite{DTLS00}.
TLS and DTLS are application-level protocols that work over TCP and UDP, respectively.
IPSec is a network-level protocol that may require a difficult  
configuration \cite{IPSec01}, which makes the protocol prone to interoperability issues, 
and specific customizations (for instance to work with 6LoWPAN \cite{IPSec02}).
Unfortunately, all these protocols are unsuitable for non-IP networks (such as the Bluetooth 
RFCOMM piconets).

\item \textbf{Difficult implementation of multicast group communications}. The standard request-response
interaction model used by CoAP supports only end-to-end message exchange. For this reason,
the IETF CoRE WG released the RFC 7390 \cite{COAP04} that defines how CoAP should support 
\textit{group communications}. As this feature leverages on IP multicast packet delivery, 
it cannot be implemented when the physical layer supports only unicast communications 
(such as in the BT and BLE network cases \cite{Bluetooth12}) or when the bridges between subnetworks 
prevent the propagation of these messages.  
Moreover, the RFC 7390 does not state how to implement \textit{secure} group communications among nodes \cite{COAP08}\cite{COAP05}. 
This is an important limitation, as the TLS and DTLS implementations currently available do not support multicast communications, 
despite some proposals were published \cite{TLS02} \cite{COAP07}. 

\end{itemize}

In such a situation, the only solution would be to 
reconfigure all the subnetworks in order to use the same addressing schema (IPv6) and to manually configure the firewall 
and the routing tables of the border routers, so as to manage packet delivery between a device 
of the first and of the third subnet.

These issues can be overcome using a peer-to-peer (P2P) protocol, working at the application level, able to support different network protocols 
through dedicated modules, thus abstracting from the link technology actually adopted and providing for automatic
bridging. This way, the mobile phone can communicate to the 6LoWPAN device (and vice versa) using a chain of bridges (the laptop,
the firewall between the first and the second subnet, the smart gateway between the second and the third subnet) in
a seamless way. This solution would have been considered unsuitable for IoT in the past, because the use of a P2P
protocol for message exchanging determines a consistent overhead compared to the use of the bare IPv4/IPv6. 
Nowadays, P2P communication became an interesting option for IoT, thanks to the availability 
on the market of small-sized embedded boards,
such as the Raspberry PI \cite{RaspberryPI01}, or the Raspberry PI-3 \cite{RaspberryPI04} that provide enough
computational power to support P2P protocols 
even on low-cost COTS hardware. 
The P2P protocol should provide the following functionalities:
\begin{itemize}
\item Host resolution by name. In a heterogeneous network, a device should be addressed by name and not by the IP address. In fact, 
IP or the Dynamic Host Configuration Protocol (DHCP) might be unavailable in the device subnetwork or the DHCP server could dynamically change the assigned IP
address \footnote{For an IP-only homogeneous network the host resolution by name can be provided by using a protocol such as ZeroConf. 
However, ZeroConf is unsuitable for a heterogeneous network due to its multicast-based design \cite{ZeroConf00}.};
\item Network protocol translation. As using IP over each link is not necessary, any protocol can be adopted. 
In the described scenario, the laptop transfers the IoT messages from the IP section of the network to the Bluetooth RFCOMM section transparently to the
application level; 
\item HTTP tunnelling. Message delivery is possible through the firewall without manual reconfiguration;
\item Routing over subnetworks. The P2P protocol must be able to deliver the messages between 
any pair of devices, even when they belong to different subnetworks (for instance, a BT mobile phone and an IEEE 802.15.4 sensor node). The P2P protocol should be
able to automatically determine the information needed for routing, and routing should be 
performed by the application level without reconfiguring the IP forwarding tables 
in the router or in the operating system.

\item Message propagation over subnetworks. If a peer needs to send the same message to all other
members of the group, it can use the multicast transmission provided by the link-layer. If multicast transmission
is not available (as in the case of BT or BLE networks), or if routers prevent the transmission of
multicast packets between the subnetworks, the P2P framework can anyway propagate the message, transmitting
a copy of the message to each peer through unicast connections in a single transaction. This is possible as 
the coordinator nodes mantain a list of all the peers currently present in the peergroup.

\end{itemize}

This paper provides two main contributions. 
The first one is EmbJXTAChord, a novel peer-to-peer (P2P) architecture specifically devised for developing 
IoT applications over heterogeneous networks. 
EmbJXTAChord integrates the good features of different sources, 
such as JXTA, EXI, CoAP, combining and augmenting them to provide a framework for the development of IoT applications
that is able to offer several nice properties, such as:
  \begin{itemize}
  \item  Service architecture based on a RESTful-API; 
  \item  Distributed and fault-tolerant service discovery; 
  \item Support for secure \textit{nested} peergroups;
  \item Routing over subnetworks transparent for the application layer, that allows to hide the presence of gateways;
  \item Support for HTTP tunnelling and NAT traversal;
  \item Availability of a distributed content management system integrated in the system (based on advertisements);
  \item Agnosticism about the network protocol that is actually used.
  \end{itemize}

  To the best of our knowledge, there is no other open-source solution that supports all these features together. 
  The implementation of the proposed architecture "mimics" the good properties of JXTA (without being JXTA) and 
  integrates several technologies that were recently developed in the IoT field. For instance, 
  EXI and a new compression protocol allow to reduce the bandwidth required for transmission,  CoAP allows the realization of a
  RESTful architecture, the new MTB enables to support new transport protocols.  In EmbJXTAChord such technologies work transparently 
for the developer, thus realizing a fully-integrated framework that simplifies the development of IoT applications.

In particular, EmbJXTAChord uses a compression algorithm that allows
to extend some interesting features borrowed from JXTA (such as distributed
and fault-tolerant resource discovery, transparent routing over subnetworks,
application protocol independence from the transport protocol) in narrowband
Wireless Sensor Networks (WSN), thus eliminating the need for dedicated
software or custom gateways. 

Moreover, EmbJXTAChord not only supports TCP/HTTP, but adds native support
for Bluetooth RFCOMM and 6LoWPAN. This paves the way to its adoption in a broad range of IoT devices
in supernetworks composed of subnets not necessarily based on IP. EmbJXTAChord solves the interoperability issues of the physical and
transport protocols, as such protocols are transparently managed by the underlying P2P overlay layer.

EmbJXTAChord also fosters security over heterogeneous networks,
adding support for secure peergroups (even nested) and for group encryption, thus allowing for unicast and multicast communication between groups of objects sharing
the same resources.

The second main contribution of the paper is jxCOAP-E, a module of EmbJXTAChord that realizes a new
version of CoAP that allows to deploy \textit{heterogeneous networks of things} consisting of
several subnetworks that are seen by the application level as a whole RESTful system, regardless of the actual 
connection topology or the available bandwidth. 
JxCOAP-E provides a distributed, fault-tolerant resource discovery paradigm and improves the CoAP multicast security 
model supporting \textit{group encryption}, that allows to create \textit{groups of objects} 
sharing the same resources, unlike UDP CoAP, which uses a centralized server for resource discovery and supports 
only unicast secure connections via DTLS.

The paper is organized as follows.  
Sect. \ref{Sec:JxtaAdvDrawbacks} recaps JXTA benefits and limitations for implementing IoT applications over heterogeneous networks, while Sect. \ref{Sec:RelatedWorks} addresses related work. 
Sect. \ref{Sec:EmbJXTAChord} describes the EmbJXTAChord features, while Sect. \ref{Sec:ExpResult} presents the 
experimental results and an assessment of the performance achievable by EmbJXTAChord using the Raspberry PI or the Raspberry PI-3.
Sect. \ref{Sec:Smart Home} deals with a scenario suitable for the proposed solution. 

Sect. \ref{Sec:TwoSmallExamples} presents two small examples of EmbJXTAChord programming, thus showing 
the effort required to the developer for using the middleware. 
Sect. \ref{Sec:PowerMeasures} provides experimental assessments of the power comsumption for a single
node in some typical use cases.

Finally, Sect. \ref{Sec:Conclusions} gives conclusions and hints for further work. 

\section {EmbJXTAChord vs JXTA}\label{Sec:JxtaAdvDrawbacks}

\subsection{Overview of JXTA}

\newcommand{\TABza} 
{
	\begin{tabular}{|c|l|l|}%
		\hline 
		   &JXTA protocol               &Description \\
		\hline
		ERP&Endpoint Routing Protocol   &It routes JXTA messages over different subnetworks, even if \\
		   &				&separated by gateways and NATs, working transparently to \\
		   &				&the upper layers of the middleware. 			     \\
		\hline
		PRP&Peer Resolver Protocol      &It is the connectionless communication protocol that allows \\
		   &				&the exchange of \textit{messages} between nodes (\textit{peers}) belonging\\
		   &				&to different subnetworks. Both unicast and multicast \\
		   &				&communications are supported.\\
		\hline
		RP &Rendezvous Protocol         &It is executed by all rendezvous peers. The protocol manages \\
		   &			        &a Distributed Hash Table (DHT) that allows to find all \\
		   &				&advertisements of the peergroup. \\
		\hline
		PDP&Peer Discovery Protocol     &It works in synergy with the Rendezvous protocol. It provides \\
		   &			        &an API that allows to find the advertisements describing the  \\
		   &				&peers, peergroups and services available within the peergroup.\\
		\hline
		PBP&Pipe Binding Protocol       &It allows a reliable connection-oriented communication between \\
		   &			        &a pair of nodes of the peergroup, regardless of the features of\\
		   &				&the underlying transport protocol. \\
		\hline
		PIP&Peer Information Protocol   &It provides statistics about network traffic among peers. \\
		   &			        & \\
		\hline
	\end{tabular}
}
\ifdefined\ACM
\begin{table}
	\tbl{}
	{      
		\TABza{Description of EmbJXTAChord protocols. These modules are reimplementations of the original JXTA 2.7 protocols.}
	}
	\label{AddedTable_I}
\end{table}
\else
\begin{table}[t]
	\setlength{\tabcolsep}{.16em}
	\scriptsize
	{
		\caption{\scriptsize {Description of EmbJXTAChord protocols. These modules are reimplementations of the original JXTA 2.7 protocols.} }
		\vspace*{-0.1in}
		\hspace*{-0.2in}
		\begin{center}
			\TABza{}
		\end{center}\label{AddedTable_I}
	}
	\vspace*{-0.2in}
\end{table}
\fi

The first version of JXTA was released by Sun Microsystem (today Oracle) in 2001, in order to allow communications over heterogeneous
networks consisting of several subnets based on different transport protocols and addressing schemes. JXTA consists of 6
protocols, whose functionalities are summarized in Tab. \ref{AddedTable_I}.

JXTA creates an overlay abstraction layer that manages all peers (i.e., nodes) through a uniform 128b address (named PeerID).

\begin{figure}[t]
  \addtolength{\abovecaptionskip}{-2.0in}
  \vspace*{-0.23in}
  \hspace*{-0.6in}
  \subfigure
  {
    \includegraphics[width=6.00in, height=2.50in]{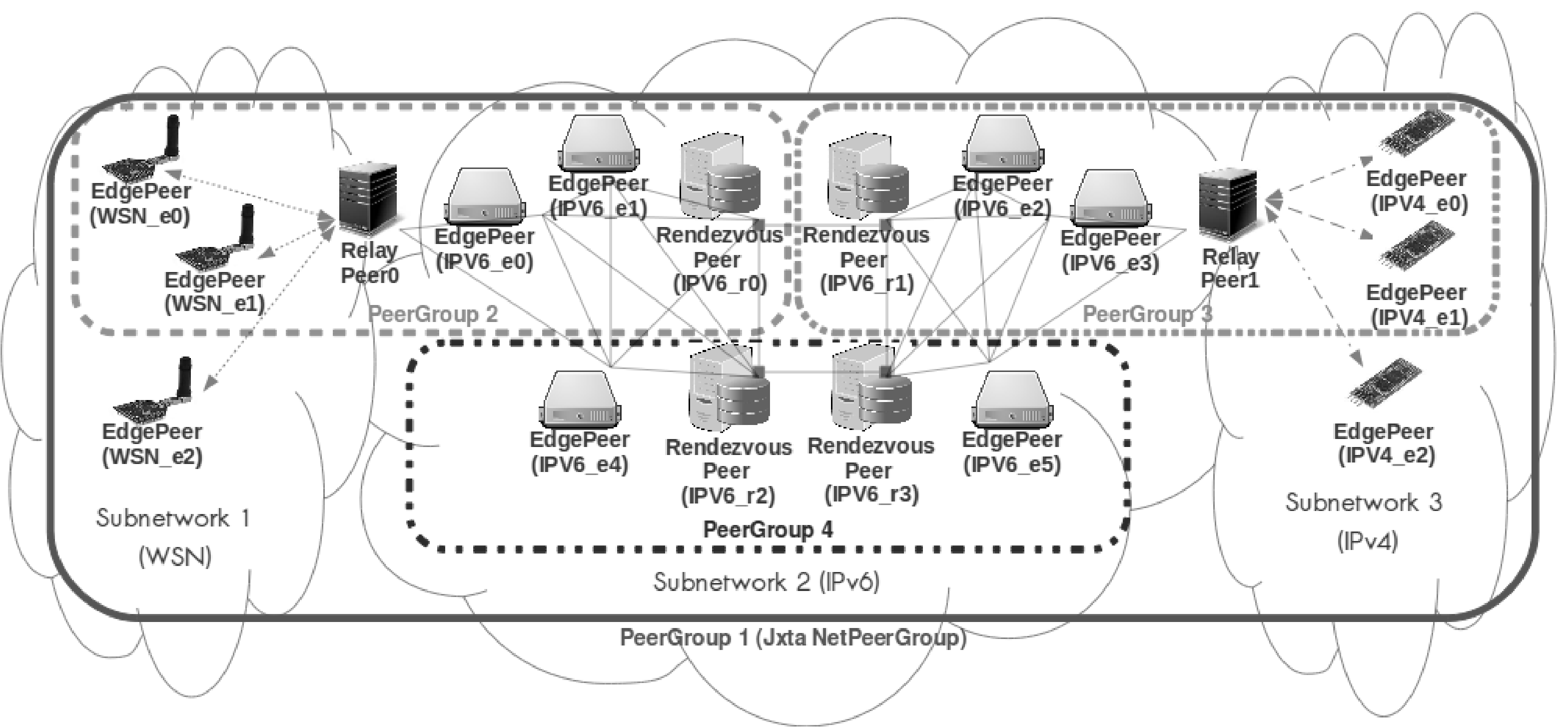} }
    \caption{Representation of a heterogeneous network using EmbJXTAChord.}
    \vspace*{-0.12in}
\label{JXTA_Scenario}
\end{figure}

The peers are organized in \textit{peergroups}. As represented in Fig. \ref{JXTA_Scenario}, peergroups can be deployed (i.e. created)
regardless of the communication technology used between nodes. Moreover, peergroups can be nested. The support for peergroups allows to protect 
resources from unauthorized accesses. For instance, in Fig. \ref{JXTA_Scenario} the peer named \textit{WSN\_e2} 
can use the resources available in the PeerGroup1 (named NetPeerGroup), while it cannot access to the resources available in the PeerGroup2. 
Conversely, the peers \textit{WSN\_e0} and \textit{WSN\_e1} can access to the resources available both in the PeerGroup1 
and in the PeerGroup2. 

Each JXTA peer can work in one of the following modes: \textit{adhoc}, \textit{edge}, \textit{rendezvous} (rdv), and \textit{relay} \cite{JXTA14}.

The \textit{edge} mode is commonly used by all nodes that work as clients, as it loads all standard JXTA protocols,
but requires an active connection to a rendezvous  peer (rdvpeer) before starting. The \textit{rendezvous} nodes also provide the functionalities
needed for mantaining operative the JXTA infrastructure, as they store part of a register, named SRDI (Shared Resource Distributed Index),
that is used to implement distributed resource discovery within the whole group \cite{JXTA08}.  The \textit{adhoc} nodes are 
resource constrained devices providing only a minimum set of functionalities. The \textit{relay} nodes are
similar to the edge nodes, but they can
propagate multicast messages among two subnetworks regardless of the support for multicast datagrams at network level. 

Two peers can communicate using virtual channels, named pipes for asynchronous unicast connections or jxta-sockets for
synchronous unicast connections. Pipes can be reliable or unreliable, whereas sockets are always reliable.
In addition, JXTA provides also propagate pipes, which ensure multicast communications within a group.
A content management service based on XML documents (xmldocs), named advertisements (adv),
is integrated in the protocol. When an advertisement  is published by a peer, it can be found 
by all the other peers of the group through the Peer discovery protocol (PDP), which provides the discovery of new 
resources (services, nodes, contents etc.) within a group of peers. 

\subsection{What EmbJXTAChord borrows from JXTA}

EmbJXTAChord inherits the following features of JXTA:

\begin{itemize}
\item \textit{Independence from the transport protocol}. EmbJXTAChord can potentially use any transport protocol, using 
dedicated modules, named Message Transport Binding (MTB) modules. 
Besides the MTB modules for TCP-IP, HTTP tunnelling (that allows the traversal of firewalls and NATs) and multicast UDP,
that were supported also in the latest version of JXTA (v2.7) \cite{JXTA09}, EmbJXTAChord integrates MTB modules for 
Bluetooth RFCOMM and 6LoWPAN communication protocols (see Sect. \ref{MessageTransportBindings}), thus supporting a wide range of devices
that are typically used in IoT applications; 

\item \textit{Support for heterogeneous networks.} EmbJXTAChord  provides routing over subnetworks 
thus allowing communications between peers belonging
to subnets using different addressing schemes and network protocols. Each peer can transparently promote itself as gateway (or sink). 
Operations as address translation or hop-by-hop delivery are automatically
performed by the EmbJXTAChord  overlay level;

\item \textit{Group security.} Peers can be grouped in peergroups, sharing a set of privileged access resources
(pipes, jxCOAP-E services). The groups can be nested, thus creating a hierarchy of privileged peers. The unicast 
and multicast communications within a group can be encrypted;

\item \textit{Group propagation.} A message can be propagated within a group even if the network level does not support 
multicast datagram propagation;

\item \textit{Distributed and fault-tolerant service discovery.} Unlike SOAP, CoAP and REST, which usually use a centralized server
for service discovery (UDDI), EmbJXTAChord  redundantly distributes all the group information in a set 
of \textit{rendezvous peers}.  This strategy, already adopted by JXTA, 
ensures high fault tolerance and allows to mantain the network consistency even under a high churn-rate  
(i.e. the rate of peers joining and leaving the group \cite{JXTA08}).   
\end{itemize}

The main engine of EmbJXTAChord is divided into 6 modules that reimplement the functionalities of the six protocols
of JXTA 2.7. Thus, these modules mantain the same names of the corresponding ones
in JXTA 2.7 (see Tab. \ref{AddedTable_I} for the functionalities). For instance, the protocol that manages the resource
discovery in EmbJXTAChord is named PDP (Peer Discovery Protocol) as in JXTA 2.7. For sake of
clearness, we will refer to these components as \textit{JXTA protocols} both in EmbJXTAChord and in JXTA 2.7 cases. 
Moreover, EmbJXTAChord mantains the message structure  of JXTA 2.7 (divided in MessageElements,
see Sect. \ref{Hyper00}). For this reason, in the rest of the paper we will refer to \textit{JXTA messages} both 
in EmbJXTAChord and in JXTA 2.7 cases.  

\subsection{Differences between EmbJXTAChord and JXTA}

 There are at least three differences between EmbJXTAChord and JXTA:

\begin{itemize}
 \item \textit{A new API, named SJXTA} (simplified JXTA) (see Sect. \ref{Sec:EmbJXTAChord}). One of the main
limitation of JXTA is its very complex Application Programming Interface (API). The SJXTA API manages 
several operations (peer, group or service discovery, pipe opening, creating or joining a new group), that are 
performed by the EmbJXTAChord engine providing few informations, thus simplifying the developer's job;  

 \item \textit{The new Hypercompression algorithm} (see Sect. \ref{Hyper00}). A second important drawback of JXTA
is its high bandwidth consumption. JXTA messages are encoded as bandwidth-consuming xmldocs, JXTA peer, peergroup, and 
pipe IDs are encoded as strings. Most items are redundant, i.e., they are repeated several times in the same
message or in multiple consecutive messages. The algorithm used to manage rendezvous peers  (named \textit{loosely-consistent 
Distributed Hash Table limited range walker})(lcDHT-LRW) \cite{JXTA08} and the source routing adopted for delivering messages over different 
subnets are bandwidth-expensive \cite{SrcRoute00}. EmbJXTAChord uses the Hypercompression algorithm, 
that reduces the bandwidth required for message transmission, working together with a Rendezvous protocol (see Sect. \ref{RendezvousProto}),
based on the more bandwidth-efficient Chord \cite{Chord00} algorithm; 
 \item \textit{The new jxCOAP-E component}.  JxCOAP-E allows the communication between server and client peers through a standardized 
(and well-known) RESTful interaction model. Moreover jxCOAP-E allows to port, in a very simple way, the applications originally written for TCP-IP homogeneous networks 
into the EmbJXTAChord hybrid environment. JxCOAP-E overcomes another important limitation of JXTA, i.e. the lack of 
a service interaction model. In fact, JXTA specifications include  an underdeveloped concept of \textit{jxta-service}, but 
nothing is said about \textit{service interaction}. This raises the need for writing application-specific source code \cite{JXTA09}. 

\end{itemize}
\section{Related work}\label{Sec:RelatedWorks}

In the last years, researchers proposed several solutions aimed to create a service
architecture over a heterogeneous network. The most interesting approaches
were the use of P2P protocols, modular middleware solutions and middleware that implements 
the Web of Things (WoT) paradigm. 

Moreover, other recent advancements in Bluetooth
Smart, Vehicular and LTE technologies can be used in the proposed
solutions, in order to integrate also this kind of subnetworks.  

About the use of P2P protocols for IoT applications, in \cite{XMPP01} XMPP (Extensible Messaging and 
Presence Protocol) was proposed as a solution for implementing a heterogeneous 
network without the need for a gateway. XMPP supports multiple transfer
protocols via extension modules, but it is unsuitable for narrowband 
networks,  as it exploits a protocol for the transmission
of binary arrays that is based on the bandwidth-greedy Base64 encoding scheme. 
Conversely, EmbJXTAChord supports a binary transfer mode that allows to optimize the bandwidth 
occupation. 

Despite a theoretical IoT model based on JXTA was presented in \cite{JXTA13}, 
there are only few works that exploit JXTA for IoT applications. 
\textit{JxSensor} is a project aimed to integrate JXTA in a WSN \cite{JXTA01}. A set of sensors is linked to a sink, managed as a virtual peer, which can be addressed using 
the common JXTA methods. JxSensor is a translation gateway 
which runs on the sink that is placed between the JXTA network and the WSN. 
The JXTA requests (responses) are delivered by the computer to (from) the gateway, where a component, named MoteAdapter, translates them into a set of
commands that can be interpreted by the WSN sensors and actuators. Only the commands for the actuators and the data gathered by the sensors are 
transmitted through the narrowband link, not the standard JXTA messages. Therefore, the sensors and actuators
can exchange data with the rest of the JXTA network through the translation gateway, but they cannot use all the JXTA functionalities.

An early version of jxCOAP-E based on JXTA 2.7, named \textit{jxCOAP}, was presented in \cite{JXCOAP00}, as a component of the Javascript runtime container \textit{jxActinium}. 
However, jxActinium is not devised for narrowband WSN. There are multiple reasons for this, such as, a wasteful use of bandwidth.
no support for Bluetooth and 6LoWPAN and the use of the old loosely-consistent 
DHT limited range walker (lcDHT-LRW) algorithm for the Rendezvous Protocol. 
In Sect. \ref{Subsec:jxCOAP} a comparison between jxCOAP-E (the version in EmbJXTAChord) and jxCOAP (the jxActinium version in \cite{JXCOAP00}) 
is provided for the sake of completeness.

\bigskip

In a modular middleware sensors and actuators  
are driven and controlled by the main server through specific drivers loaded from a cloud. The main server 
also converts, before transmission, the directives of the middleware into a set of short commands in a format 
understandable by the sensors or actuators.
For example, SenseWrap exploits modules named Virtual Sensors in order
to communicate 
to sensors and actuators \cite{IOTMidd00}.  
The Virtual Sensors hide 
the hardware and the network protocol that are effectively used
for communication, performing protocol conversion and exposing to the clients a uniform
interface based on TCP/UDP-IP. 
The configuration is realized through the Zeroconf protocol.  
Unfortunately, the work in \cite{IOTMidd00} does not provide information about the security policies 
for the middleware. Conversely, EmbJXTAChord can directly work over other network protocols 
than IP, without any message encapsulation or conversion into IP packets, and does not need Zeroconf 
for autoconfiguration. 

The solution proposed in \cite{IOTMidd01}, named Smart Home Gateway, exploits a central server
running a modular system based on OSGi (Open Service Gateway initiative), in order to manage the whole set of connected devices. 
When a new device is connected to the home network, the smart gateway automatically loads from a server
in the cloud the correct driver (such as a new OSGi component) for the related Controller Device. 
A Controller Device is a modem able to communicate to a subset of devices exploiting a specific 
communication protocol (e.g., X10, Insteon, ZigBee).   
The work assumes that each Controller Device can be connected directly to the smart gateway, 
as no support is provided for hop-by-hop delivery between subnetworks. As a consequence, a failure in 
the smart gateway immediately affects the whole system. Furthermore, the smart gateway manages the whole 
security protocol.
Conversely, EmbJXTAChord supports hop-by-hop delivery between subnetworks based on
different network protocols and also allows to create smart environments in which a rendezvous  
peer failure does not determine the failure of the whole system. 

Moreover, unlike IP-based solutions such as Smart Home Gateway or SenseWrap, EmbJXTAChord
provides support for both unicast and multicast secure communications, also implementing
several access levels through peergroups. 

\bigskip

Web-of-Things is a paradigm that uses the architecture, the protocols and the services used in the Web in order to discover, 
manage and integrate smart objects into the global Internet \cite{SOAPWS03}. In the WoT vision, sensors and actuators should be managed using
a simple Web browser. In the last years, many frameworks were presented in this field, aimed to efficiently interconnect devices 
belonging to different subnetworks.
In general, each of these frameworks includes a web server, that runs one or more 
web services representing the state of sensors and actuators by means of a RESTful API, 
a supervisor that runs an execution engine, thus elaborating the gathered data,
 and one or more \textit{smart gateways}
that interface sensors and actuators using a lightweight protocol.

\bigskip

The most used protocols for WoT communication are CoAP \cite{COAP01}, MQTT\cite{IP00_Repl_00}\cite{IP00_Repl_02}, 
SMQ\cite{SMQ01} \cite{SMQ02} and ActiveMQ \cite{ActiveMQ00} (see Tab. \ref{AddedTable_III}). In the development of EmbJXTAChord, we have chosen to create
the jxCoAP-E component (i.e. CoAP over JXTA) because CoAP offers several advantages over other IoT protocols.
For instance, comparing with the Message Queue Telemetry Transport (MQTT) \cite{IP00_Repl_02}\cite{IP00_Repl_00},  
Simple Message Queries (SMQ) \cite{SMQ01} \cite{SMQ02} and ActiveMQ \cite{ActiveMQ00} protocols,  CoAP provides a request/response communication model 
that does not require a central broker. 

However, CoAP, SMQ and MQTT are based on IP, so they inherit all the limitations about connectivity in the heterogeneous 
networks described in Sect. \ref{Sec:Introduction}. As a result, differently from jxCOAP-E, they cannot leverage 
on features such as routing over subnetworks, peergroups or secure multicast connections.
Moreover, the central broker used by MQTT is a potential point of failure and it can become 
overloaded when too nodes are active at the same time. Conversely, jxCOAP-E does not require a centralized broker 
and allows to use simultaneously more servers on different peers within a peergroup.

\newcommand{\TABzc} 
{
	\begin{tabular}{|l|l|l|l|l|l|l|}%
		\hline 
				       &jxCOAP-E          &CoAP       &MQTT &SMQ &ActiveMQ\\
		\hline
		  Supported transport  &TCP,UDP,HTTP,     &UDP        &TCP  &TCP &TCP, UDP,\\
		  protocols            &RFCOMM,SICS       &IP         &IP   &IP  &HTTP. XMPP\\
				       &                  &           &     &    &Websockets\\
		\hline
		  Secure transport     &TLS over JXTA     &DTLS       &TLS  &TLS &HTTPS,\\
		  protocols for unicast&MTBs (TCP,HTTP,   &           &     &    &TLS   \\
		  communications       &RFCOMM,SICS)      &           &     &    &Secure\\
				       &		  &           &     &    &WebSockets\\
		\hline
		  Support for secure   &AES group         &-          &-    &-   &- \\
		  nested peergroups    &encryption        &           &     &    &  \\
		\hline
		  Communication	       &Request-Response  &Request-Response &Publish-Subscribe &Request-Response &Publish-Subscribe\\
		  model		       &Publish-Subscribe &Publish-Subscribe&                  &Publish-Subscribe&\\
		\hline
		  One-to-one           &Supported         &Supported  &-    &Supported&-\\
		  communications       &                  &           &     &         &\\
		\hline
		  Multicast	       &Supported         &Only if multicast          &-    &-    &Supported\\ 		
		  communications       &                  &UDP available \cite{COAP04}&     &     &         \\
		\hline
		  Secure multicast     &Supported (AES    &-          &-    &-               &Supported\\ 		
		  communications       &group encryption) &           &     &                &\\
	        \hline
		  Central broker       &No                &No         &Yes  &Yes             &Yes\\
		  required             &                  &           &     &                &\\
		\hline
		  Distributed	       &Supported         &-          &-    &-               &Yes\\
		  service discovery    &		  &           &     &                &\\
		\hline
		  Routing over         &Supported         &-          &-    &-               &-\\
		  subnetworks	       &                  &           &     &                &\\
		\hline
		  In-band	       &Supported         &-          &-    &Only topic      &-\\
		  compression	       &                  &           &     &names           &\\
		\hline
		  NAT	               &Supported         &-          &-    &Supported       &Supported\\
		  traversal	       &(via JXTA PRP)    &           &     &(via WebSockets)&(via WebSockets)\\
		\hline
	\end{tabular}
}
\ifdefined\ACM
\begin{table}
	\tbl{}
	{      
		\TABzc{Comparison between jxCOAP-E and other IoT communication protocols.}
	}
	\label{AddedTable_III}
\end{table}
\else
\begin{table}[t]
	\setlength{\tabcolsep}{.16em}
	\scriptsize
	{
		\caption{\scriptsize {Comparison between jxCOAP-E and other IoT communication protocols.} }
		\vspace*{-0.1in}
		\hspace*{-0.3in}
		\begin{center}
			\TABzc{}
		\end{center}\label{AddedTable_III}
	}
	\vspace*{-0.2in}
\end{table}
\fi

\bigskip 

Node-RED \cite{FRED03}\cite{FRED02} and WoTKit processor \cite{WotKit01} are WoT platforms
that were developed to efficiently interconnect devices belonging to different subnetworks (see Tab. \ref{AddedTable_IV}).
They provide  an execution engine that runs programs where software modules and devices are represented 
as entities connected by wires. WoTKit provides the Dashboard, a browser-based interface aimed to manage and run the programs. 
FRED (a Frontend for Node-RED) \cite{FRED01} is a commercial version of Node-RED (produced by SenseTecnic) that supports the simultaneous 
execution of multiple flows in a multiuser environment.

EVRYTHNG \cite{EVRYTHNG00} is a commercial platform (provided by the Evrythng company) 
aimed to integrate whatever physical object in the cloud. 
Evrythng associates each physical object to a Web Object that allows to remotely control the item
using a RESTful interface.
Evrythng provides also the THNGHUB gateway \cite{EVRYTHNG02}, that allows to 
integrate into the cloud non-IP based devices (such as Bluetooth or ZigBee sensors). 

The WoTKit, Node-RED, FRED and THNGHUB gateways support several transport protocols for the communication with
end-devices (see Tab. \ref{AddedTable_IV}). However, gateways send the gathered data to the execution engine 
using IP-based protocols (such as MQTT or WebSockets). As a consequence, these frameworks are 
affected by all the limitations already described in Sect. \ref{Sec:Introduction} when they are used in a multi-hop heterogeneous network
or when the support for nested secure peergroups is necessary. 
Moreover EVRYTHNG and THNGHUB are fully dependent on a connection to the servers mantained by the Evrythng company.
Conversely, EmbJXTAChord supports non-IP transport protocols but, differently from THNGHUB, it is able to create 
independent cloud architectures.

\bigskip


\newcommand{\TABzd} 
{
	\begin{tabular}{|l|l|l|l|l|l|l|l|l|}%
		\hline 
			               &EmbJXTA    &WoTKit       &Node-RED     &FRED      &Evrythng   &SenseWrap  &Smart Home &HPS   \\ 
			               &Chord      &\cite{WotKit01}&\cite{FRED03}&\cite{FRED01}&\cite{EVRYTHNG00}&\cite{IOTMidd00}&Gateway\cite{IOTMidd01}&\cite{Bluetooth11}\\
		\hline
		  Supervisor           &Required   &ActiveMQ   &MQTT       &MQTT       &Cloud      &SenseWrap   &Central OSGi&HTTP Proxy\\
		  Entity               &only       &broker     &broker     &broker     &managed    &server      &server      &Server\\
		                       &to deploy  &           &           &           &by the     &            &            &      \\
				       &peergroups &           &           &           &company    &            &            &      \\ 
		\hline
		  Embedded devices     &RaspPI     &RaspPI     &RaspPI     &RaspPI     &Every dev. &Every dev.   &-            &Every dev.\\
		  that can be used     &   	   &Arduino    &Arduino    &Arduino    &able to run&able to run  &             &able to run\\
		  as gateway nodes     &           &           &           &           &ThngHub    &Virt. Sensors&             &BLE stack\\ 
		\hline
		  Protocols/           &BT,ZigBee  &BT,ZigBee, &BT,ZigBee, &BT,ZigBee  &BT,ZigBee, &BT,ZigBee   &X10,ZigBee,  &BLE  \\
		  interfaces           &Wi-Fi      &Wi-Fi,     &Wi-Fi,     &Wi-Fi,     &Wi-Fi,     &Wi-Fi       &Insteon,     &GATT/ATT   \\
		  for end-devices      &RaspPI     &           &ZWave,     &ZWave,     &ETH        &Sun SPOT    &uPNP         &      \\
		  supported            &GPIO       &           &RaspPI     &RaspPI     &           &            &             &      \\
				       &           &           &GPIO       &GPIO       &           &            &             &      \\
		\hline
		  Communication        &JXTA       &ActiveMQ   &MQTT       &MQTT       &MQTT,      &Proprietary &-(OSGi server&GATT to    \\
		  protocols            &(TCP,TLS,  &           &(TCP/TLS)  &(TCP/TLS)  &HTTP,      &(over TCP)  &is directly  &HTTP       \\
		  (gateway nodes-      &BT,SICS,   &           &WebSockets &WebSockets  &TCP/TLS,  &            &connected    &translation\\
		  to-supervisor)       &HTTP tunn.)&           &           &           &WebSockets &            &to sensors)  &           \\
		\hline
		  Communication        &JXTA       &HTTP       &MQTT       &MQTT       &HTTP,      &Proprietary &HTTP,        &HTTP,      \\
		  protocols 	       &(TCP/TLS,  &           &(TCP/TLS)  &(TCP/TLS)  &HTTPS      &(over TCP)  &HTTPS        &HTTPS      \\
		  (supervisor-         &BT,SICS,   &           &WebSocket  &WebSocket  &REST       &            &REST         &REST       \\
		  to-user)             &HTTP tunn.)&           &HTTP       &HTTP       &           &            &             &           \\
		\hline
		  Publish/Subscribe    &Supported  &Supported  &Supported  &Supported  &Supported  &Supported   &Supported    &Supported\\
		  communication	       &(jxCOAP-E) &(ActiveMQ) &(MQTT)     &(MQTT)     &(MQTT,     &(proprietary&(via OSGi    &(GATT    \\
		  model		       &           &           &           &           &CoAP,      &protocol)   &Event        &notify   \\
				       &           &           &           &           &WebSocket, &            &Manager)     &followed by\\
				       &           &           &           &           &HTTP)      &            &             &HTTP     \\
				       &           &           &           &           &           &            &             &POST)    \\
                \hline
		  Multiple users       &Supported  &Supported  &Not        &Supported  &Supported  &No          &Engine in    &No        \\
		  on a single          &           &           &supported  &           &           &Execution   &OSGi server  &Execution \\
		  execution server     &           &           &           &           &           &Engine      &supports     &Engine    \\
				       &           &           &           &           &           &provided    &all users    &provided  \\
		\hline
		  Secure nested        &Supported  &-          &-          &-          &-          &-           &-            &-     \\
		  peergroups           &           &           &           &           &           &            &             &      \\
		\hline
		  Every peer can       &Automatic  &-          &-          &-          &-          &-           &-            &-     \\
		  become a gateway     &           &           &           &           &           &            &             &      \\
		\hline
		  Unicast comm.        &Supported  &-          &Supported  &Supported  &-          &-           &-            &-     \\
		  between              &(via       &           &           &           &           &            &             &      \\
		  end-devices  	       &JXTA PBP)  &           &           &           &           &            &             &      \\
		\hline
		  Multicast	       &Supported  &-          &-          &-          &-          &-           &-            &-     \\ 		
		  communications       &(via       &           &           &           &           &            &             &      \\
				       &JXTA RP)   &           &           &           &           &            &             &      \\
		\hline
		  Secure multicast     &Supported  &-          &-          &-          &-          &-           &-            &-     \\ 		
		  communications       &(via Group &           &           &           &           &            &             &      \\
				       &Encryption)&           &           &           &           &            &             &      \\
	        \hline
		  Service	       &Distributed&Centralized&Centralized&Centralized &Hubs discov&Virtual   &OSGi driver  &Centralized\\
		  discovery            &(jxCOAP-E  &           &           &            &locally    &Sensors   &registers in &(Discovery \\
				       &servers    &           &           &            &via mcast. &discoved  &the Central  &performed  \\
				       &discovered &           &           &            &Connect. to&using     &Server the   &by the  \\
				       &via        &           &           &            &cloud req. &ZeroConf  &new devices  &GATT    \\   
				       &JXTA PDP)  &           &           &            &for auth.  &          &             &central \\ 
				       &           &           &           &            &and sync   &          &             &node)   \\ 
		\hline
		  Routing over         &Supported  &-          &-          &-          &-          &-           &-            &-     \\
		  subnetworks	       &           &           &           &           &           &            &             &      \\
		\hline
		  In-band	       &Supported  &-          &-          &-          &-          &-           &-            &-     \\
		  compression	       &           &           &           &           &           &            &             &      \\
		\hline
		  NAT traversal	       &Supported  &-          &Only via   &Only via   &-          &-           &-            &-     \\
		  in unicast           &           &           &WebSockets &WebSockets  &           &            &             &      \\
		  communication.       &           &           &           &           &           &            &             &      \\
	        \hline
		  Platform             &-          &Dashboard  &Dashboard  &Dashboard  &Dashboard  &-           &WebServices  &WebServices \\	     
		  accessible           &           &           &           &           &           &            &RESTful      &RESTful \\
                  via web browser      &           &           &           &           &           &            &API          &API\\
		\hline
	\end{tabular}
}
\ifdefined\ACM
\begin{table}
	\hspace{-0.6in}
	\tbl{}
	{      
		\TABzd{Comparison between EmbJXTAChord and some alternative IoT middleware solutions.}
	}
	\label{AddedTable_III}
\end{table}
\else
\begin{table}[t]
	\setlength{\tabcolsep}{.16em}
        \tiny
	{
		\caption{\scriptsize {Comparison between EmbJXTAChord and some alternative IoT middleware solutions.} }
		\begin{center}
			\TABzd{}
		\end{center}\label{AddedTable_IV}
	}
	\hspace*{-0.5in}
	
\end{table}
\fi

\begin{figure}[t]
  \addtolength{\abovecaptionskip}{-2.0in}
  \vspace*{-0.23in}
  \hspace*{-0.4in}
  \subfigure
  {
    \includegraphics[width=5.50in, height=1.75in]{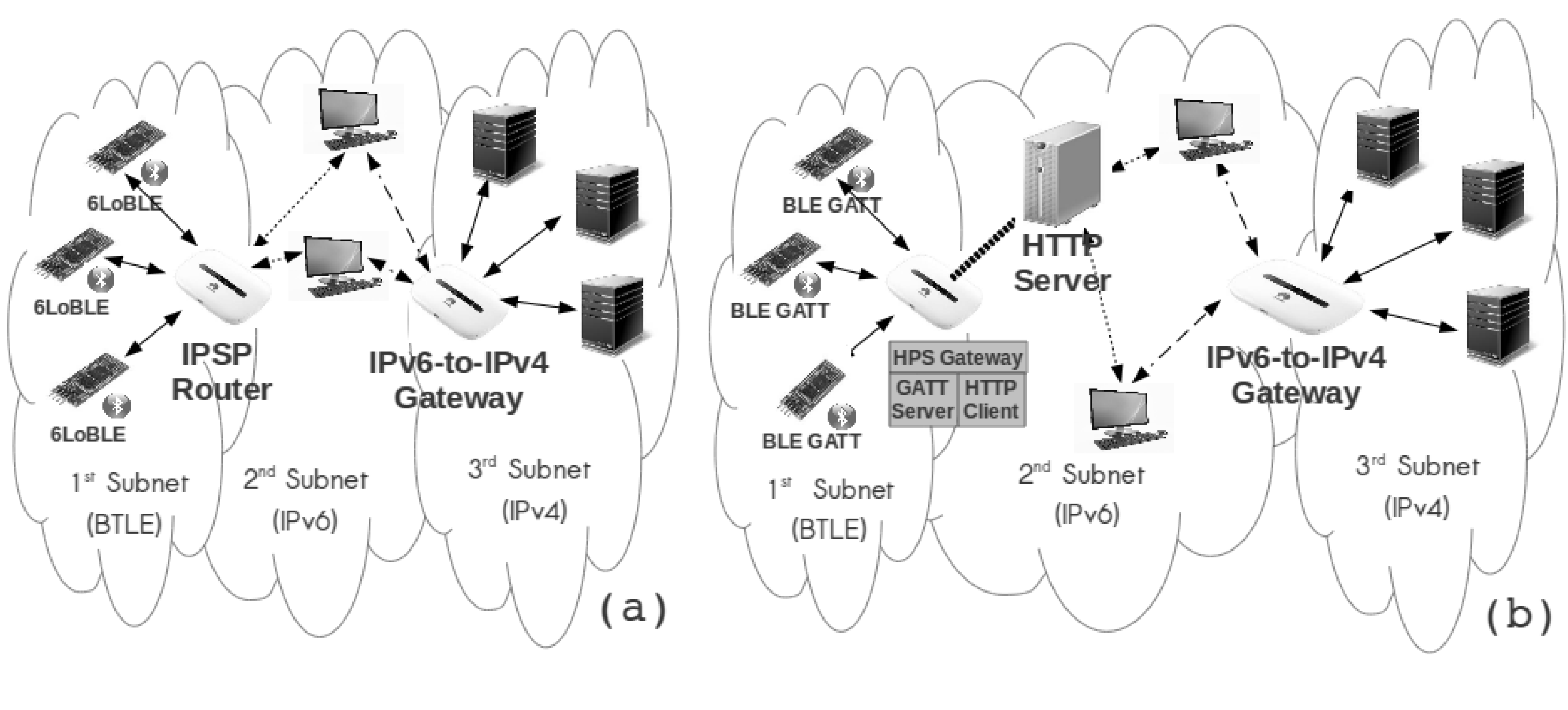} }
    \caption{Representation of a heterogeneous network that includes BLE nodes (a) using the IPSP Router
and 6LoWPAN over BLE (b) using the HPS gateway.}
    \vspace*{-0.12in}
\label{BLE_Scenario}
\end{figure}

About the Bluetooth hybrid networks, in 2010 the Bluetooth Special Interest Group (SIG) released the version 4.0 of the BT specifications,
thus defining a new standard named Bluetooth Low Energy (BLE, also known as Bluetooth Smart). 
From v4.2, Bluetooth Smart introduced new features such as HPS (HTTP Proxy Service) \cite{Bluetooth11}
and IPSP (Internet Protocol Support Profile) \cite{Bluetooth10}\cite{Bluetooth12}. The HPS gateways provide to HTTP-to-GATT 
translation, thus allowing to read/write some features of the BLE sensors (named GATT \textit{Characteristics})
using a RESTful interface. IPSP \cite{Bluetooth10}\cite{Bluetooth12} allows to implement a 6LoWPAN network over BLE (6LoBLE). A BLE node
(named \textit{Router}) provides to reroute the 6LoWPAN packets coming from the other BLE nodes
to the external (non-Bluetooth) subnetwork.

Unfortunately, IPSP supports only IPv6. Moreover, no support for routing over 
subnetworks is provided. As a consequence, both IPSP and HPS are unable to overcome 
the issues described in Sect. \ref{Sec:Introduction}.
For instance, in Fig. \ref{BLE_Scenario} a peer in the third subnetwork (IPv4-based) 
could not communicate to any BLE device in the first subnetwork.  The only way to overcome
this issue would be to install an IPv4/IPv6 gateway between the second and the third subnetwork 
and to reconfigure the routing tables in order to deliver the packets to/from the 
IPSP or HPS gateway. EmbJXTAChord, instead, is able to replace the functionalities of an 
IPv4/IPv6 gateway thus routing the messages between every node of the three subnetworks 
without these reconfiguration steps. 

It is important to point out that EmbJXTAChord could work also over 6LoBLE, 
using the Message Transport Binding for IPv6. However, this solution would not be ubiquitous, because IPSP can be used 
only if it is supported by the operating system (for instance, IPSP is not supported by the Microsoft operating systems 
older than Windows 10 \cite{Bluetooth13} and by the versions of BlueZ (Linux) older than v5.0). 
In order to measure the performance of EmbJXTAChord using BLE, we used the
RaspberryPI-3 that integrates a chipset that is compliant with the v4.1 of 
the Bluetooth standard \cite{Broadcom1}. The results are shown in Sect. \ref{Sec:ExpResult}.

\bigskip

Other interesting enhancements for hybrid networks are related to the V2X (Vehicle-to-Everything)
paradigm. The communication between vehicles and roadside units (RSUs) can be realized
using DSRC (Dedicated Short Range Communication, a wireless technology based on the WAVE
standard developed by the IEEE 1609 WG) \cite{VANET00}, or LTE cellular 
technology \cite{VANET08}.  
3GPP Release 12 introduced new specifications for low-power devices with low bandwidth requirements,
such as \textit{LTE-Direct} (also known as \textit{ProSe}) \cite{VANET04} \cite{VANET07}, aimed to promote LTE as a general 
solution for M2M communications. LTE-Direct defines procedures for device discovery 
and communication, thus allowing to group several nearby mobile phones into a cluster, 
controlled by a \textit{clusterhead} (i.e. a  mobile phone acting as a master). This way, LTE
communications can work even when the base station is not available \cite{VANET04}.  
3GPP Release 14 allows to use LTE-Direct as alternative for DSRC, or to connect the mobile phone 
with the vehicle (C-V2X, cellular V2X paradigm) \cite{VANET09}.  

EmbJXTAChord is not a competitor for LTE and DSRC, as it works at the application layer, 
while the previously cited technologies mainly work at the PHY and MAC layers. 
Rather, EmbJXTAChord is a framework that can be advantageously integrated 
with LTE and V2X technologies. 

As LTE-Direct is interoperable with Wi-Fi Direct, but not with
BLE or IEEE 802.15.4, EmbJXTAChord can be used to integrate 
this kind of devices. 
Moreover LTE clusters cannot be nested, while EmbJXTAChord supports
secure nested peergroups, thus allowing the transmission of a
message only to a subset of the nodes in the LTE network. 

EmbJXTAChord needs only the name of a peer to establish
a connection, regardless of the change of the assigned IP address,
that is a frequently occurring event in DSRC vehicular networks. 

Finally, LTE and DSRC do not define any
standard solution for content discovery and caching, despite it was observed that
these functionalities can be advantageously used in vehicular networks \cite{VANET08}. 
Conversely, EmbJXTAChord makes available for the applications 
its own technology based on the Peer Discovery and Rendezvous protocols.

\section{EmbJXTAChord features}\label{Sec:EmbJXTAChord}
\begin{figure}
\addtolength{\abovecaptionskip}{-0.6in}
\centering
\includegraphics[width=3.9in, height=2.85in]{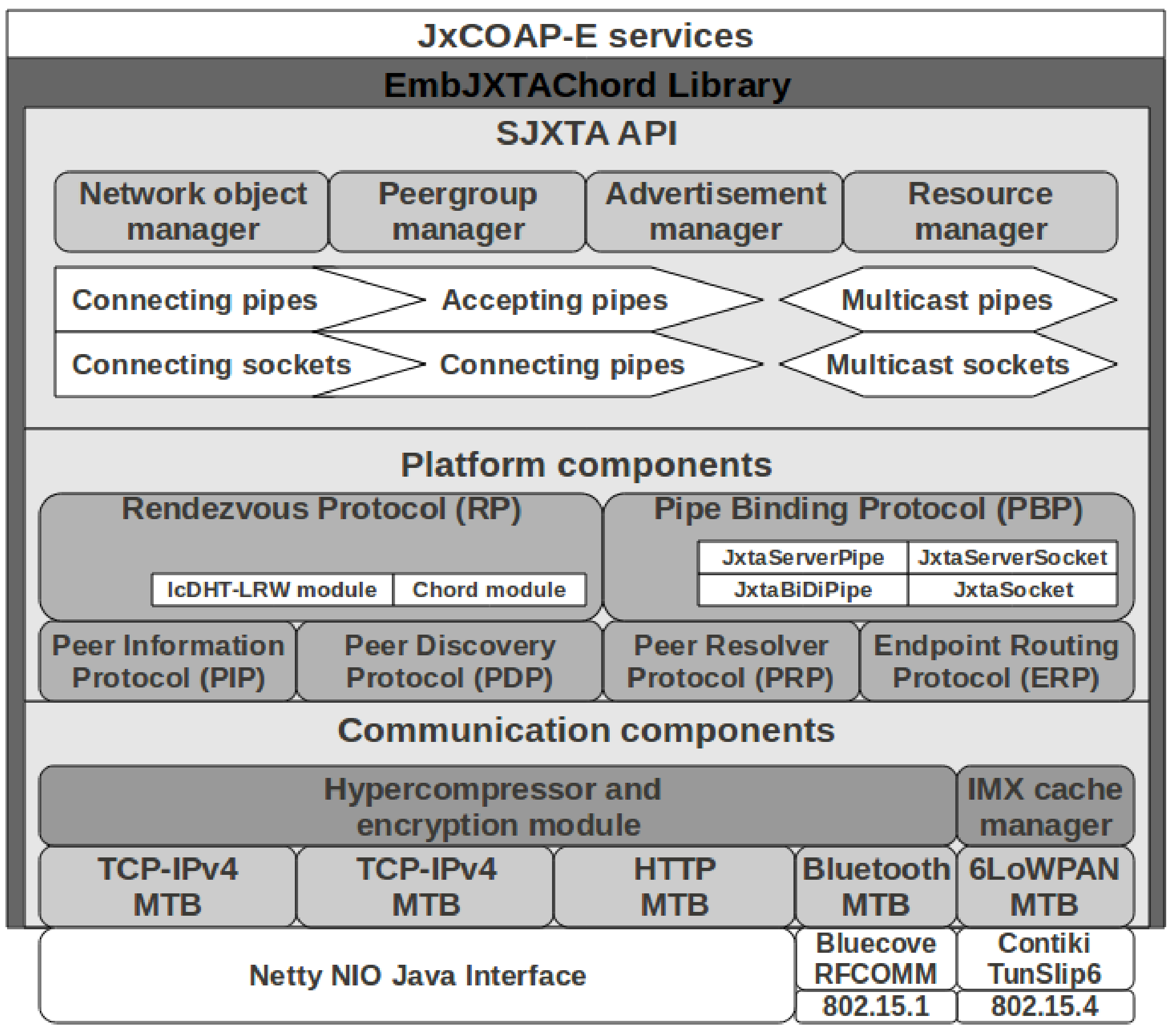}
\caption{The EmbJXTAChord architecture.}
\vspace*{-0.1in}
\label{Fig_Arch}
\end{figure}

Fig. \ref{Fig_Arch} shows the EmbJXTAChord architecture. 
EmbJXTAChord provides a new API, named SJXTA (Simplified JXTA), aimed to simplify the application development. 
The idea is that the developers should provide only the minimum set of information needed for a given operation. 
The following new concepts are introduced:

\begin{itemize}
\item \textit{The network objects} (netobj). These are Java objects containing
references to all the service instances and to all the configuration
parameters of a group. When starting a new JXTA session, the developer initializes a standard
netobj related to the JXTA NetPeerGroup \cite{JXTA09}. The only parameters
required are the \textit{PeerName}, an optional \textit{PeerProperties} string, and a 64b mapped 
\textit{Options} parameter. 
Each group is associated to a netobj, therefore SJXTA automatically provides a new child netobj each time a new child 
group is created or is joined by a peer. Using the netobj methods, the developer can read and configure
all the pieces of information about the current JXTA instance (such as peername, PeerID etc.), and can access
to all the resources available in the group;

\item \textit{Advertisement generation}. Whenever a peer creates a new resource (peergroup, peer, pipe, 
socket or service), a new advertisement is generated and published. In JXTA 2.7, these complex operations are 
delegated to the developer. Conversely, in EmbJXTAChord the advertisements related to the
new built resources are automatically generated and published by the SJXTA level; 

\item \textit{Connecting and accepting pipes, connecting and accepting sockets}. In SJXTA sockets and pipes are addressable 
through their names, therefore there is no need for looking up the relevant advertisement  within the group. When it is necessary to create
a virtual channel allowing the connection by other peers, the developer instantiates a new \textit{AcceptingSocket} (that provides
an \ttfamily accept() \rmfamily method). The only parameters required for this operation are the netobj of the related
peergroup, the \textit{SocketName} and (optionally) a \textit{SocketProperties} string.  
At the remote side, when it is necessary to access an accepting socket, whose name is known in the group, the developer creates a 
\textit{ConnectingSocket} object and calls its \ttfamily connect(SocketName) \rmfamily method. SJXTA looks for an 
advertisement  with the given \textit{SocketName} and then establishes the connection. 
SJXTA sockets support only synchronous data transmission.  
Conversely, accepting and connecting pipes also support asynchronous data transfer (i.e. through a \textit{listener});
\item \textit{Multicast pipes and multicast sockets}. These objects allows to propagate data towards all peers of the group;
\item \textit{Resource discovery}. Any EmbJXTAChord resource is addressed by name, as the SJXTA
level transparently provides to look up the related advertisement. Moreover 
SJXTA provides methods for \textit{resource discovery} that return a name list of the resources that are compliant with a set of
requirements. The developer can run the look up operations using a first filter on the resource type (\textit{peer}, \textit{group} or 
\textit{pipe}), a second filter on a regular expression on its name, and a third filter on its properties. 
\end{itemize} 

\subsection{The new Message Transport Binding modules}\label{MessageTransportBindings}

While JXTA 2.7 only supports TCP-IP, UDP-IP or HTTP-tunnelling connections, EmbJXTAChord introduces two new Message Transport Binding (MTB)
modules, aimed to support Bluetooth and IEEE 802.15.4 WSN networks. 

The Bluetooth Message Transport Binding exploits Bluecove \cite{Bluetooth01}, a Java development library implementing JSR-82 specifications \cite{Bluetooth02}\cite{BNEP01} 
that supports unicast connections to BT devices through a wide range of protocol
stacks: Microsoft, Widcomm or Bluesoleil (under Windows) and BlueZ (under Linux) \cite{Bluetooth03}.  
The mandatory protocol RFCOMM (serial port emulation) is used, thus supporting up to 60 simultaneous connections
and ensuring the widest compatibility \footnote{ 
Despite Bluetooth might be supported also 
using BNEP (network emulation), this is not a general solution, as some operating systems, such as Windows XP, do not support the piconet 
Group Ad-hoc Network  role \cite{Bluetooth00} or multiple simultaneous connections.}.  
Each device is addressable through its BT name or 96b address (\ttfamily btspp://00228372FFC0 \rmfamily
is an example of JXTA EndpointAddress for a BT device).

The SICSLOWPAN Message Transport Binding was designed for supporting WSNs based on the IEEE 802.15.4 standard. 
It is used, for instance, by the ZigBee protocol, but since many available ZigBee stacks are commercial, API-incompatible or with unmantained sources 
(such as FreakZ or Open-ZB \cite{ZigBee04}), here a workaround was adopted.  
If the target EndpointAddress is of the \ttfamily sicslowpan://[ipv6addr] \rmfamily type, EmbJXTAChord establishes a connection
creating an IPv6 tunnel to the serial port of the local
mote. This is a device, equipped with a microcontroller and an IEEE 802.15.4-compliant transmitter, that runs a bridge application based 
on Contiki (an operating system for tiny devices that supports RPL routing and 6LoWPAN transport protocols) 
\cite{Contiki01_Repl00} \cite{RPL01} \cite{RPL02}.  
This solution ensures a wide hardware support, as any Contiki-compatible mote can be used.
SICSLOWPAN MTB configures the mote at startup, requesting its link-local IPv6
address, which is registered in the peer advertisement  in order to allow communication exchanges with
other nodes of the WSN. Moreover, the MTB runs an algorithm for segmentation and
reassembly that allows for multicast propagation of JXTA messages over the WSN,
thus overcoming the limitation on the maximum size of the IP packets in 6LoWPAN
(i.e., not larger than 1280B \cite{6LoWPAN04}).

\subsection{The rendezvous protocol}\label{RendezvousProto}

\begin{figure}[t]
	\addtolength{\abovecaptionskip}{-1.5in}
	\hspace*{-0.62in}
	\vspace*{-0.16in}
	\subfigure { \label{ChordRdv:a}\includegraphics[width=5.9in, height=3.4in]{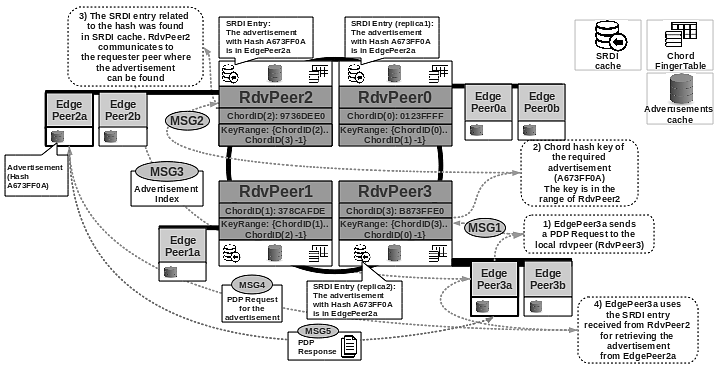} }
	\caption{A representation of advertisement discovery using Chord RP protocol.}
	\label{ChordRdv:merge}
\end{figure}

EmbJXTAChord replaces the bandwidth-greedy lcDHT-LRW rendezvous  protocol with an alternative implementation 
based on Chord \cite{Chord00}, an algorithm based on the theory of consistent hashing. 
This solution, which was successfully tested in JXTACh \cite{JXTAChord00}, is able to effectively 
improve the look up time of the advertisements and the bandwidth usage, as it entails less network traffic.
Unlike lcDHT-LRW, Chord does not require that the local Rendezvous PeerView (RPV)  
(that in Chord is named \textit{fingertable}) be exchanged between rendezvous  nodes when no peer joins 
or leaves the group
\footnote{Under JXTA 2.7 each rdvpeer selects, every $T_{LRW}$ seconds, a \textit{random sized} subset of the other rendezvous 
nodes (randomly chosen) where to send its own copy of the current RPV (RPV-exchange). This process is aimed to obtain the convergence
of the RPV tables of all the rdvpeers in the group (see \cite{JXTA08} for further details).}
(only small messages for \textit{fingertable stabilization} and \textit{predecessor checking} are 
periodically exchanged \cite{Chord00}).

Fig.\ref{ChordRdv:merge} shows the advertisement discovery process using the Chord Rendezvous Protocol. In EmbJXTAChord, 
each advertisement is characterized by a 128b hash value (key), but here in Fig.\ref{ChordRdv:merge} and in the case described in
this Section, for a simpler representation, we assume that 32b hash values are used.   

Each peer (rendezvous  or edge) manages a local cache containing the advertisements recently discovered/published and whose
lifetime is not yet expired. In Fig. \ref{ChordRdv:merge} each rdvpeer is characterized by:

\begin{itemize}
 \item a 32b \textit{Chord identifier} (128b IDs are used in the real implementation) that is computed when the peer joins 
the group by hashing the peername and the endpoint address;
 \item a \textit{key range}, that defines the range of 32b keys (128b in the real implementation) for which the rdvpeer is
responsible in Chord circular domain. 
\end{itemize}

Each rdvpeer is connected to one or more edgepeers (for instance, in Fig.\ref{ChordRdv:merge} EdgePeer2a and EdgePeer2b
are linked to RdvPeer2). 

When an edgepeer (\textit{publishing edgepeer}) needs to publish a new advertisement, the following steps are performed:

\begin{itemize}
 \item The publishing edgepeer stores the advertisement in its own local cache and sends a PDP query to the local rdvpeer (\textit{publishing rdvpeer});
 \item The publishing rdvpeer computes the hash key of the new advertisement and performs a Chord look-up operation \cite{Chord00} aimed to determine the 
\textit{target rdvpeer}, i.e. the peer responsible for the hash key. For instance, in the
scenario shown in Fig.\ref{ChordRdv:merge}, if the 32b hash of the advertisement is \ttfamily A673FF0A, \rmfamily  the target rdvpeer is RdvPeer2 because 
it is responsible for the key range \ttfamily \{9736DEE0..B873FFDF\}\rmfamily;
 \item The publishing rdvpeer contacts the target rdvpeer, that stores a new SRDI entry \textit{(advertisement hash, JXTA PeerID of the storing node)} in its
own local SRDI cache (this new entry is named \textit{advertisement index} in JXTA terminology \cite{JXTA08});
 \item The publishing rdvpeer contacts also the two rdvpeers closest to the target rdvpeer in the Chord circular domain, thus registering two replicas 
of the SRDI entry, in order to implement JXTA fault-tolerant strategy \cite{JXTA08}. For instance, in Fig.\ref{ChordRdv:merge} the
two replicas of the SRDI entry are stored in the cache of RdvPeer1 and RdvPeer3.   
\end{itemize}

It is important to observe that advertisements are never transferred through the Chord domain. Only SRDI entries need to be
transferred when a new rdvpeer joins or leaves the group. Fig. \ref{ChordRdv:merge} shows the steps performed when a peer (in the example
\textit{EdgePeer3a}) starts the discovery process of an advertisement:

\begin{itemize}
 \item EdgePeer3a contacts (MSG1) the local rdvpeer (\textit{discovering rdvpeer}, RdvPeer3 in the example);
 \item The discovering rdvpeer computes the hash key of the required advertisement (\ttfamily A673FF0A \rmfamily in the example) and performs 
a Chord lookup operation, thus determining the \textit{target rdvpeer} responsible for such key (RdvPeer2). It was demonstrated \cite{Chord00}
that the lookup operation requires, with high probability, $O(log_{2} N)$ hops within the Chord domain, if
all the fingertables of the N rdvpeers are already stabilized;
 \item The discovering rdvpeer contacts (MSG2) the target rdvpeer that looks for the SRDI entry related to the hash
in its own SRDI cache (in the example in Fig. \ref{ChordRdv:merge} the entry is immediately found);
 \item The target rdvpeer returns (MSG3) the found SRDI entry (i.e. the advertisement index)
to the discovering edgepeer (EdgePeer3a);
 \item The index returned to EdgePeer3a contains the ID of the peer \textit{where} the advertisement
is stored. Finally, EdgePeer3a contacts EdgePeer2a (MSG4), thus retrieving the document (MSG5). 
\end{itemize}
 
If the Chord fingertable of some rdvpeers is not stabilized yet, or if some rdvpeer is unreachable, the discovering rdvpeer
might not find the necessary SRDI entry in the cache of the target rdvpeer. In such a case, a strategy based on
a limited-range walker \cite{JXTA08} is performed, thus looking for the replicas of the SRDI entry 
in the near rdvpeers of the Chord domain. 

The cost for mantaining DHT consistency is evaluated in Sect. \ref{Chord_DHT_Costs}.

\subsection{Hypercompression algorithm}\label{Hyper00}
EmbJXTAChord uses a new binary message format. Each message consists of
several namespaces (ns), each of which contains several \textit{MessageElements} (msgelem), featured by 
a name, a MIME type 
\footnote{EmbJXTAChord and JXTA 2.7 indicate the content of a MessageElement
(\textit{text/plain, application/octet-stream} etc..) using the media types defined by
the MIME (Multipurpose Internet Mail Extensions) standard \cite{MIME00}.},
a content and (optionally) a signature.
EmbJXTAChord inherits such message structure from JXTA 2.7 (version 1.0 of the
JXTA specifications \cite {JXTA09}). However, EmbJXTAChord and JXTA encode
messages using two different binary formats, therefore a peer that runs
EmbJXTAChord cannot communicate with a peer that runs JXTA 2.7. 

The binary format used by JXTA 2.7 suffers from several drawbacks:

\begin{itemize}
 \item \textit{
XML and string contents are uncompressed}. In JXTA 2.7, the bandwidth is mainly used for names, MIME
types and xmldocs, which are transmitted in text form; 
 \item \textit{JXTA PeerIDs are transmitted as long strings}. Each 128b ID requires bytes for the prefix (for example \ttfamily urn:jxta:cbid-\rmfamily), 
followed by 32 characters; 
 \item \textit{Endpoint addresses are uncompressed}. Addresses as \ttfamily tcp://x.y.z.w:port \rmfamily are transmitted as 
strings.  In some cases, this can be expensive. For instance, an IPv6 address (128b, i.e. 16B in binary
format) may require more than 40B if transmitted as an UTF-8 string; 
 \item \textit{The elements belonging to the same message are often redundant}. IDs and addresses for peers and peergroups 
can be repeated several times in the same message. 
\end{itemize}

\begin{figure}
\centering
\includegraphics[width=4.7in, height=1.7in]{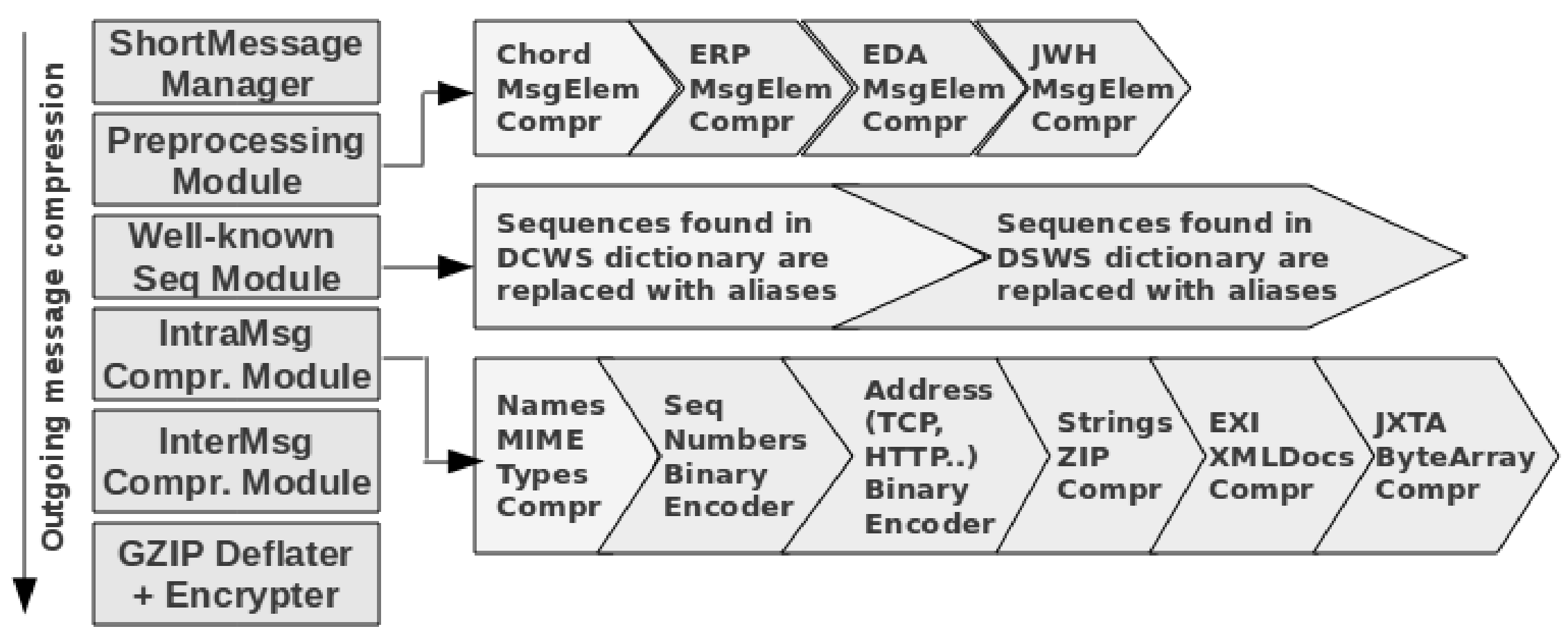}
\caption{The architecture of the EmbJXTAChord compressor manager}

\vspace*{-0.12in}
\label{Fig2:a}
\end{figure}
EmbJXTAChord exploits a Compressor Manager (CM) (see Fig. \ref{Fig2:a}) that works cutting out both \textit{intramessage} and \textit{intermessage}
redundancies. It intercepts all outgoing (or incoming) messages, transparently compressing (uncompressing) them.

In the first compression step (S1) the outgoing JXTA message is passed to a \textit{preprocessing module}, that deals with  
some large MessageElements (ChordWalker (CWL), EndpointRouterMessage (ERP), EndpointDestinationAddress 
(EDA) and JxtaWireHeader (JWH)), splitting them into a sequence of single partial ``subelements'' (pipe, group or peer IDs, network 
addresses, 128b keys) in the form \ttfamily prefix://id\rmfamily, that are ready to be compressed by the next stages.

In the second step (S2), the system scans the content of all XML and string elements, 
looking for any sequence in the form \ttfamily prefix://seq\rmfamily, thus creating a \textit{dictionary of custom well-known sequences} (DCWS)
that depends on the specific message. All sequences found in the content of each msgelem are replaced with an alias consisting
of a control char $CH_{1}$ + the 8b ID identifying in DCWS the replaced custom sequence. 
Next, the operation is repeated using a \textit{dictionary of standard well-known sequences} (DSWS) whose
items are standardized and known a priori by all peers. 
The sequences found in the content of each msgelem are replaced with an alias $CH_{2}$ + the 8b ID identifying in DSWS the replaced item. 
For each message, the DCWS is increasingly reordered and transmitted using a differential format,
while the DSWS is not transmitted.

In the third step (S3), the CM performs the \textit{intramessage compression} through the following operations:
\begin{itemize}
 \item S3a: The pair (\textit{name, MIME type}) is compressed into a 4B stream. The first byte is a control char $CH_{3}$ declaring some
compression options for the msgelem, the second byte identifies the MIME type, the third byte is an alias ID referred to 
a dictionary of \textit{well-known msgelem names} and the fourth one is a 8b suffix code; 
 \item S3b: Hexadecimal numbers (such as IDs, custom and well-known sequences, etc.) are packed in binary form;
 \item S3c: IPv4, IPv6 and BT addresses are binary-packed; 
 \item S3d: Strings are compressed using the deflate algorithm (gzip). EmbJXTAChord exploits a gzip dictionary containing several
sequences that appear frequently in JXTA data stream (such as \ttfamily urn:jxta:cbid-\rmfamily, \ttfamily jxta://cbid-\rmfamily..);
 \item S3e: The ByteArray elements are compressed using the deflate algorithm (gzip);
 \item S3f: The xmldocs are compressed through OpenEXI, an open-source and free implementation for Java
of the EXI encoder \cite{EXI00}\cite{EXI02}. An optimized schema, named jxta.xsd, is used to improve the efficiency of the EXI+GZIP combination.
\end{itemize}

The compression of hexadecimal numbers and addresses is performed 
in two steps (s3b, s3c) because of the different
binary formats used. For the numbers, the format contains some control bytes and the binary representation of the hexadecimal value.
Special codes may be used to indicate that a well-known subsequence was contained in the main sequence before compression 
and the original position (the CM implements this function using a \textit{dictionary of well-known binary subsequences} (DBS)
that is similar to DSWS).   
For the addresses, a more complex, protocol-specific, binary format is used, that is able to indicate also the network prefix 
(\ttfamily tcp://, btspp:// \rmfamily etc.) and the position of dots and colons.

In the fourth step (S4), the CM performs the \textit{intermessage compression} (IMX) \footnote{As this step is computationally expensive, it can be advantageously used
only for the slowest channels (it is used for SICS MTB, but not for TCP, HTTP and BT MTBs).}. The receiver and the sender mantain a IMX cache containing the $c_{X}$
xmldocs and the $c_{S}$ strings recently sent (received) more often to (from) the other side. The cache is updated whenever
a new message is sent (received) and increasingly reordered on the basis of the 64b content hash. When the transmitter finds in the local IMX cache 
the msgelem to send, the latter is replaced with a short 8b code $K_{C}$ in the transmitted packet. The receiver gets the missing 
msgelem from the $K_{C}$-th position in its own local IMX cache. Special codes are provided for registering/purging elements, thus mantaining the
caches of both sides coherent.    
  
The final step (S5) consists in the compression of the whole CM chain output through a 2nd-level gzip deflater. The result is encrypted (if it 
is required, as explained in Sect. \ref{Security}) and finally transmitted on the channel. 
An assessment of the performance measured for the Compressor Manager is provided in Sect. \ref{CM_Bench}.

\subsection{jxCOAP-E services}\label{ServiceInter}

As said in Sect.\ref{Sec:Introduction}, EmbJXTAChord provides jxCOAP-E, a modified version of CoAP specifically devised 
to work over the JXTA communication components. JxCOAP-E allows to merge the features of CoAP 
(compactness and RESTful interaction) and JXTA (support for hybrid networks).

\medskip

EmbJXTAChord leverages on the service discovery architecture defined by JXTA specifications \cite{JXTA09},
 thus allowing a server peer to provide \textit{custom JXTA services} to the other 
client peers of the group. 
The communication to a JXTA service is provided by the Peer Resolver Protocol (PRP). When the server starts a new service,
a new resolver listener is registered in the PRP module. A client can access to a service provided by a server  
(which must belongs to the same group), through the PRP method \ttfamily ResolverQuery()\rmfamily, providing the server \textit{PeerID}, 
the \textit{ServiceName}, and the \textit{ServiceParam} string. The server performs the operation required 
and answers using the method \ttfamily ResolverResponse()\rmfamily. 
The interaction model is stateless and connectionless, as message delivery
through PRP is unreliable\cite{JXTA09}.

\medskip

In order to ensure that all peers in the group are informed about the availability of a new service, the server peer publishes 
a \textit{module specification advertisement} (MSA), containing the name and description of the service. 

Differently from JXTA 2.7, that do not define rules on the interaction model used for services, thus leaving this complex task 
to the developer, EmbJXTAChord exploits the RESTful interaction model provided by CoAP.  

\medskip

When a service is deployed, a new instance of a \textit{jxcoap \footnote{We hereby refer to \textit{jxcoap} for properties and concepts that are common to
both the implementations, i.e., to the jxCOAP proposed in \cite{JXCOAP00} and to the jxCOAP-E proposed in this paper.} virtual server} 
is started and then bound to a new \textit{JXTA low level service} (jxllservice), whose module specification advertisement is published within the group.
When a node needs to access a service, it runs a new instance of a \textit{jxcoap virtual client} (vclient), 
which encapsulates each CoAP request into a PRP message that is sent to the jxllservice of the server peer. 
The resources provided by the jxcoap server are addressed through strings named \textit{jxURI} \cite{JXCOAP00}. The client 
interacts with the server through an API consisting of 4 operations (jxGET, jxPOST, jxPUT and jxDELETE) that 
require a jxURI specifying the target resource. 

\medskip

The access to a jxcoap service consists of 3 steps. First, the client looks for the MSA advertisement  using the SJXTA
resource discovery API. If the peer is an authorized member of the group the server belongs to, 
the advertisement 
is found, thus retrieving the PeerID of the server  and initializing a new virtual client instance (\textit{binding}).
EmbJXTAChord  replicates on multiple rendezvous  peers the reference to the node that stores 
the service advertisement  (see Sect. \ref{RendezvousProto}), 
thus ensuring it can be found even if some rendezvous  peers are unreachable. 
Next, the client retrieves from the jxcoap server the list of the available resources through a jxGET operation to
the standard jxURI \textit{``./well-known/core''}. Finally, the client accesses all the resources provided by the
bound server. An assessment of jxCOAP-E performance 
is in Sect. \ref{Subsec:jxCOAP}. 

\subsection{Secure peergroups}\label{Security}

EmbJXTAChord supports a new feature named \textit{AES group encryption}, that allows the creation of secure peergroups.
This feature overcomes  two limitations of JXTA 2.7:

\begin{itemize}
 \item In JXTA 2.7 multicast communications and PRP message exchange cannot be protected \cite{JXTA16};
 \item JXTA 2.7 supports TLS-based unicast connections (\textit{secure pipes}) between any pair of peers, even if the underlying 
transport protocol is not TCP. However, as the JXTA specifications do not define a Certification Authority Service,
it is up to the developers to manage the transmission of the X.509 certificates (public keys) \cite{JXTA14}.
\end{itemize}

EmbJXTAChord protects the PRP and multicast communications within a group in the following way. 
When a rdvpeer (named \textit{group owner}) creates a new peergroup, it can optionally state a \textit{Group Traffic
Encryption Key} (GTEK), which is used by an AES-128 encrypter integrated in the 2nd level gzip deflater 
(step S5 in Sect. \ref{Hyper00}). Hence, only the peers that know such a password can join the group, find the
resource advertisements and decode the PRP messages. GTEK may be a Preshared key, or it can be requested by the
edge to the group owner via TLS before joining the child group (Certificate-based scheme). 

EmbJXTAChord introduces also a new version of peergroup advertisement that is used by
TLS for retrieving the public keys needed to establish a connection. 
When a new edgepeer is joining a new group, it
looks for the peergroup advertisement that includes the PeerID and the X.509 certificate of the group owner. 
Next, the edge contacts the owner requesting the group fingertable, thus providing its peer advertisement  and X.509 certificate. 
In this way, the group owner mantains a \textit{central keystore} with a list of all the public keys of the connected peers (\textit{group members}), while each
member mantains a \textit{peripheral keystore} with at least the public key of the group owner. Each pair of peers in the 
group can use the keystore in the group owner  to acquire the peer advertisement (and thus the public key) of the recipient,
thus being able to establish a secure unicast TLS connection.

\subsection{EmbJXTAChord in an ISO/OSI network stack}

\begin{figure}[t]
  \addtolength{\abovecaptionskip}{-2.0in}
  \hspace*{-0.20in}
  \vspace*{-0.16in}
  \subfigure
  {
    \includegraphics[width=4.80in, height=2.90in]{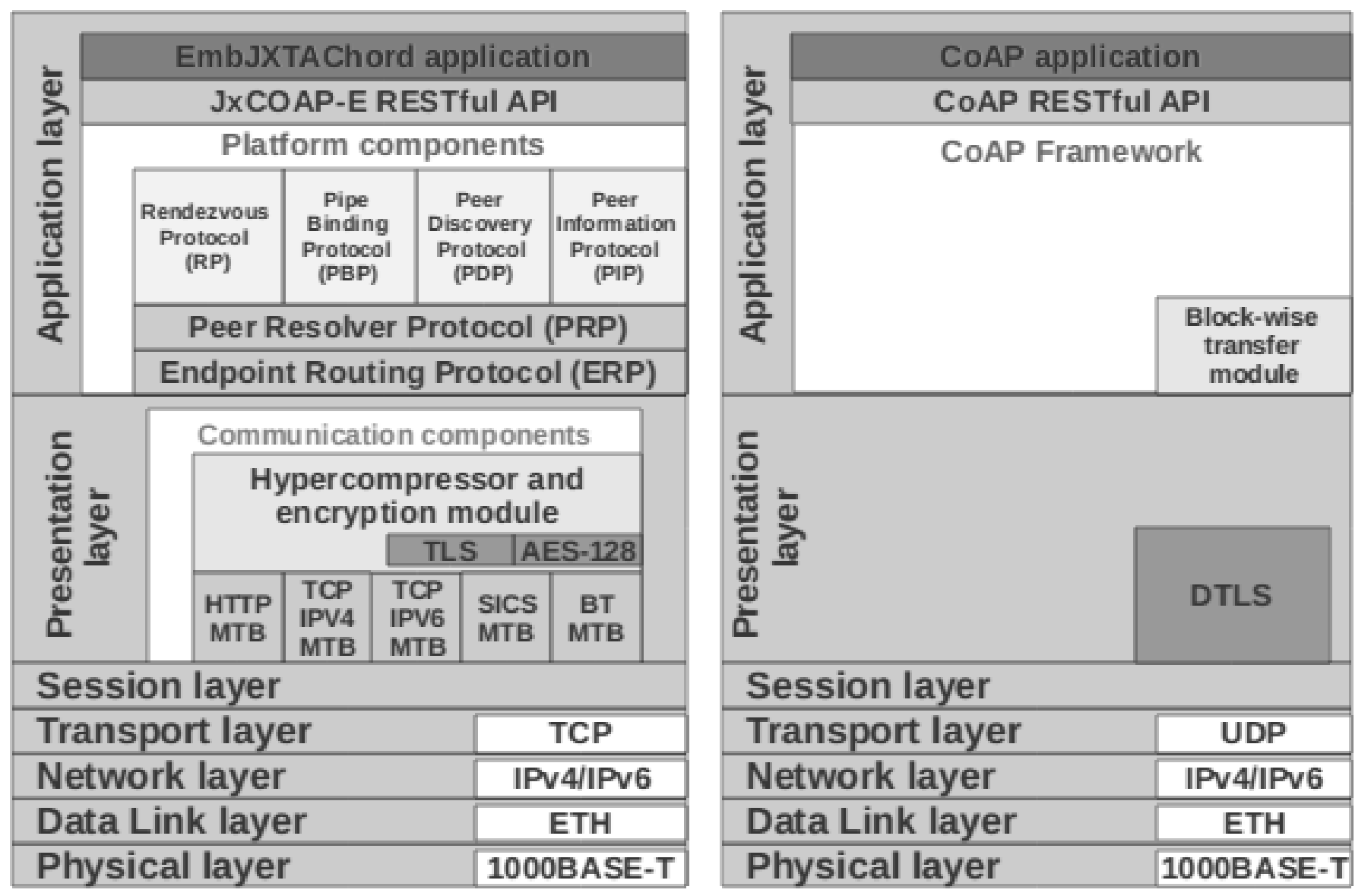} }
    \caption{A comparison between jxCOAP-E and classic UDP CoAP ISO/OSI network stacks.}
\label{OSINetworkStack_Comp}
\end{figure}

Fig. \ref{OSINetworkStack_Comp} shows a comparison between a novel architecture using jxCOAP-E over EmbJXTAChord
and a classic RESTful architecture based on CoAP over UDP. The comparison is represented using two ISO/OSI                                                                                         
network stacks.
The stack assumes that the architectures work over an IPv4-based Ethernet network, therefore the physical,
data-link and network layers are identical in both cases. 

The transport layer (OSI layer 4) is different because CoAP works over an unreliable protocol such as UDP 
whereas jxCOAP-E can use several reliable protocols (TCP, HTTP, 6LoWPAN, Bluetooth RFCOMM). 

The session layer (OSI layer 5) does not contain any component in both cases, because in Fig.\ref{OSINetworkStack_Comp} 
we consider only the single-application case. 

The presentation layer (OSI layer 6) for jxCOAP-E contains the EmbJXTAChord \textit{communication components}
(i.e. the Hypercompressor and the Message Transport Binding modules). They provide to encrypt/decrypt 
and to compress/uncompress JXTA messages. These components also provide some \textit{overlay functionalities}
that allow to use a uniform addressing scheme based on 128b JXTA PeerIDs, thus making transparent to the 
upper layer the transport protocol that is actually used. In the UDP CoAP case, the presentation layer 
provides only to encrypt messages through the DTLS protocol. 

The application layer (OSI layer 7) for jxCOAP-E contains the EmbJXTAChord \textit{platform components}. 
They are the six modules of the framework (see Tab.\ref{AddedTable_I}) that provide the
applications (and to the jxCOAP-E services) with the typical functionalities of the JXTA framework (resource discovery,
unicast and multicast communications, traffic information). 

In the case of UDP CoAP, the OSI layer 7 is
much simpler. It consists of a RESTful service framework (such as Californium \cite{COAP02}) that 
implements the request/response and publish/subscribe CoAP interaction models. 
As CoAP works over an unreliable and connectionless transport protocol (UDP), the RESTful service 
framework includes a module for \textit{block-wise transfer} \cite{COAP03} of large payloads, which provides
functionalities such as segmentation, reassembly and retransmission. JxCOAP-E does not need such
functionalities, as it works over reliable transport protocols.

\subsection{Support for heterogenous networks}

\begin{figure}[t]
  \addtolength{\abovecaptionskip}{-2.0in}
  \hspace*{-0.42in}
  \vspace*{-0.16in}
  \subfigure
  {
    \includegraphics[width=5.50in, height=2.90in]{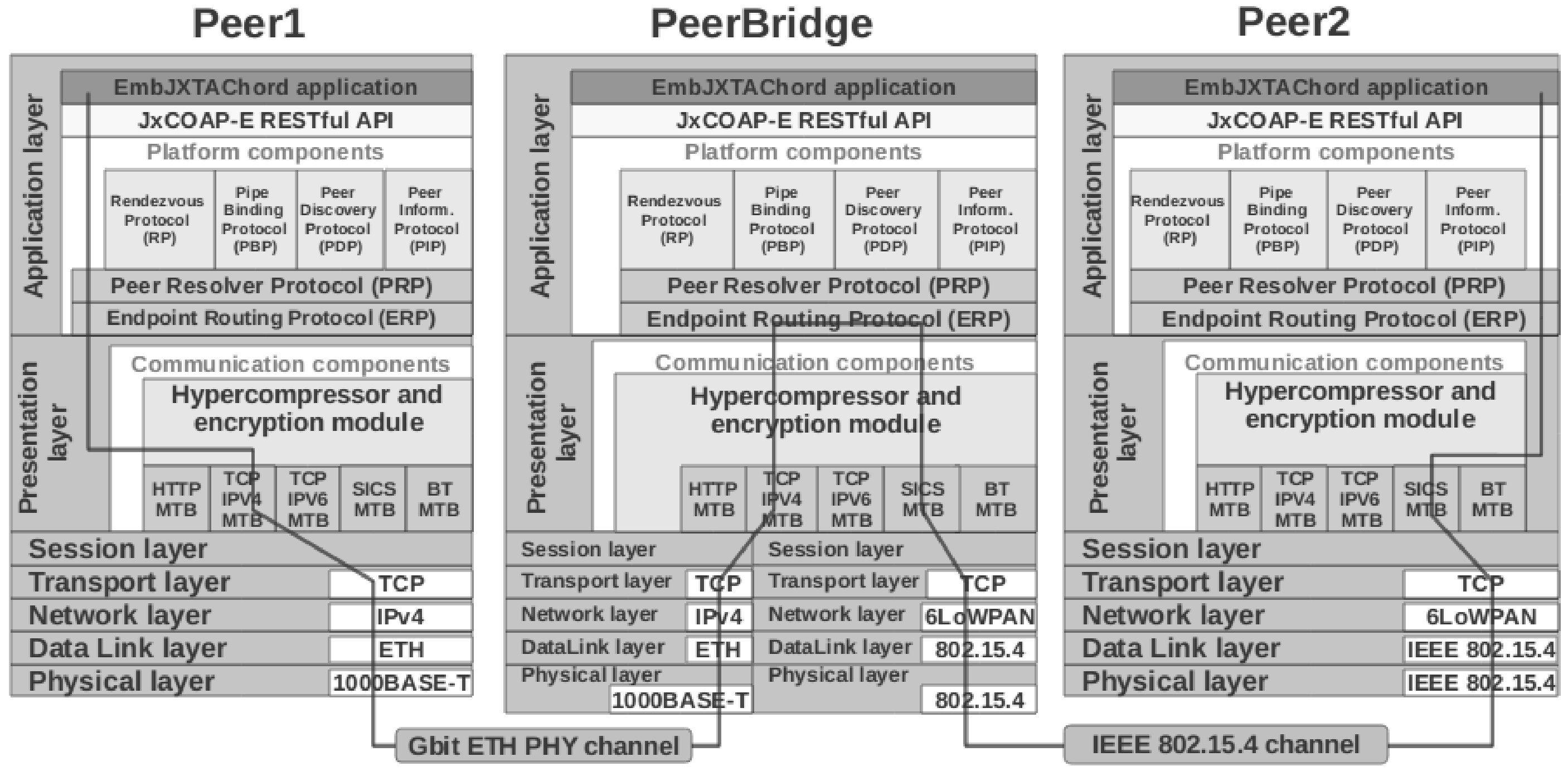} }
    \caption{A representation of the communication between two peers belonging to different subnetworks, using OSI-inspired network stacks. \textit{PeerBridge} acts
	     as a gateway between the Ethernet network (IPv4) and the IEEE 802.15.4 WSN network (6LoWPAN).}
\label{OSINetworkStack}
\end{figure}

Fig. \ref{OSINetworkStack} shows how EmbJXTAChord works over different transport protocols in a heterogeneous network.
The \textit{PeerBridge} node acts as a gateway between \textit{Peer1} (connected 
through an Ethernet link using IPv4) and \textit{Peer2} (connected through an IEEE 802.15.4 link using 6LoWPAN). 
The Endpoint Routing Protocol (ERP) determines (transparently to the application layer) that the PeerBridge node
can be used as a gateway to reach Peer2 (this information is made available to all peers in the group
using a route advertisement). 

Once PeerBridge promoted itself to the gateway role, ERP automatically provides
to uncompress/decode the messages coming from the Ethernet source node. Next it encodes, compresses and
reroutes them to the IEEE 802.15.4 destination node. 

The main advantage of using EmbJXTAChord in this scenario is that routing over subnetworks is automatically performed 
at the application level, without requiring the reconfiguration of the routing tables in the switches
\footnote{The configuration can be manually performed by the sysadmin, or using tools such as the \textit{Router
Advertisement Daemon} (radvd) that leverages on the IPv6 Neighbour Discovery Protocol (NDP) \cite{NDP01}. 
However, NDP is available only for the IPv6 subnetworks. Furthermore, radvd 
does not support multiple hops over three or more subnetworks. Finally, radvd is 
available only for Linux and its installation and configuration are not
user-friendly tasks.},
the installation of a \textit{border-router} between the IPv4 and 6LoWPAN subnetworks \cite{6LoWPAN05} or
the configuration of proprietary software for IPv4-to-IPv6 tunnelling \cite{Teredo01}.

%

\section{Experimental results}\label{Sec:ExpResult}

The performance of EmbJXTAChord was tested under three respects. Sect. \ref{EmbJXTAChord_vs_JXTA_Comp}
compares the performance of EmbJXTAChord and JXTA 2.7. Sect. \ref{Performance_over_Heterogenous_Networks}
measures the performance of an unicast connection (\textit{connecting pipe}) between two peers, thus
demonstrating that the proposed framework can work on hybrid narrowband networks (Bluetooth, 6LoWPAN), making transparent
the presence of a gateway through routing over subnetworks. 
Sect. \ref{Subsec:jxCOAP} deals with the performance of jxCOAP-E over homogeneous and 
heterogeneous networks, thus demonstrating that EmbJXTAChord can work on low-cost COTS hardware such as 
the RaspberryPI and RaspberryPI-3 boards with latency times that are acceptable for a wide range of applications.

\subsection{EmbJXTAChord vs JXTA 2.7}\label{EmbJXTAChord_vs_JXTA_Comp}

Some tests aimed to compare the performance of EmbJXTAChord and JXTA 2.7
were performed. Sect. \ref{SubSec:ExecutionTimes} measures the overhead determined 
by EmbJXTAChord encryption and compression respect to JXTA 2.7, in order to verify
if the proposed framework is unsuitable for using on RaspPI or RaspPI-3 boards.

Sect. \ref{SubSec:Null-app-overhead} measures the network overhead determined by 
Chord periodic message exchange, thus paving the way for the tests in Sect. \ref{PipeMessageOverhead} 
and Sect.\ref{MulticastPipeMessageOverhead} that measure the message overhead in unicast and 
multicast connections. All these trials demonstrate the effectiveness of the Hypercompression algorithm in 
reducing the bandwidth waste respect to JXTA 2.7.

Finally, Sect. \ref{CM_Bench} measures the performance of the Hypercompression algorithm
in function of the enabled compression schemes.

\subsubsection{Execution times of common operations}\label{SubSec:ExecutionTimes}

The latency values for some common EmbJXTAChord operations 
in an Ethernet network were measured using a modified version of the JxtaBench 
project \cite{JXTA15}. 

Six testbed configurations (TB1-TB6) were considered.  
In the first one (TB1), the peer0 (server) was configured as \textit{rendezvous} and the peer1 (client) was configured as
\textit{edge}. In the second one (TB2), both peers were configured as rendezvous. The peers were personal computers (PC)
with an AMD Phenom II 3.0 Ghz CPU, running KUbuntu OS v15.10 and Oracle JRE1.7.
All machines worked in single-core mode.
In TB3-TB4 and in TB5-TB6 the peer roles were the same of
TB1-TB2, respectively, but with a different hardware. The TB3-TB4 tests were
performed using two Raspberry PI model B+, based on a single-core SoC with
a 850 MHz ARMv11 CPU and 512 MB RAM \cite{RaspberryPI01}. The TB5-TB6 tests were
performed using two Raspberry PI-3 model B based on the BCM2837 SoC, integrating 
four 1.2 Ghz ARMv8 CPUs and 1 GB RAM \cite{RaspberryPI04}. All RaspPI tests were
performed using Raspbian OS and Oracle JRE1.8 for the 32b ARM processors.

\newcommand{\TABa}
{
\begin{tabular}{||c c||c|c||c|c||c|c||} 
\hline
   &Testbed configuration        &TB1      &TB2     &TB3      &TB4      &TB5      &TB6        \\
\hline
   &server node                  &rdvpeer  &rdvpeer &rdvpeer  &rdvpeer  &rdvpeer  &rdvpeer    \\
\hline
   &client node                  &edgepeer &rdvpeer &edgepeer &rdvpeer  &edgepeer &rdvpeer    \\
\hline
   &CPU                          &AMD      &AMD     &ARM11    &ARM11    &ARM11    &ARM11      \\
   &                             &Phenom2  &Phenom2 &(RaspPI  &(RaspPI  &(RaspPI-3&(RaspPI-3  \\
   &                             &(PC)     &(PC)    &mod. B+) &mod. B+) &mod. B)  &mod. B)    \\
\hline
   &Cores used for testing&1(3Ghz)  &1(3Ghz) &1(850Mhz)&1(850Mhz)&4(1.2Ghz)&4(1.2Ghz)  \\
\hline
\hline
\multicolumn{8}{|c|}{\textbf{Startup operations}} \\
\hline
\hline
\multicolumn{8}{|c|}{EmbJXTAChord (compression and group encryption enabled, Chord)} \\
\hline
op1 &start(server node)(*)       &1312     &1312    &32501    &32501    &10040    &10040 \\
op2 &deploy CustomPeerGrp(*)     &525      &525     &1377     &1377     &708      &708 \\
op3 &create a new pipe(*)        &16       &16      &358      &358      &88       &88 \\
\hline
op4 &start(client node)          &1654     &1671    &26313    &27029    &6155     &6691 \\
op5a&init NetPeerGroup (rdv)     &-        &2009    &-        &27869    &-        &5597 \\
op5b&join NetPeerGroup (edge)    &2318     &-       &37452    &-        &17798    &- \\
op6a&join CustomPeerGroup (rdv)  &-        &2766    &-        &22866    &-        &6604 \\
op6b&join CustomPeerGroup (edge) &16743    &-       &24795    &-        &17243    &- \\
\hline
\multicolumn{8}{|c|}{JXTA 2.7 (compression and group encryption disabled, lcDHT-LRW)} \\
\hline
op1 &start(server node)(*)       &1179     &1179    &29678    &29678    &11812    &11812 \\
op2 &deploy CustomPeerGrp(*)     &39       &39      &860      &860      &224      &224 \\
op3 &create a new pipe(*)        &2        &2       &136      &136      &14       &14 \\
\hline
op4 &start(client node)          &1548     &1320    &31549    &27534    &6056     &5394 \\
op5a&init NetPeerGroup (rdv)     &-        &752     &-        &3928     &-        &3062 \\
op5b&join NetPeerGroup (edge)    &1647     &-       &28799    &-        &16620    &- \\
op6a&join CustomPeerGroup (rdv)  &-        &1119    &-        &5695     &-        &3494 \\
op6b&join CustomPeerGroup (edge) &16145    &-       &17019    &-        &16440    &- \\
\hline
\hline
\multicolumn{8}{|c|}{\textbf{Common operations}} \\
\hline
\hline
\multicolumn{8}{|c|}{EmbJXTAChord (compression and group encryption enabled, Chord)} \\
\hline
op7 &discovery other peers       &13       &32      &1220     &1292     &134      &255 \\
op8 &discovery remote pipe adv   &18       &27      &712      &861      &141      &230 \\
op9 &bind remote pipe            &26       &42      &1347     &1459     &189      &271 \\
\hline
\multicolumn{8}{|c|}{JXTA 2.7 (compression and group encryption disabled, lcDHT-LRW)} \\
\hline
op7 &discovery other peers       &10       &11      &680      &290      &51       &95 \\
op8 &discovery remote pipe adv   &6        &10      &255      &283      &57       &60 \\
op9 &bind remote pipe            &16       &13      &714      &385      &69       &82 \\
\hline
\end{tabular} 
}
\ifdefined\ACM
  \begin{table}
  \tbl{Latency values for startup and common operations (ms). The values were measured on the client node, except for the ones marked with (*) that
were measured on the server node.}
  {      
      \TABa{}
  }
  \label{Table_I}
  \end{table}
\else
  \small
  \begin{table}[t]
  \scriptsize
  {
      \caption{\scriptsize {Latency times for startup and common operations (ms)} }
      \vspace*{-0.1in}
      \setlength{\tabcolsep}{.16em}
      \begin{center}
      \TABa{}   
      \vspace{-12pt}
      \footnotetext{The values were measured on the client node, except for the ones marked with (*) that were measured on the server node.
}
      \end{center}
      \label{Table_I}
  }
  \end{table}
\fi
\newcommand{\TABb}
{
\begin{tabular}{|c||c|c|c|c|c||} 
  \hline  
		       & EmbJXTA   & EmbJXTA   & EmbJXTA   & EmbJXTA   &JXTA \\
		       & Chord     & Chord     & Chord     & Chord     &2.7  \\
  \hline
		       & Compr.    & Compr.    & Compr.    & Compr.    & Compr.    \\
		       & Config C0 & Config C1 & Config C2 & Config C3 & Config C4 \\
  \hline
  \hline
  Rdv protocol         & Chord & Chord     & Chord     & Chord     & LRW \\
  Large MsgEl comp.(S1)& yes   & yes       & no        & no        & no  \\
  Well-known seq.(S2)  & yes   & yes       & yes       & no        & no  \\
  String comp.(S3d)    & yes   & yes       & yes       & no        & no  \\
  ByteArray comp.(S3e) & yes   & yes       & yes       & no        & no  \\
  XML EXI comp.(S3f)   & yes   & yes       & no        & no        & no  \\
  Intermsg comp.(S4)   & yes   & no        & no        & no        & no  \\
  GZIP deflater(S5)    & yes   & yes       & no        & no        & no  \\
  \hline
\end{tabular}
}
\ifdefined\ACM
    \begin{table}
    \tbl{Compressor configurations}
    {      
	\TABb{}
    }
    \label{Table_ComprCfg}
    \end{table}
\else
    \begin{table}[t]
    \setlength{\tabcolsep}{.32em}
    \scriptsize
    {
	\caption{\scriptsize {Compressor configurations} }
        \hspace*{-0.1in}
	\vspace*{-0.1in}
	\begin{center}
	\TABb{}
	\end{center}
	\label{Table_ComprCfg}
    } 
    \vspace*{-0.1in}
    \end{table}
    \normalsize
\fi

Finally, all the tests were repeated with JXTA 2.7 (using lcDHT-LRW, 
without compression) in order to evaluate the overhead 
determined by EmbJXTAChord. 
During the test, the remote node created an accepting pipe and a new custom peergroup. The local
node (edge or rendezvous ) first joined the NetPeerGroup and the custom peergroup, 
then looked up for the other peers advertisements and pipes and, finally, connected to
the remote pipe.
Observing the execution times for the most common operations, that are shown in Tab. \ref{Table_I}, 
some conclusions can be reached:

\begin{itemize}
\item Some operations on the server side (op1-3) and on the client side (op4-6) require
very long times (expecially on RaspPI, TB3-4). They might be still acceptable for 
some cases, as these operations are executed only once at
startup. The reason for these long times is that each JXTA module can
serve only the requests of a single group, therefore deploying or joining a
new peergroup requires the JVM loads a new instance of all the active
JXTA modules;
\item Some startup operations are longer for the edgepeers than for the rdvpeers.
The edgepeers run NetPeerGroup joining (op5b) and CustomPeerGroup
joining (op6b), that are performed contacting a local rdvpeer  
in order to register the new edgepeer, publish its own advertisement and 
retrieve the advertisement related to the custom peergroup using RP and PDP
protocols (see Sect. \ref{RendezvousProto}). 
The rdvpeers run NetPeerGroup initialization (op5a) and CustomPeerGroup 
joining (op6a), that only require to initialize and stabilize their own RPV. 
They do not need to find the advertisement index of the custom peergroup 
using the RP protocol, as some SRDI entries are automatically transferred 
in their own cache when they join the Chord group (Chord key transfer, see \cite{Chord00});
\item The execution times of some common operations (op7-9) are higher for
EmbJXTAChord than for JXTA 2.7. This is reasonable because of the
overhead due to EmbJXTAChord compression and group encryption. The
overhead for the op7-9 was lower than 20ms for PC (TB1-TB2), 1.1s for
RaspPI (TB3-TB4) and 200ms for RaspPI-3 (TB5-TB6);
\item For the rdvpeers, the execution times of NetPeerGroup initialization (op5a)
and CustomPeerGroup joining (op5b) are longer in EmbJXTAChord than
in JXTA 2.7, expecially on RaspPI (TB3-TB4). This is mainly due to the
initialization and stabilization of the 128b Chord Fingertable that is computationally 
expensive (see Sect. \ref{RendezvousProto}).
\end{itemize}

In conclusion, EmbJXTAChord can be used for the common operations also 
on low-cost COTS hardware such as RaspPI and RaspPI-3. The
higher overhead comparing with JXTA 2.7 is the price to pay for the enhancements
in compression and security. However, the use on the RaspPI is suitable only
if the application does not require low startup times. Otherwise, the multicore
RaspPI-3 (TB5-6) is preferable.

\subsubsection{Average DHT management overhead}\label{SubSec:Null-app-overhead}

In the second test, a number $r$ of computers, with $r=\{2,4,6,8,10\}$,
equipped with the same CPU used for the measurements described in
Sect. \ref{SubSec:ExecutionTimes}, were connected to the same LAN, with
the aim of measuring the \textit{DHT management overhead}, i.e., the
average bandwidth occupation due to the periodic message exchanges of
Chord \cite{Chord00} or lcDHT-LRW \cite{JXTA08}.

All the computers were configured as \textit{rendezvous} peers. Five
configurations for the compressor manager (C0-C4, see Tab. \ref{Table_ComprCfg})
were used. The first one (C0) exploits all EmbJXTAChord features,
whereas the latter (C4) is equivalent to JXTA 2.7
exploiting the lcDHT-LRW protocol.

The k-th rendezvous  peer ($k=\{0..(r-1)\}$) was started at the time instant
$t^{start}(k) = k \cdot 60s$.
The rendezvous  peers sent two types of Chord service messages: \textit{Chord
stabilize} with periods $T_{stab}=8s$ and \textit{Chord predecessor-checking}
with periods $T_{pred.chk}=4s$, respectively.
Moreover, they performed a \textit{fix-fingers} operation every $T_{fix.fing}=4s$
(these messages are required for checking that the rendezvous  nodes are still alive or
for the updating of Chord fingertable in the nodes, see \cite{Chord00} for more details).
For JXTA 2.7, lcDHT-LRW was configured so that RPV exchanges occurred
every $T_{LRW}=8s$.

The Tab. \ref{Table_II} shows the average bandwidth overhead
$\phi_{dht-mgm}(r)=\Phi_{n.trx.bytes}/\Delta T$
where $\Phi_{n.trx.bytes}$ is the number of bytes sent by rdv0 to the
other rendezvous  peers during the observation time $\Delta T=60 s$ between
two following joining events. 

When EmbJXTAChord is used, increasing the number of rdvpeers determines higher values 
of management overhead. It is interesting to observe that this may be different for JXTA 2.7
(see for instance the case r=10) because of the randomness determined by the RPV-exchange operation  
periodically performed by lcDHT-LRW algorithm \cite{JXTA08} (see Sect.\ref{RendezvousProto}).

In the best case the average bandwidth overhead was
reduced from 10.794 KB/s (JXTA 2.7) to 0.203 KB/s (EmbJXTAChord).
It is interesting that the value for $\phi_{dht-mgm}(r)$ measured for
JXTA 2.7 sometimes can be lower than the one measured for EmbJXTAChord when all
the compression schemes are disabled, as the lcDHT-LRW algorithm
is more efficient than Chord when few rendezvous  peers are used
\cite{JXTAChord00}. However, when all the compression
schemes are enabled, EmbJXTAChord always outperforms JXTA 2.7, as it
requires less bandwidth.

\subsubsection{Average pipe message overhead}\label{PipeMessageOverhead}

\newcommand{\TABc}
{
  \begin{tabular}{|c|c||c|c|c|c|c||} 
    \hline  
    & Nr rdv & EmbJXTA   & EmbJXTA   & EmbJXTA   & EmbJXTA  & JXTA      \\
    & peers  & Chord     & Chord     & Chord     & Chord    & 2.7       \\
    & (r)    & Compr.    & Compr.    & Compr.    & Compr.   & Compr.    \\
    &        & Config C0 & Config C1 & Config C2 & Config C3& Config C4 \\
    \hline
    \hline
    $0 \rightarrow all$ & 2 & 0.201 & 0.245 & 0.858 & 1.468 & 1.119 \\
    $0 \rightarrow all$ & 4 & 0.201 & 0.391 & 1.112 & 1.496 & 2.431 \\
    $0 \rightarrow all$ & 6 & 0.215 & 0.533 & 1.841 & 2.766 & 3.721 \\
    $0 \rightarrow all$ & 8 & 0.203 & 0.289 & 1.014 & 1.353 & 10.794 \\
    $0 \rightarrow all$ &10 & 0.182 & 0.358 & 1.180 & 1.566 & 1.606 \\
    \hline
  \end{tabular}
}
\ifdefined\ACM
    \begin{table}
    \tbl{Average DHT management overhead (KB/s)}
    {      
	\TABc{}
    }
    \label{Table_II}
    \end{table}
\else
    \small
    \begin{table}[t]
    \setlength{\tabcolsep}{.32em}
    \scriptsize
    {
	  \caption{\scriptsize {Average DHT management overhead (KB/s)} }
	  \vspace*{-0.1in}
      \begin{center}
	  \TABc{}
      \end{center}
      \label{Table_II}
    }
    \vspace*{-0.25in}
    \end{table}
\fi
\newcommand{\TABd} 
{
  \begin{tabular}{|c|c||c|c|c|c|c||}%
    \hline  
    & Nr rdv & EmbJXTA   & EmbJXTA   & EmbJXTA   & EmbJXTA  & JXTA      \\
    & peers  & Chord     & Chord     & Chord     & Chord    & 2.7       \\
    & (r)    & Compr.    & Compr.    & Compr.    & Compr.   & Compr.    \\
    &        & Config C0 & Config C1 & Config C2 & Config C3& Config C4 \\
    \hline
    \hline
      $0 \rightarrow 1$ & 4 & 0.097 & 0.176 & 0.529 & 0.532 & 1.930  \\
      $0 \rightarrow 2$ & 4 & 0.000 & 0.000 & 0.000 & 0.000 & 0.286 \\
      $0 \rightarrow 3$ & 4 & 0.104 & 0.215 & 0.582 & 0.964 & 0.215 \\
    \hline
  \end{tabular}
}
\ifdefined\ACM
    \begin{table}
    \tbl{Average DHT management overhead between two rendezvous  nodes (KB/s)}
    {      
	\TABd{}
    }
    \label{Table_III}
    \end{table}
\else
    \begin{table}[t]
    \setlength{\tabcolsep}{.32em}
    \scriptsize
    {
      \caption{\scriptsize {Average DHT management overhead between two rendezvous  nodes (KB/s)} }
      \vspace*{-0.1in}
      \begin{center}
      \TABd{}
      \end{center}
      \label{Table_III}
    }
    \vspace*{-0.2in}
    \end{table}
\fi
\newcommand{\TABe} 
{
  \begin{tabular}{|c|c|c||c|c|c|c|c||}%
      \hline
                         &                 &payload& EmbJXTA   & EmbJXTA   & EmbJXTA   & EmbJXTA   & JXTA \\
			 &	           &size   & Chord     & Chord     & Chord     & Chord     & 2.7  \\
                         &                 &       & Compr.    & Compr.    & Compr.    & Compr.    & Compr.    \\
                         &                 &       & Config C0 & Config C1 & Config C2 & Config C3 & Config C4 \\
      \hline
      \hline
       $\gamma$          &$0 \rightarrow 3$&256B   & 2.081 &  2.363 & 3.353  & 4.790  & 4.859 \\
	                 &$0 \rightarrow 3$&1024B  & 1.326 &  1.408 & 1.589  & 1.949  & 1.964 \\
                         &$0 \rightarrow 3$&10240B & 1.035 &  1.041 & 1.063  & 1.074  & 1.096 \\
      \hline
       $\Phi_{msg.size}$ &$0 \rightarrow 3$&256B   &   533B &   605B &   858B &  1226B &  1244B\\
                         &$0 \rightarrow 3$&1024B  &  1358B &  1442B &  1627B &  1996B &  2012B\\
                         &$0 \rightarrow 3$&10240B & 10599B & 10661B & 10892B & 11002B & 11228B\\
    \hline
  \end{tabular}
}
\ifdefined\ACM
    \begin{table}
    \tbl{Transmitted-message-vs-payload-size ratio $\gamma$ and average message size $\Phi_{msg.size}$ (B) for accepting/connecting pipes (4 rendezvous peers )}
    {      
	\TABe{}
    }
    \label{Table_IV}
    \end{table}
\else
    \begin{table}[t]
    \setlength{\tabcolsep}{.32em}
    \scriptsize
    {
      \caption{\scriptsize {Transmitted-message-vs-payload-size ratio $\gamma$ and average message size $\Phi_{msg.size}$ (B) for accepting/connecting pipes (4 rendezvous peers )} }
      \vspace*{-0.1in}
      \begin{center}
	  \TABe{}
      \end{center}
      \label{Table_IV}
    }
    \vspace*{-0.2in}
    \end{table}
\fi
\newcommand{\TABf} 
{
  \begin{tabular}{|c|c|c||c|c|c|c|c||}%
    \hline
                         &                 &payload& EmbJXTA   & EmbJXTA   & EmbJXTA   & EmbJXTA   & JXTA \\
                         &                 &size   & Chord     & Chord     & Chord     & Chord     & 2.7  \\
                         &                 &       & Compr.    & Compr.    & Compr.    & Compr.    & Compr.    \\
                         &                 &       & Config C0 & Config C1 & Config C2 & Config C3 & Config C4 \\
    \hline
    \hline
       $\gamma$          &$0 \rightarrow 3$&256B   & 2.747 & 3.388 & 7.013 & 10.373 & 9.636 \\
                         &$0 \rightarrow 3$&1024B  & 1.478 & 1.622 & 2.492 & 3.338 & 3.159 \\
			 &$0 \rightarrow 3$&10240B & 1.052 & 1.069 & 1.153 & 1.223 & 1.215 \\
    \hline
       $\Phi_{msg.size}$ &$0 \rightarrow 3$&256B   & 703B   & 867B   &  1795B &  2655B &  2467B\\
			 &$0 \rightarrow 3$&1024B  & 1514B  & 1661B  &  2552B &  3418B &  3235B\\
		         &$0 \rightarrow 3$&10240B & 10776B & 10946B & 11815B & 12533B & 12451B\\
    \hline
  \end{tabular}
}
\ifdefined\ACM
    \begin{table}
    \tbl{Transmitted-message-vs-payload-size ratio $\gamma$ and average message size $\Phi_{msg.size}$ (B) for multicast pipes (4 rendezvous peers )}
    {      
	\TABf{}
    }
    \label{Table_V}
    \end{table}
\else
    \begin{table}[t]
    \setlength{\tabcolsep}{.32em}
    \scriptsize{
	  \caption{\scriptsize {Transmitted-message-vs-payload-size ratio $\gamma$ and average message size $\Phi_{msg.size}$ (B) for multicast pipes (4 rendezvous peers )} }
	  \vspace*{-0.1in}
      \begin{center}
	  \TABf{}  
      \end{center}\label{Table_V}
    } 
    \vspace*{-0.2in}
    \end{table}
    \normalsize
\fi
As JXTA adds some information to the data transmitted, a
third test was performed in order to measure the \textit{additional
message overhead} $\Delta m$ added
to the payload when connecting or accepting pipes are used. A
group of 4 rendezvous peers (rdvpeer 0,1,2,3) 
was deployed letting them join the group, in sequence, with a
60s offset from each other.
When all 4 rdvpeers joined the group,
the Chord predecessor of rdvpeer0 was rdvpeer1 and the successor was
rdvpeer3. 
Since then, for a duration of $\Delta T=240s$, the average \textit{DHT management overhead} 
$\phi_{dht-mgm (0 \rightarrow n)}=\Phi_{n.trx.bytes}/\Delta T$ for the messages sent 
from rdvpeer0 to rdvpeer-n ($n=1,2,3$) was measured (Tab. \ref{Table_III})
\footnote{Using Chord protocol (C0,C1,C2,C3 configuration), there was no
traffic towards rdvpeer2 as
it was neither a predecessor nor a successor of rdvpeer0.}.
As a further trial, the rdvpeer0 was connected to an \textit{accepting
pipe} created by rdvpeer3, after all the 4 rdvpeers had joined.
The rdvpeer0 sent $N$ messages containing $\Phi_{payload.size}$ bytes
(randomly generated each time) to rdvpeer3,
and then closed the pipe.
The parameter \textit{transmitted-message-vs-payload-size ratio} 
$\gamma$ is defined as:
\begin{equation}\label{eq01_1}
\gamma=\frac{\Phi_{n.trx.bytes} - \phi_{dht-mgm (0 \rightarrow 3)} \Delta T}{N \Phi_{payload.size}} = \frac{\Phi^{*}_{n.trx.bytes}}{N \Phi_{payload.size}}
\end{equation}
where $\Phi_{n.trx.bytes}$ is the number of bytes sent by rdvpeer0 (measured by Wireshark) 
during $\Delta T = 240s$ since the pipe connection and $\phi_{dht-mgm (0 \rightarrow 3)}$
is the overhead due to DHT management measured in the previous trial.

$\Phi^{*}_{n.trx.bytes}$ is the number of bytes due to message transmission, therefore the
\textit{average size of the transmitted messages} $\Phi_{msg.size}$ 
can be defined as $\Phi^{*}_{n.trx.bytes}/N$. 

Replacing in eq. \ref{eq01_1}, the following condition holds:

\begin{equation}\label{eq01_2}
\gamma = \frac{\Phi_{msg.size}}{\Phi_{payload.size}}  
\end{equation}

where $\gamma$ 
is the \textit{transmitted-message-vs-payload-size ratio} (if 
$\gamma < 1$, $\gamma^{-1} > 1$ measures the \textit{compression efficiency} 
of the algorithm).

In the performed tests, the rdvpeer0 transmitted $N=25000$ messages with
a payload of $\Phi_{payload.size}=256,1024,10240B$,
using the 5 configurations reported in Tab. \ref{Table_ComprCfg}.
The measured values for $\gamma$ and $\Phi_{msg.size}$ are reported in
Tab. \ref{Table_IV} 
\footnote{As TCP is reliable, the \textit{isReliable} option for JXTA pipes was always disabled before tests.
For JXTA 2.7 data (C4 compr. mode), when lcDHT-LRW algorithm is used, the rdvpeer3
can receive RP messages not only from rdvpeer0 but from every rendezvous peer of the group. As a
consequence, a different formula is used ($\Phi_{msg.size}=(\Phi_{n.trx.bytes}-\phi_{dht-mgm (all \rightarrow 3)})/N$).
$\Phi_{n.trx.bytes}$ and $\phi_{dht-mgm (all \rightarrow 3)}$ were measured using Wireshark at rdvpeer3.}.
Using 256B payloads, the $\gamma$ values measured for JXTA 2.7
and for EmbJXTAChord are respectively 4.86 and 2.08 (for resulting average
message sizes respectively of 1244B and 533B). 
Hence, JXTA 2.7 appears to be unsuitable for the smallest messages as the
overhead is too high with respect to the payload.
Using 1024B payloads, the $\gamma$ values measured for JXTA 2.7
and for EmbJXTAChord are respectively 1.96 and 1.32 (for resulting average
message sizes respectively of 2012B and 1358B).
Using 10240B payloads, the measured $\gamma$ values are respectively
1.096 and 1.035 (for average message sizes of 11228B and
10599B, respectively). In conclusion, EmbJXTAChord improves the bandwidth occupation
in all tested cases.

\subsubsection{Average multicast pipe message overhead}\label{MulticastPipeMessageOverhead}
   
The fourth test measured the overhead related to \textit{multicast pipes} 
(mpipe). EmbJXTAChord supports the multicast delivery of messages using the \textit{Chord walker 
propagation}, whereas JXTA 2.7 exploit the LRW walker \cite{JXTAChord00}\cite{JXTA08}.  
When a rdvpeer needs to propagate a content through its group, it sends the message to its successor 
which, in turn, retransmits rdvpeer-by-rdvpeer  through the whole Chord circular domain. 
Next, each rendezvous peer  retransmits the message to its connected edgepeers, thus realizing the 
propagation of the content through the whole group. 
 
The trial was performed under the same conditions of the third test. The rdvpeer0 created a mpipe, the rdvpeers 1-3 joined 
the group in sequence with a $60s$ offset from each other, and once Chord fingertable 
stabilization occurred (so that the successor of rdvpeer0 became rdvpeer3), 
the rdvpeer3 started to receive the data sent in multicast by rdvpeer0. The rdvpeer0 propagated 
$N=25000$ messages with a payload of $\Phi_{payload.size}=256,1024,10240B$, using the 5 configurations 
in Tab. \ref{Table_ComprCfg}. About the traffic between rdvpeers (0,3), the
\textit{average size of the transmitted messages} $\Phi_{msg.size}$ and the \textit{transmitted-message-vs-payload-size 
ratio} $\gamma$ can be found through eq. (\ref{eq01_2}) and (\ref{eq01_1}). 

A comparison between the values for $\gamma$ reported in Tab.\ref{Table_IV} and Tab.\ref{Table_V} shows
that the overhead is larger for multicast pipes than for unicast ones. 
Moreover, the values for $\gamma$  in Tab.\ref{Table_V} show
that EmbJXTAChord improves the bandwidth occupation also when multicast pipes are used.

\subsubsection{Performance of the compression algorithm}\label{CM_Bench}
\begin{figure}[t]
	\vspace*{-0.20in}
	\hspace*{-0.4in}
	\includegraphics[width=5.5in, height=2.7in]{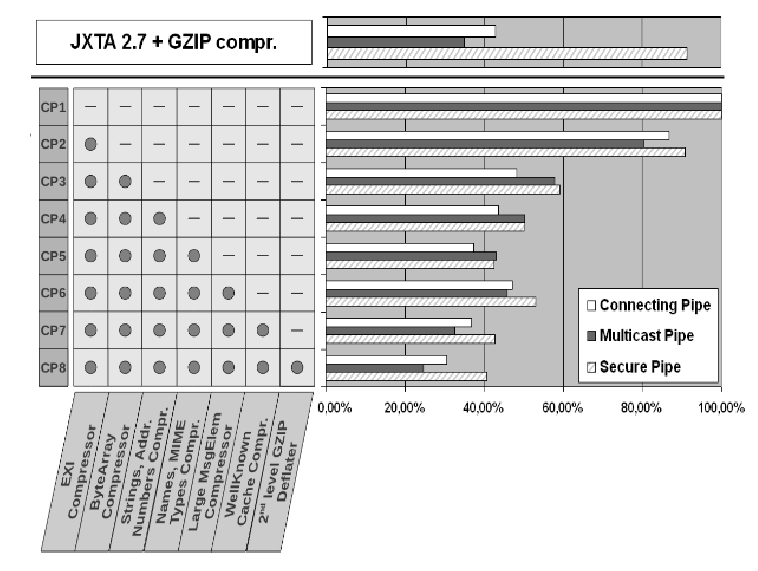} 
	\caption{Compression ratio (\%) with different compression schemes.}
	\label{Fig3}
	
	\vspace*{-0.15in}
\end{figure}
The fifth test was aimed to measure the performance of the Compressor Manager (CM)
using different compression schemes.
A single 1024B message was created, consisting of 128 sequences, each made up of a random byte \textit{x} 
repeated 8 times. The message was sent through a unicast, a multicast
and a secure pipe between two rendezvous  peers.
The sizes of the uncompressed message were 2006B, 3443B and 2473B, respectively.

Fig. \ref{Fig3} shows the message sizes, measured enabling 
or disabling the different compression schemes  (compression configurations CP1-CP8).  The larger size reduction is provided 
by the EXI compressor (S3f) and by the compression of strings, addresses and numbers (s3b-s3d). 
With all the compression features enabled, the message can be compressed down to 31\% (unicast pipe), 
25\% (multicast pipe) and 41\% (secure pipe) of the original size. 
On secure pipes the CM shows the worst performance, as 
compression is affected by the AES-128 encryption and by the cipher-block chaining (CBC) padding \cite{AES00}.
Moreover, the messages produced by EmbJXTAChord are shorter than the ones that could be obtained  
applying a gzip deflater to JXTA messages thus demonstrating the 
effectiveness of the Hypercompression algorithm.  

\subsection{Performance on Bluetooth-IEEE 802.15.4 heterogeneous network}\label{Performance_over_Heterogenous_Networks}

\begin{figure}[t]
	\centering
	\includegraphics[width=4.2in, height=1.42in]{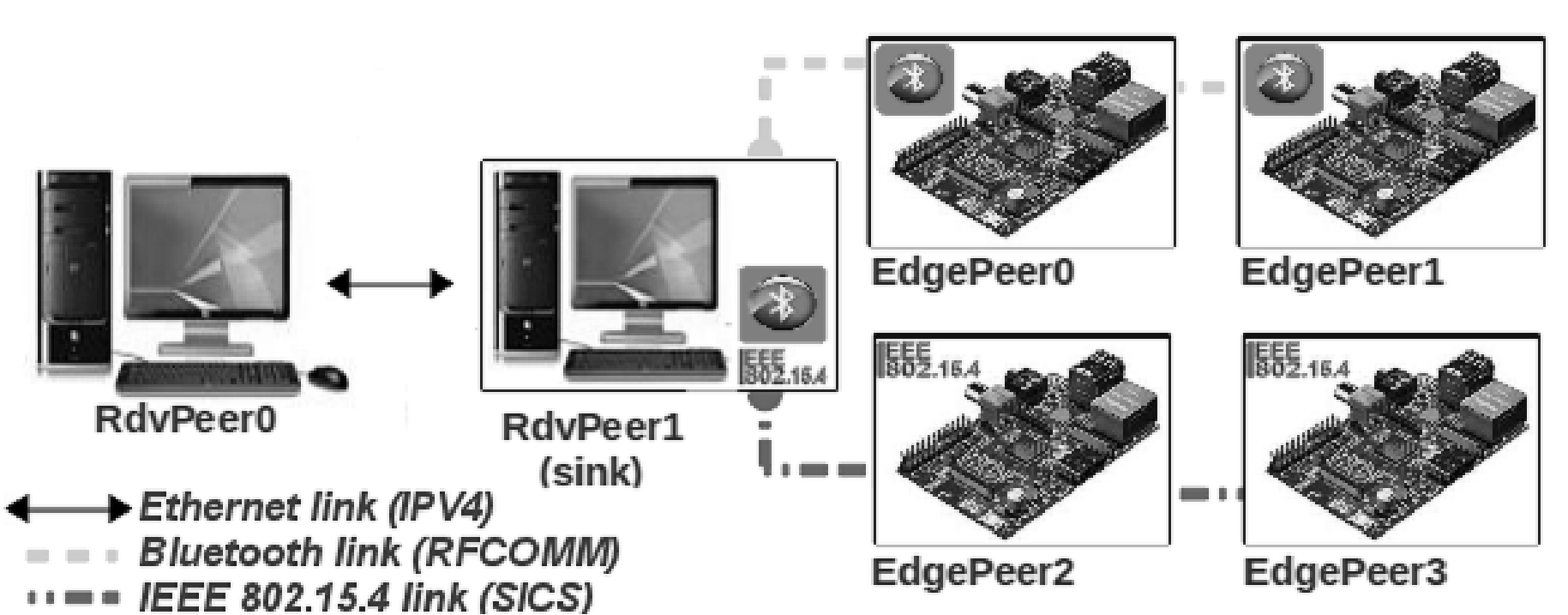}
	\caption{The topology of the heterogeneous network used for testing.}
	\label{Fig4}
	
	\vspace*{-0.08in}
\end{figure}

In order to show that EmbJXTAChord  can work on a narrowband architecture consisting of 
embedded devices, a test was performed on a heterogeneous network consisting of one PC (rdvpeer0),
that communicates using the IPv4 protocol with another PC (rdvpeer1) acting
as a sink and four RaspPI devices (edgepeers 0,1,2,3), connected to rdvpeer1
using RFCOMM over Bluetooth (BT v2.1) and 6LoWPAN,  respectively (see Fig. \ref{Fig4}). 
The rdvpeers (0,1) were linked through a Gigabit Ethernet wire. 
The edgepeers (0,1) were connected to rdvpeer1 through two USB BT adapters based on a Silicon Cambridge transmitter.  
The edgepeers (2,3) used two boards STM Dizum MB950 \cite{STM32W00} equipped with a 
M3 Cortex CPU, 256 KB RAM and an IEEE 802.15.4 transmitter (the operating system was Contiki 2.7 with NullMAC layer
\footnote{ContikiMAC is not available for MB950, therefore NullMAC is the default option for this microcontroller unit. }).
\newcommand{\TABg} 
{
	\begin{tabular}{|c||c|c|c||c|c|c|c||c|c|c|c|}%
		\hline 
		link            &TLS&sender&recv&1KB&2KB&4KB&10KB&1KB&2KB&4KB&10KB\\
		\hline
		&   &	 &    &\multicolumn{4}{|c||}{Raspberry edgepeers}&\multicolumn{4}{|c|}{PC edgepeers} \\
		\hline
		BT $\rightarrow$BT                &-  &edgepeer0&edgepeer1&4.53&8.63&17.08&37.50&54.45&65.17&128.51&135.02\\
		BT $\rightarrow$BT                &yes&edgepeer0&edgepeer1&2.34&4.02& 8.21&13.82&35.38&44.44& 94.95& 81.70\\
		BT $\rightarrow$BT$\rightarrow$ETH&-  &edgepeer0&rdvpeer0&2.57&4.17&10.94&33.32&59.56&65.33&101.81&124.30\\
		BT $\rightarrow$BT$\rightarrow$ETH&yes&edgepeer0&rdvpeer0&1.87&3.47& 6.57&11.23&33.25&56.79& 79.76& 77.56\\
		\hline
		\hline
		SICS$\rightarrow$SICS                &-  &edgepeer2&edgepeer3&0.60&1.67&2.77&3.98&0.67&2.10&3.31&4.42\\
		SICS$\rightarrow$SICS                &yes&edgepeer2&edgepeer3&0.12&0.27&0.51&0.74&0.48&1.07&2.02&2.33\\ 
		SICS$\rightarrow$SICS$\rightarrow$ETH&-  &edgepeer2&rdvpeer0&0.52&1.08&2.05&1.55&0.54&1.62&3.16&3.54\\
		SICS$\rightarrow$SICS$\rightarrow$ETH&yes&edgepeer2&rdvpeer0&0.24&0.53&0.99&1.03&0.45&0.82&1.83&1.11\\
		\hline
		\hline
		SICS$\rightarrow$BT                  &-  &edgepeer2&edgepeer0&0.63&1.87&2.24&3.16&0.66&1.99&3.26&3.40\\ 
		SICS$\rightarrow$BT                  &yes&edgepeer2&edgepeer0&0.25&0.92&0.80&1.95&0.35&1.06&1.11&2.27\\
		\hline
	\end{tabular}
}
\ifdefined\ACM
\begin{table}
	\tbl{Transfer rate measured on heterogeneous network (KB/s)}
	{      
		\TABg{}
	}
	\label{Table_VI}
\end{table}
\else
\begin{table}[t]
	\setlength{\tabcolsep}{.16em}
	\scriptsize
	{
		\caption{\scriptsize {Transfer rate measured on heterogeneous network (KB/s)} }
		\vspace*{-0.1in}
		\begin{center}
			\TABg{}
		\end{center}\label{Table_VI}
	}
	\vspace*{-0.2in}
\end{table}
\fi

\newcommand{\TABm} 
{
	\begin{tabular}{|c||c|c|c||c|c|c|c||}%
		\hline 
		link            &TLS&sender&recv&1KB&2KB&4KB&10KB\\
		\hline
		&   &	 &    &\multicolumn{4}{|c||}{Raspberry-3 edgepeers}\\
		\hline
		BLE $\rightarrow$BLE                &-  &edgepeer0&edgepeer1&49.1&91.1&126.3&120.9\\
		BLE $\rightarrow$BLE                &yes&edgepeer0&edgepeer1&40.8&79.2& 94.8&103.8\\
		BLE $\rightarrow$BLE$\rightarrow$ETH&-  &edgepeer0&rdvpeer0 &43.5&86.1&112.2&115.3\\
		BLE $\rightarrow$BLE$\rightarrow$ETH&yes&edgepeer0&rdvpeer0 &34.3&76.2& 81.3& 93.2\\
		\hline
	\end{tabular}
}
\ifdefined\ACM
\begin{table}
	\tbl{Transfer rate measured for RaspberryPI-3 using Bluetooth Low Energy (KB/s)}
	{      
		\TABm{}
	}
	\label{Table_VI_B}
\end{table}
\else
\begin{table}[t]
	\setlength{\tabcolsep}{.16em}
	\scriptsize
	{
		\caption{\scriptsize {Transfer rate measured for RaspberryPI-3 using Bluetooth Low Energy (KB/s)} }
		\vspace*{-0.1in}
		\begin{center}
			\TABm{}
		\end{center}\label{Table_VI_B}
	}
	\vspace*{-0.2in}
\end{table}
\fi

In the first trial, the edgepeer1 created an accepting socket that was used by the edgepeer0 to send
$N=1000$ messages using a direct BT connection. Each payload contained $\Phi_{payload.size}=1024,2048,4096,10240B$, randomly chosen. 
The same procedure was repeated establishing a direct SICS  connection between the edgepeer3 and the edgepeer2.
In the second trial, the edgepeer0 (edgepeer2) sent data via BT (via SICS) to rdvpeer1 which 
acted as a \textit{sink} for rdvpeer0.
In the third trial, the data exchange was between the edgepeers (2,0) (the rdvpeer1 acted as a \textit{bridge} between BT and SICS subnets, 
thus allowing communication between nodes not directly linked at physical level). 
The three trials were repeated using a secure socket based on the TLS protocol.
Finally, the whole experiment was repeated replacing the RaspPI boards with computers, which were configured as in Sect. \ref{SubSec:ExecutionTimes}.
Tab. \ref{Table_VI} shows that the transfer rates $\beta=N \Phi_{payload.size}/T_{trx}$, where
$T_{trx}$ is the time needed to send all the messages, heavily depend on the 
nodes computational power, due to the compression and
encryption overhead. A fair interpretation of the performance on 6LoWPAN 
must take into account that the throughput of the Contiki devices at the application level is consistently
lower than the raw theoretical data rate of the IEEE 802.15.4 radio interface (256Kb/s). For example, in a work that
measured the real end-to-end performance of some motes connected to a PC through a serial port, 
a real throughput between 32 and 70 Kb/s (i.e. 4 and 8.75 KB/s) for 
various models of transceivers is reported \cite{ZigBee05}. 

The best throughput is achieved using larger payloads. This suggests to pack multiple short messages into a single large one, to optimize the performance. 

Finally, some trials were repeated using some RaspberryPI-3 equipped with a Bluetooth Smart transmitter (v4.1). Tab. \ref{Table_VI_B} shows
that RaspPI-3 with BLE largerly outperforms the old RaspPI with the old Bluetooth v2.1.

\subsection{jxCOAP-E performance}\label{Subsec:jxCOAP}

The last experiment, made up of several trials, was aimed to measure the performance
of a server based on jxCOAP-E under several aspects.

\subsubsection{Scalability test}\label{Subsec:Scalability} In the first trials, the scalability of the jxcoap server was measured. 
First a configuration consisting of an rendezvous  PC \textit{server} and 
$s=[1..7]$ PCs (edgepeers) was deployed. Each edgepeer ran $v=10$ jxcoap virtual client instances 
(see Sect. \ref{ServiceInter}) for $T=180s$, therefore the rendezvous  server responds to 
$n=s*v \in [10..70]$ virtual clients in total  during the trial.
All the peers were equipped with the same AMD CPU used in Sect. \ref{SubSec:ExecutionTimes} and
connected via Ethernet.
Each vclient sent a jxGET request containing a parameter $\eta$ to a \textit{coap} service \footnote{ 
In this experimental section we refer to UDP CoAP for the original version proposed by CoRE WG and to \textit{coap} for concepts
that are common to UDP CoAP, jxCOAP and jxCOAP-E. } 
running on the server, then waited for a response before repeating the operation. All the trials were performed 
using the CoAPBench benchmark, modified to work over jxCOAP-E \cite{COAP02}.

Four services were used: \textit{HelloWorld}, \textit{Fibonacci}, \textit{Newton} and \textit{SortSquareRoot}.
The first service responded to a client request with a ``Hello world'' string. The second one 
returned 
a string containing the $\eta$-th number of the Fibonacci sequence. The third service
returned 
an array containing the square roots of the first $\eta$ integer numbers, determined by the server using the Newton 
method iterated $\xi=25000$ times and encoded as 8B double precision float numbers. 
The fourth service received a xmldoc from the client containing $\eta$ float numbers and
replied with another xmldoc containing the related square roots, determined by the server using its CPU floating unit
and encoded as strings, increasingly sorted through the bubble-sort algorithm. 
When the same value of $\eta$ is used, the size of the uncompressed payload included in the
\textit{coap} response is larger for the SortSquareRoot service than for the Newton one.  

\begin{figure}[t]
	\addtolength{\abovecaptionskip}{-1.0in}
	\subfigure
	{
		\hspace*{-0.52in}
		\includegraphics[width=5.4in, height=1.2in]{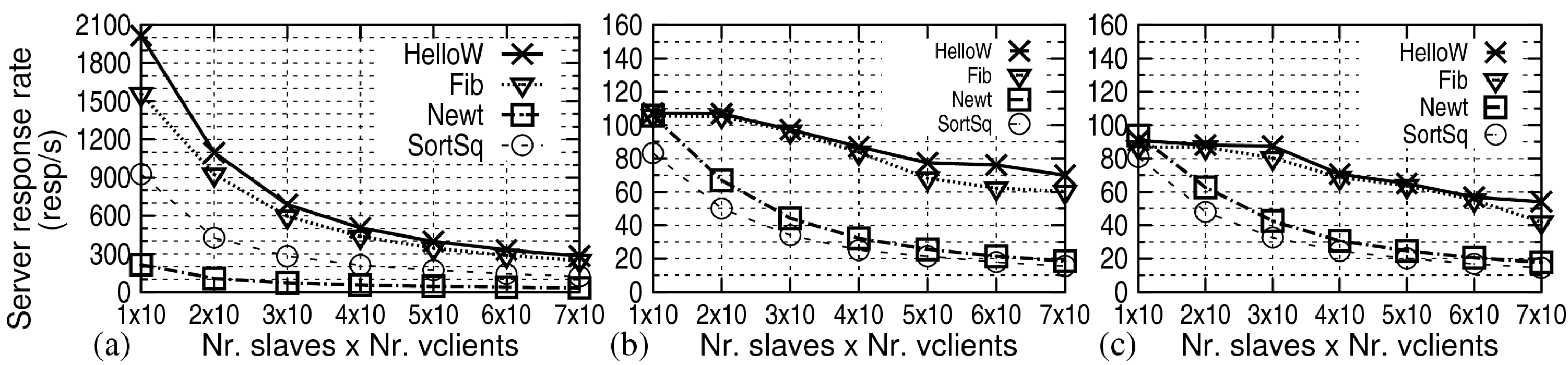} }
	\caption{Scalability measures for the jxcoap server. The graph shows the response rate measured by each edgepeer during the test. (a) Using UDP without compression. (b) Using jxCOAP-E (all compression
		schemes enabled). (c) Using jxCOAP-E within a group with compression and AES-128 encryption enabled.}
	\vspace*{-0.15in}
	\label{Fig4:merge}
\end{figure}

As these trials were aimed to measure the scalability of the \textit{coap} server, only the number $s=[1..7]$ of
connected edgepeers varied, while all the requests were made with $\eta=25$.    
All trials were repeated using UDP CoAP, jxCOAP-E within the NetPeerGroup and 
jxCOAP-E within a peergroup secured by AES-128 (see Sect. \ref{Security}). 
Fig. \ref{Fig4:merge}(a) shows that using UDP CoAP the SortSquareRoot service scaled better than
the Newton service, as the second one requires a higher elaboration time. 
Conversely, Fig. \ref{Fig4:merge}(b-c) show that using jxCOAP-E the SortSquareRoot service was outperformed
by the Newton service, as the first one entails a larger overhead due to EXI compression.   
The effect of AES-128 encryption on the jxCOAP-E response rate was minor.

\begin{figure}[t]
	\addtolength{\abovecaptionskip}{-1.0in}
	\subfigure
	{
		\hspace*{-0.53in}
		\includegraphics[width=5.6in, height=2.5in]{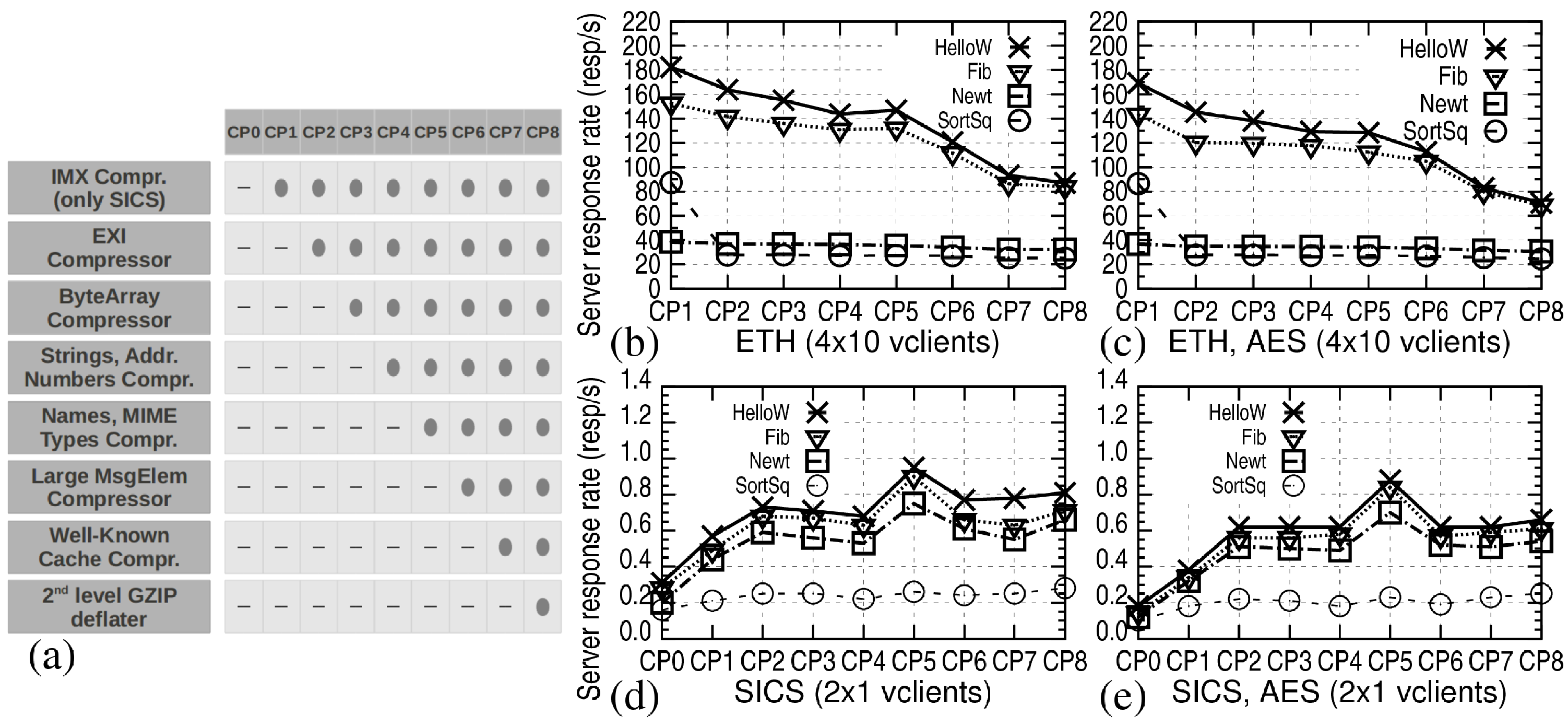} }
	\caption{Scalability measures for the jxcoap server. The graph shows the average response rate measured by each edgepeer, using several compressor configurations. (a) The compressor configurations used for the
trials. (b-c) Response rate for ETH trials without and with AES-128 encryption. (d-e) Response rate for SICS trials without and with AES-128 encryption.}
	\vspace*{-0.15in}
	\label{Fig4B:merge}
\end{figure}

Fig. \ref{Fig4:merge} also shows that the response rate provided by the UDP CoAP server was much higher
than the one provided by the jxCOAP-E server. This is a reasonable result, because EmbJXTAChord 
manages the complex structure of the JXTA messages \cite{JXTA09}, made up of several namespaces and
MessageElements (see Sect.\ref{Hyper00}), together with the operations for compression and encryption.  
In order to measure the overhead introduced by compression, a further experiment was performed. In an Ethernet 
network, the jxCOAP-E server was connected to $s=4$ edgepeers, each running $v=10$ jxcoap vclient instances. 
Each vclient connects to the \textit{HelloWord, Fibonacci, Newton} and \textit{SortSquareRoot} services for
$T=180s$, using $\eta=25$ for all requests. Each trial was performed without and with AES-128 encryption. 
The experiment was repeated multiple times, each time changing the enabled compressor schemes (CP1..CP8) 
accordingly to the table in Fig.\ref{Fig4B:merge}(a).
 
Fig.\ref{Fig4B:merge}(b-c) show that the measured response rates (without and with AES encryption) increased
more and more when multiple compression schemes were disabled, because of the lower overhead determined by the compression.
The largest improvements were measured for the \textit{HelloWorld} and \textit{Fibonacci} services, that are characterized 
by the shortest elaboration times. Obviously, disabling multiple compression schemes determines a higher bandwidth
consumption, but in the Ethernet case this does not significatively affect the performance. 

Conversely, when a narrowband network is used, the compression schemes implemented by EmbJXTAChord
can improve both the response rate and the bandwidth utilization. In order to prove this assertion, the previous experiment was
repeated using a SICS network made up of a single jxCOAP-E server and $s=2$ edgepeers, each running 
$v=1$ virtual clients. Fig. \ref{Fig4B:merge}(d-e) show the measured response rate. It is interesting
to observe that in some trials the default CP8 configuration (all compression schemes enabled) is outperformed by
the CP5 configuration (S1,S2,S5 compression schemes disabled, see Sect.\ref{Hyper00}). This happens because
the CP5 configuration in some cases (see Fig.\ref{Fig3}) can provide low size in transmission (but not the
\textit{lowest} possible size) together with low compression overhead. The CP5 configuration can be used as an alternative to 
the default CP8 configuration when the application needs to optimize the response rate
and the number of nodes connected to the sink peer is low ($s=2$ in the described test). 
Otherwise, if the minimization of the bandwidth required for transmission
is mandatory as many nodes communicate at the same time to the sink, the default 
CP8 configuration is preferable.   

In conclusion, EmbJXTAChord is suitable for all the cases where the advantages described in Sect.\ref{Sec:JxtaAdvDrawbacks}
make acceptable for the developer the larger overhead caused by compression and encryption. 
In order to allow the optimization of the transmission strategy used for each link, 
EmbJXTAChord adds some \textit{feature bits} to the message that signal to the recipient node
the compression schemes that are currently in use.

\begin{figure}[t]
	\addtolength{\abovecaptionskip}{-1.5in}
	\hspace*{-0.62in}
	\vspace*{-0.16in}
	\subfigure { \label{Fig_RTT_jxCOAP_PC___jxActComp}\includegraphics[width=5.9in, height=3.2in]{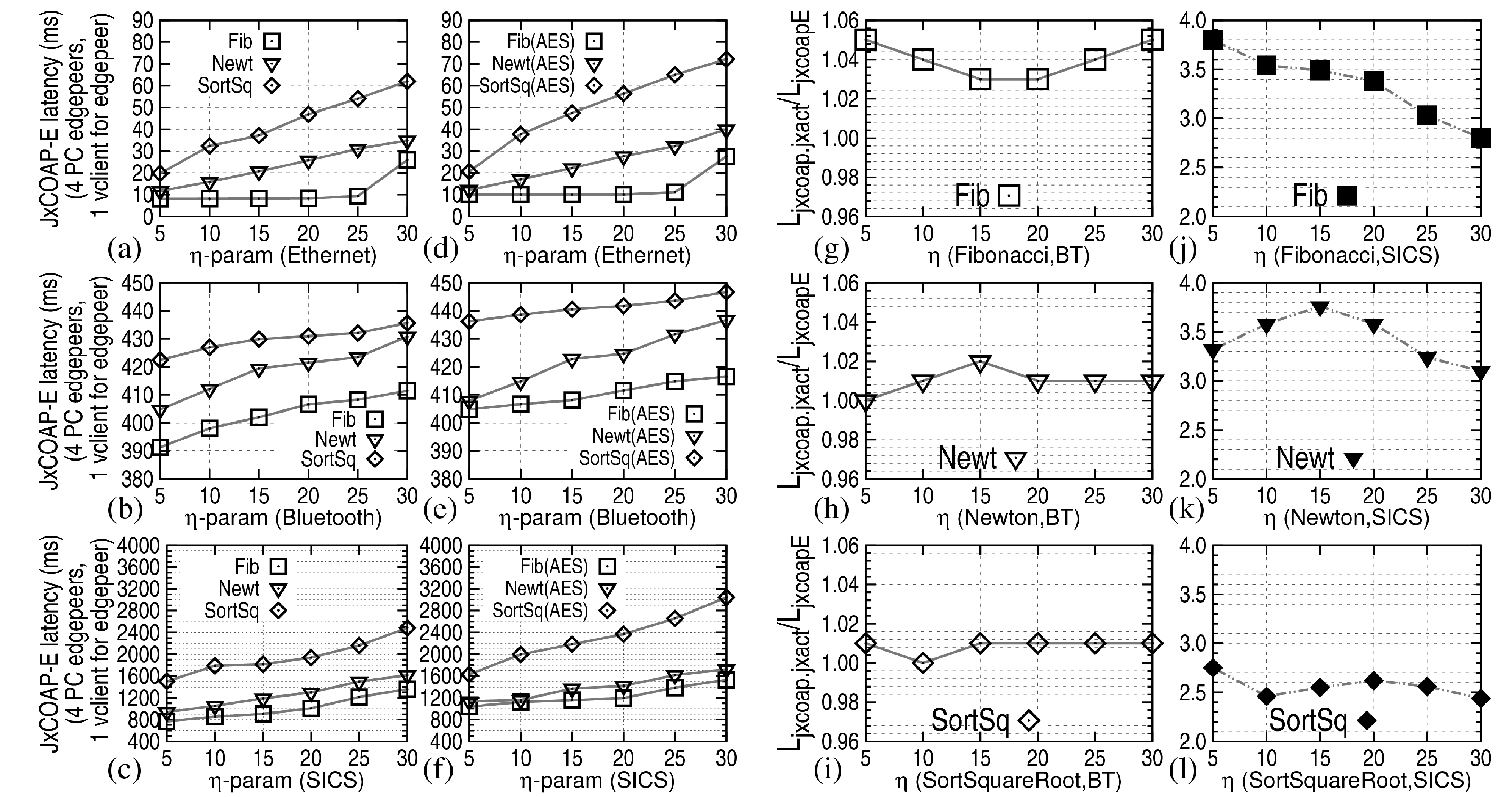} }
	\caption{(a-f) Latency values measured for jxCOAP-E on a homogeneous network using 1 rendezvous server and 4 PC edgepeers without and with AES encryption; 
		 (g-l) Performance comparison between jxCOAP-E and jxCOAP under SICS and BT networks.}
	\label{SuperFigure1}
\end{figure}

\begin{figure}[t]
	\addtolength{\abovecaptionskip}{-1.5in}
	\hspace*{-0.62in}
	\vspace*{-0.16in}
	\subfigure { \label{Fig_RTT_jxCOAP_RaspPI_RaspPI3}\includegraphics[width=5.9in, height=3.2in]{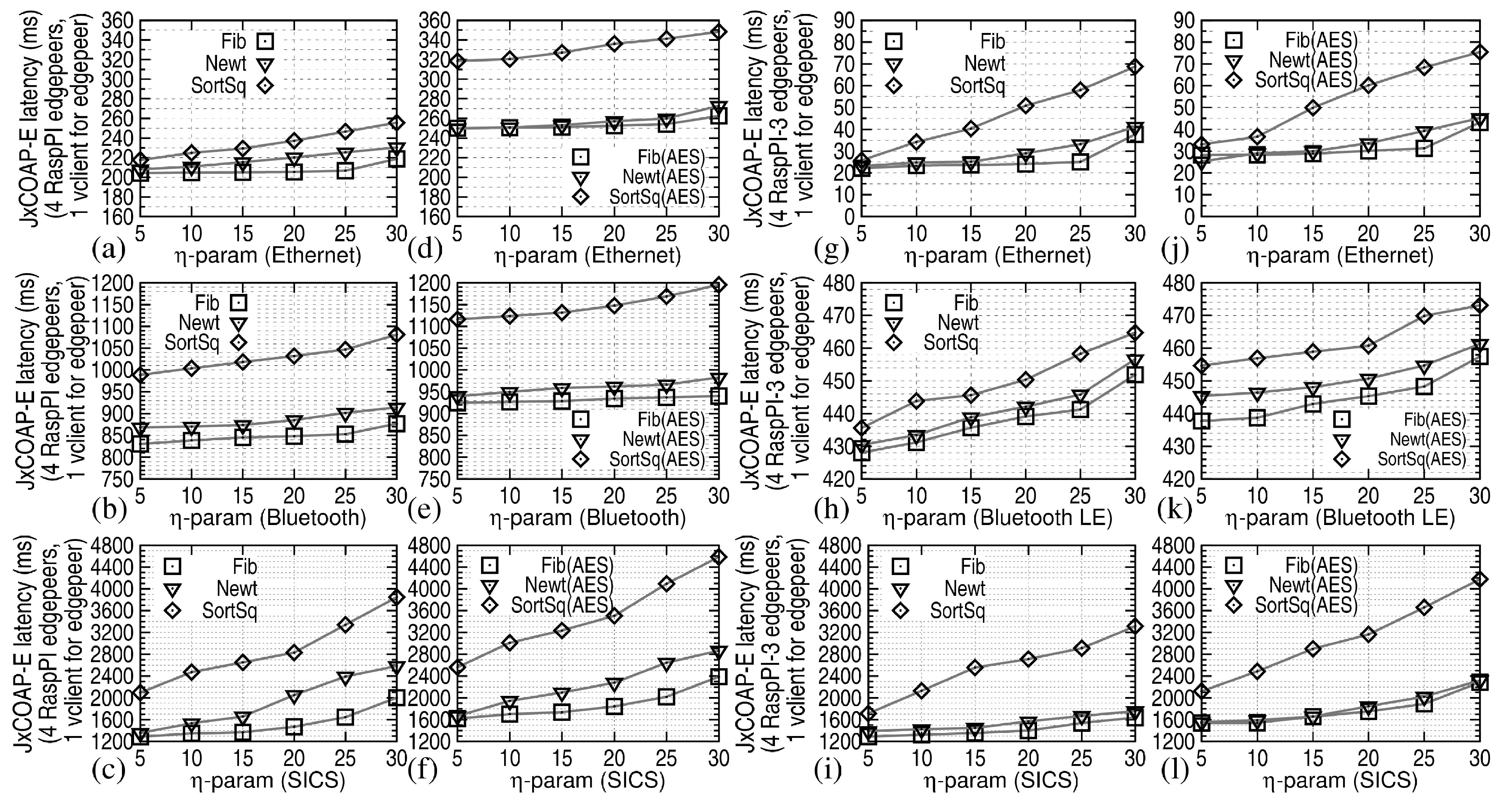} }
	%
	\caption{Latency values measured for jxCOAP-E on a homogeneous network (a-f) using 1 rendezvous server and 4 RaspPI edgepeers without and with AES encryption; 
		 (g-l) using 1 rendezvous server and 4 RaspPI-3 edgepeers without and with AES encryption.}

	\label{SuperFigure2}
\end{figure}

\begin{figure}[t]
	\addtolength{\abovecaptionskip}{-1.5in}
	\hspace*{-0.55in}
        \vspace*{-0.21in}
	\subfigure { \label{Fig_RTT_UDPCoapFactors}\includegraphics[width=5.5in, height=0.9in]{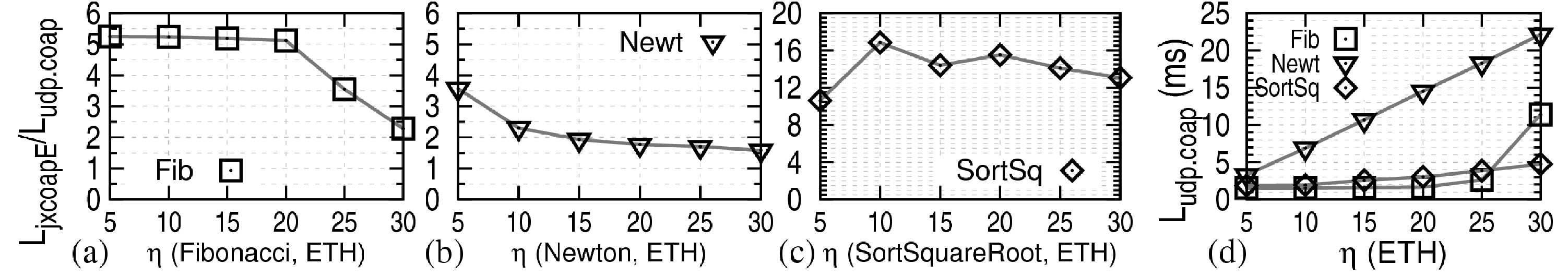} }
	\caption{(a-c) Performance comparison between jxCOAP-E and UDP CoAP over an Ethernet-based homogeneous network (1 rendezvous server, 4 edgepeers) (d) The latency values determined
for UDP CoAP over an Ethernet network. }
	\label{SuperFigure3}
\end{figure}

\subsubsection{jxCOAP-E latency over a homogeneous network (no encryption)}\label{jxCOAP_TestLabel0}

The following tests were aimed to measure the average round-trip latency ($\mathcal{L}(\eta)$) of the jxcoap requests sent to the Fibonacci, Newton or SortSquareRoot
service over a homogeneous network. Such a metric, that refers to  all clients, can be determined in function of $\eta$, averaging all the measured RTT (round-trip
times) of the jxcoap requests received by the server (the average is done overally, i.e. regardless of which of the $n$ vclients sent the request).

A PC rendezvous  \textit{server} was connected to $s=4$ edgepeer PCs, each running $v=1$ vclient (PC-only configuration). Hence, the server 
responded to $n=s*v=4$ vclients in all. Each test was performed for $T=180s$,  using the values in $S=\{5,10,15,20,25,30\}$ for the $\eta$ parameter.  
The tests were performed on a network made up of one server and of four vclients linked through Ethernet (ETH), Bluetooth (BT v2.1)  or IEEE 802.15.4 (SICS) links.
The trials were performed using jxCOAP-E within the standard NetPeerGroup.

The whole experiment was performed using the PC-only configuration and, later, replacing the 4 PCs with 4 RaspPI model B+ boards (RaspPI configuration)
and with 4 RaspPI-3 model B boards (RaspPI-3 configuration). 
\footnote{The \textit{coap} \textit{block-wise transfer mode} \cite{COAP03} was disabled during tests, as 
RFCOMM, 6LoWPAN and TCP are reliable and connection-oriented protocols. For SICS tests only, the intermessage compression (see Sect. \ref{Hyper00}) was used with the parameters
$c_{S}=30$ and $c_{X}=10$. In this experiment and in the following ones, the Hypercompressor was configured to use all
compression schemes (CP8 default configuration). }. 
The CPU and OS were the same described in Sect. \ref{SubSec:ExecutionTimes}.  The trials related to the RaspPI-3 
configuration were performed using Bluetooth transmitters compatible with BLE v4.1. 

Fig. \ref{SuperFigure1}(a-f), Fig. \ref{SuperFigure2}(a-f) and Fig. \ref{SuperFigure2}(g-l) show the latency values measured respectively 
for PC-only, RaspPI and RaspPI-3 configurations. 
The slope changes in the $\mathcal{L}$ curves are mainly determined by the time required for compression (which varies on $\eta$) and
by the compression ratio that, not being constant, affects the payload size and thus the transmission time. 
The SICS values are affected also by the overhead due to the IMX cache manager (see Sect. \ref{Hyper00}), as the CM must calculate the hash of 
each xmldoc or string msgelem before transmission.

The results show that using ETH, BT or BLE $\mathcal{L}(\eta)<1.2s$ for all configurations. Moreover, $\mathcal{L}(\eta)<0.5s$
for BLE and RaspPI-3, thus ensuring a responsive behaviour for a wide set of applications (home management and automation, ambient assisted living, health monitoring).  
The latencies measured for SICS, instead, are significantly higher.  This makes the use of EmbJXTAChord over SICS networks suitable only 
for applications that do not require a reactive behaviour (i.e. smart metering, data gathering, enviromental monitoring).  
Performance under SICS can be improved 
tailoring the IMX cache parameters and the compression configuration on the basis of the traffic 
generated by the specific application.

\subsubsection{Comparison between jxCOAP-E and UDP CoAP over an Ethernet-based homogeneous network}

These trials compared the latency values for the three jxcoap services to the ones measured for a UDP CoAP server, using Ethernet and PC-only
configuration. Fig. \ref{SuperFigure1}(a) and Fig. \ref{SuperFigure3}(d) show the $\mathcal{L}_{jxcoapE}$ and $\mathcal{L}_{udp.coap}$ times for each value of $\eta$, while  
Fig. \ref{SuperFigure3}(a-c) show the ratio $\rho(\eta)=\mathcal{L}_{jxcoapE}(\eta)/\mathcal{L}_{udp.coap}(\eta)$.   
Fig. \ref{SuperFigure3}(d), which refers to UDP CoAP, indicates that the Newton service requires the highest elaboration time, 
followed by the SortSquareRoot and the Fibonacci ones.  
Despite this, the $\mathcal{L}_{jxcoapE}$ times detected for the Newton service are always lower than the ones detected for the SortSquareRoot
service, thus indicating that the jxCOAP-E latency depends more on the compression and delivery times (which are affected in turn 
by the payload size and by the channel bandwidth) than on the elaboration time. The SortSquareRoot service requires more time for
compression than the Newton service, as the payload returned in the response by the first one is a xmldoc that is EXI+GZIP compressed, 
while the payload returned by the second one is a ByteArray that is only GZIP compressed (see Sect. \ref{Hyper00}).  

\subsubsection{Encryption overhead}

\newcommand{\TABi} 
{
	\begin{tabular}{|c||c|c|c||c|c|c||c|c|c||c|c|c||}%
		\hline 
		            &\multicolumn{3}{|c||}{Ethernet ($\beta^{ETH}$)}&\multicolumn{3}{|c||}{Bluetooth ($\beta^{BT}$)}&\multicolumn{3}{|c||}{SICS ($\beta^{SICS}$)}&\multicolumn{3}{|c||}{Hybrid ($\beta^{BTSICS}$)}\\
		\hline
		            &Fib   &Newt  &SortSq&Fib  &Newt &SortSq&Fib   &Newt  &SortSq&Fib   &Newt  &SortSq\\
		\hline
		PC          &19.0\%& 7.5\%&17.6\%&1.8\%&1.0\%& 2.7\%&23.3\%&11.6\%&19.1\%&17.6\%&7.57\%&14.7\%\\
		RaspPI      &22.1\%&17.8\%&41.3\%&9.8\%&8.4\%&11.5\%&24.3\%&18.6\%&21.9\%&18.4\%&10.2\%&38.5\%\\
		RaspPI-3(*) &22.5\%&15.0\%&17.5\%&1.6\%&2.2\%& 2.8\%&24.3\%&18.1\%&20.4\%&12.3\%& 9.4\%&32.6\%\\
		\hline
		\multicolumn{13}{||l||}{(*) BLE v4.1 used for Bluetooth testing} \\
		\hline
	\end{tabular}
}
\ifdefined\ACM
\begin{table}
	\tbl{Average $\mathcal{L}$ overhead values (\%) determined by the AES-128 encryption on homogeneous and heterogeneous networks. }
	{      
		\TABi{}
	}
	\label{Table__VII}
\end{table}
\else
\begin{table}[t]
	\setlength{\tabcolsep}{.16em}
	\scriptsize
	{
		\caption{\scriptsize {Average $\mathcal{L}$ overhead values (\%) determined by the AES-128 encryption on homogeneous and heterogeneous networks.}}
		\vspace*{-0.1in}
		\begin{center}
			\TABi{}
		\end{center}\label{Table__VII}
	}
	\vspace*{-0.20in}
\end{table}
\fi

In order to measure the overhead determined by AES encryption, the trials described in Sect. \ref{jxCOAP_TestLabel0} were repeated 
within an AES-128 secure peergroup. 
Fig. \ref{SuperFigure1}(d-f), Fig. \ref{SuperFigure2}(d-f) and Fig. \ref{SuperFigure2}(j-l) show the $\mathcal{L}_{jxcoapE.AES}(\eta)$ times measured when AES-128 encryption is enabled.
Assuming that $\mathcal{L}_{jxcoapE.AES}(\eta)=\mathcal{L}_{jxcoapE}(\eta) + \Delta \mathcal{L}_{el}(\eta) + \Delta \mathcal{L}_{trx}(\eta)$, where $\Delta \mathcal{L}_{el}(\eta)$ is
the elaboration time needed for encryption and $\Delta \mathcal{L}_{trx}(\eta)$ is the additional time needed for the transmission of the larger payloads
due to the use of secure peergroups, the overhead caused by AES on a single trial was defined as 
\begin{equation}\label{eq05}
\alpha(\eta)=\frac{\mathcal{L}_{jxcoapE.AES}(\eta)}{\mathcal{L}_{jxcoapE}(\eta)} - 1=\frac{\Delta \mathcal{L}_{el}(\eta) + \Delta \mathcal{L}_{trx}(\eta)}{\mathcal{L}_{jxcoapE}(\eta)}
\end{equation}
The $\alpha (\eta)$ values,  measured using for $\eta$ the values in $S = \{5,10,15,20,25,30\}$,  were averaged thus obtaining 
the $\beta = avg_{(\eta \in S)}{\alpha (\eta)}$ values reported in Tab.\ref{Table__VII}. 

Tab.\ref{Table__VII} shows that the overhead values are mostly below 25\%.   
For instance, using RaspPI-3 configuration, the latency values measured for 
the SortSquareRoot service on the BLE network were $\mathcal{L}(\eta=20)=450.37$ms without encryption and 
$\mathcal{L}(\eta=20)=460.72$ms with encryption (overhead of 10.35ms, 2.30\%). The low overhead 
measured for BLE makes this very suitable for applications requiring responsiveness.  

The trials on SICS experienced the highest overhead. For instance, using PC configuration, the latency values
measured for Newton service on the SICS network was $\mathcal{L}(\eta=10)=1053.24$ms without encryption
and $\mathcal{L}(\eta=10)=1161.54$ms with encryption (overhead of 108.30ms, 10.28\%). 

Using RaspPI configuration, the latency values measured for 
the SortSquareRoot service on the SICS network were $\mathcal{L}(\eta=20)=2830.63$ms without encryption and 
$\mathcal{L}(\eta=20)=3503.99$ms with encryption (overhead of 673.36ms, 23.79\%). The values measured
for the Fibonacci service on SICS were $\mathcal{L}(\eta=20)=1468.47$ms without encryption and 
$\mathcal{L}(\eta=20)=1841.96$ms with encryption (overhead of 373.49ms, 25.43\%).

These overhead values are determined by the additional elaboration time $\Delta \mathcal{L}_{el}(\eta)$ 
for AES encryption and by the additional transmission time $\Delta \mathcal{L}_{trx}(\eta)$. 
In fact, the use of a secure peergroup increases the message size as additional
data (such as the PeerGroupID) must be transmitted in some MessageElements.
Moreover, $\Delta \mathcal{L}_{trx}(\eta)$ is affected also by CBC padding that
can determine a large overhead in a WSN because of the small size of the IEEE 802.15.4 
frames \cite{AES00}. The measured latency values indicate that the use of EmbJXTAChord in a WSN 
is suitable only with fast processors and for applications that do not require the transmission 
of large payloads.

\subsubsection{Comparison between jxCOAP-E and jxCOAP}

These trials compared the performance of the jxCOAP-E version integrated in EmbJXTAChord and of the early jxCOAP version 
based on JXTA 2.7 integrated in jxActinium \cite{JXCOAP00}.
Using the PC-only configuration, in order to measure the performance of jxCOAP, 
all compression schemes were disabled and all latencies were measured again for BT and SICS links.
Next, the ratio $\sigma(\eta)=\mathcal{L}_{jxcoap.jxact}(\eta)/\mathcal{L}_{jxcoapE}(\eta)$ was computed for each value of $\eta$. 
Fig. \ref{SuperFigure1}(j-l) show that on the slow SICS links jxCOAP-E largerly outperforms jxCOAP with latency improvements  between 144\% and 280\% ($\sigma^{SICS}(\eta)$ is bounded between 2.44 and 3.80). 
Moreover Fig. \ref{SuperFigure1}(g-i) show that jxCOAP-E outperforms jxCOAP also on the BT links,
with latency improvements between 1\% and 6\% ($\sigma^{BT}(\eta)$ is bounded between 1.00 and 1.06).  
In all the cases, the reduction of the transmission time compensates for the overhead introduced by the compression.

\subsubsection{Latency over a hetereogeneous network made up of Bluetooth and 6LoWPAN subnetworks}
\begin{figure}[t]
	\addtolength{\abovecaptionskip}{-1.8in}
	\hspace*{-0.55in}
	\vspace*{-0.10in}
	\subfigure { \includegraphics[width=5.60in, height=2.50in]{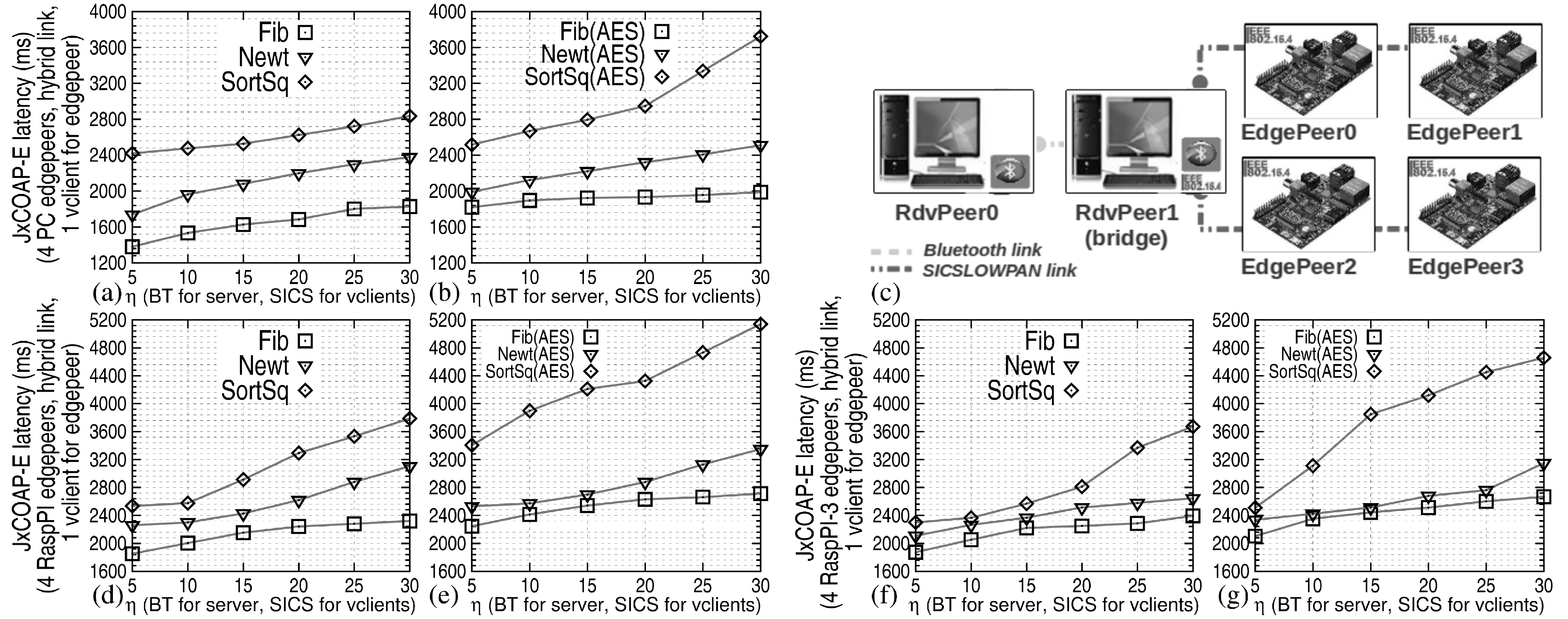} }  
	\caption{Latency values measured for a BT-SICS hybrid network (a-b) using 1 rdv server and 4 PC edgepeers without and with AES encryption; 
		 (d-e) using 1 rdv server and 4 RaspPI edgepeers without and with AES encryption; (f-g) using 1 rdv server and 4 RaspPI-3 edgepeers 
		 without and with AES encryption; (c) the hybrid network architecture used for testing.}
	\label{SuperFigure4}
\end{figure}
A PC rendezvous  \textit{server} (rdvpeer0) was connected through a BT link to a second PC rendezvous (rdvpeer1) that was in turn 
connected to $s=4$ edgepeer PCs, each running $v=1$ vclients, through SICS wireless links
(Fig. \ref{SuperFigure4}(c)).
In this way, rdvpeer1 worked as a \textit{bridge} between the BT and SICS subnetworks. The latencies of the jxcoap requests were
measured for the Fibonacci, Newton and SortSquareRoot services, using jxCOAP-E within the standard NetPeerGroup and within an AES-128 secure peergroup. 
The experiment was performed using the PC, RaspPI and RaspPI-3 configurations.  
The results show that EmbJXTAChord is able to ensure a seamless communication between the server working in the BT subnetwork 
and the vclients working in the SICS subnetwork. Fig. \ref{SuperFigure4} shows that the $\mathcal{L}_{jxcoapE}$ times
are in most cases higher than the ones measured for the homogeneous network, because of the overhead due to 
the network translation performed by the bridge. 
The $\beta^{BTSICS}$ overhead values measured using AES-128 encryption are reported in Tab. \ref{Table__VII}. 

\subsection{The cost of Chord consistent hashing}\label{Chord_DHT_Costs}

\newcommand{\TABj} 
{
	\begin{tabular}{|c|c|c|c|c|c|}%
		\hline
			Number of&Messages&Single msg  &Overall msg &Single msg  &Overall msg      \\
			nodes (N)&per rdvpeer &size (B)    &size (B)    &size (B)    &size (B)\\
		                 &($log_{2}N$) &(compressed)&(compressed)&(uncompressed)&(uncompressed)\\
		\hline
			8	&3	&479	&1437	&1732	&  5196 \\
			64	&6	&479	&2874	&1732	&  10392\\
			128	&7	&479	&3353	&1732	&  12124\\
			2048	&11	&479	&5269	&1732	&  19052\\
			4096	&12	&479	&5748	&1732	&  20784\\
			16384	&14	&479	&6706	&1732	&  24248\\
		\hline
	\end{tabular}
}
\ifdefined\ACM
\begin{table}
	\tbl{Overall size (B) of the messages required to each of the N rdvpeers for updating Chord fingertable when a new rdvpeer joins the group. }
	{      
		\TABj{}
	}
	\label{Table_IX}
\end{table}
\else
\begin{table}[t]
	\setlength{\tabcolsep}{.16em}
	\scriptsize
	{
		\caption{\scriptsize {Overall size (B) of the messages required to each of the N rdvpeers for updating Chord fingertable when a new rdvpeer joins the group. }}
		\vspace*{-0.1in}
		\begin{center}
			\TABj{}
		\end{center}\label{Table_IX}
	}
	\vspace*{-0.20in}
\end{table}
\fi

As described in Sect. \ref{RendezvousProto}, each rdvpeer of the peergroup implements the Chord
DHT algorithm, mantaining in memory its own fingertable. 

In \cite{Chord00} it was demonstrated that joining or leaving a new rdvpeer requires 
to update the informations of other $O(log_{2} N)$ rdvpeers, each of which
needs in turn, with high probability, $O(log_{2} N)$ Chord messages to update 
its own fingertable.
Using EmbJXTAChord, each of the messages needed for updating the fingertable 
(named \textit{find\_successor} requests in Chord terminology) contains a 128b \textit{start field} 
together with other informations about the sender node.

In our experiments, it was measured that the transmitted size of these messages was 479B (when all
compression schemes were enabled) and 1732B (without compression). 
Tab.\ref{Table_IX} shows that the overall size of the messages transmitted by each rdvpeer of the group
when a new joining/leaving event occurs, assuming that they are in number of $O(log_{2} N)$. The table
shows that compression allows to mantain very low the transmission cost even with a high number of rdvpeers.  

When a new rdvpeer joins the group, it is necessary also to transfer part of
the SRDI entries into the cache of the new member. Each SRDI entry requires further 32B (256b)
(128b for the hash value of the advertisement and 128b for the PeerID of the storing node). Unfortunately, the overall size 
of the SRDI update message is not predictable  a priori as it depends on 
the number of SRDI entries (i.e. on the number of stored advertisements).

\section{A typical scenario for a smart home}\label{Sec:Smart Home} 

In order to show the advantages of using JXTA in the field of IoT, let us consider a scenario consisting in 
a small residence made up of some three-storey houses. Some services are shared by all the tenants of the
residence. 
Moreover, each house contains some smart objects, sensors and actuators. 
The smartphone of each member of the family can be used for person authentication at the family's house entrance.
While a webcam, connected to a small PC acting as the house Authentication Peer (AP), acquires the face image of a family member, in their  
pocket the smartphone runs a small application, based on EmbJXTAChord, that transmits to the AP the face biometric template through a BT connection. 
A robust face authentication algorithm that works using small-sized templates can be found, for instance, in \cite{FACERFID00}.  
Once the AP has recognized the identity of the family member, it sends some commands (using
jxCOAP-E) to all the smart devices in the house, thus opening the main door 
and customizing several parameters of the environment (temperature and lightning level in the rooms, 
kind of music to play etc.). 

Sensors and actuators can be connected to one or multiple smart peers, each one made up of a 
Raspberry PI-3 \cite{RaspberryPI04} or of a Raspberry PI-Zero \cite{RaspberryPI03_ACM} 
connected via USB to a 6LoWPAN transmitter board. 
In this way, all the sensor boards can access all the functionalities of the whole peergroup (including the services
provided by laptops, by PCs connected via Ethernet or by BT devices such as mobile phones) leveraging on the hop-by-hop delivery
and on the routing over subnetworks. No proprietary hardware for bridging ETH/BT/SICS is required, because 
the single rendezvous  peer of the floor acts as a gateway between the Bluetooth or 
IEEE 802.15.4 subnetworks  and the rest of system.
Another advantage is that the GPIO port of the Raspberry can be directly connected to digital sensors and actuators \cite{RaspberryPI02}.

All the smart devices, sensors, actuators and domestic appliances in a house  belong to the same HomePeerGroup (HPG), protected using AES-128 encryption. 
Each smart device in the group can use the REST-ful interface for coordinating its own operations with the activities currently performed by
the other devices. For instance, each appliance can read the power available at the moment, using this information to defer the starting time of a job.  

If several HPGs exist, one for each house, a device in a house  cannot access or see the services or devices installed 
in another house.  
However, all devices installed in the residence can be grouped into the Residence Peergroup (RPG), in order to share some of their own functionalities. 
For example, in each of the houses of the residence there is a smart monitor able to connect to the entryphone in order to see who is at the residence entrance. 
In order to preserve confidentiality about the visitors of each house, the entryphone and the smart monitor create a secure socket via TLS  
before starting the video connection. Moreover, any two members of the RPG use their smart monitors to communicate through a secure connection.

In the proposed architecture, a new peer can be added to the system without any user intervention.  
Once a peer has published an advertisement  within the group, all its functionalities are exploitable 
by all the peers of the same group, even if they are not in the same subnetwork. 
%
\section{Server and client sample programs}\label{Sec:TwoSmallExamples}

About the scenario described in Sect. \ref{Sec:Smart Home}, a server-side sample program is shown in Listing \ref{Src1}.  
The rendezvous server (whose JXTA name is \textit{PeerServer}) deploys a custom peergroup named \textit{ChildPeerGroup0}.
Next, in the custom peergroup a new jxCOAP-E service named \textit{SJXTA\_CoapService0} is created. The communication
is protected using the preshared-key \textit{GROUP\_PASSW} (0000). 

\lstdefinestyle{customc}{
  belowcaptionskip=1\baselineskip,
  breaklines=true,
  captionpos=b,                    
  language=C,
  showstringspaces=false,
  keywordstyle=\fontfamily{pcr}\fontseries{mc}\scriptsize\bfseries,
  basicstyle=\fontfamily{pcr}\fontseries{mc}\scriptsize\bfseries,
  commentstyle=\fontfamily{pcr}\fontseries{mc}\scriptsize\bfseries,
}

\lstset{escapechar=@,style=customc,caption={The source code of the server-side program.},label={Src1},linewidth=15cm}
\begin{lstlisting}[language=C]
package sjxta_main; 

import sjxta.*; 
import sjxta_classes.*; 

public class FaceAuthentication extends JxCoapResourceBase 
{ 
    public FaceAuthentication(String name) 
    { 
	super(name); 
    } 

    public void handleGET(JxCoapExchange exchange) 
    { 
	// Here is the code for face authentication and response 
        // to the client peer
    } 
} 

public class Main 
{ 
  private static final String GROUP_PASSW = "0000"; 

  public static void main(String[] args) throws java.io.IOException 
  { 
     SJXTA_NetObj MyNetObj = new SJXTA_NetObj();
     SJXTA_NetObj nObj_MyChildPeerGroup = new SJXTA_NetObj();
     SJXTA_AuthInitializer_PSE MyAuthInit = null;

     // Server configured as rendezvous peer

     int ErrCode = SJXTA_NetObj.Init (MyNetObj, 
                                      "PeerServer", 
                                      SJXTA_CONST.RENDEZVOUS); 

     if (ErrCode==0) // No errors. NetPeerGroup is now associated to 
     { 	             // the network object named MyNetObj

        ErrCode = SJXTA_NetObj.StartTheConnection(MyNetObj, 0); 

	if (ErrCode==0) // No errors.
	{
          // Deploy the child peergroup (protected by AES-128 using GROUP_PASSWD) 
	  MyAuthInit = new SJXTA_AuthInitializer_PSE(GROUP_PASSWD); 
	  
          ErrCode = SJXTA_PeerGroup.CreateNewPeerGroup(MyNetObj, 
                                                       nObj_MyChildPeerGroup,
                                                       "ChildPeerGroup0", 
                                                       MyAuthInit, 0); 

	  if (ErrCode==0) // New group created successfully. The new group
	  {     	  // is now associated to the network object 
			  // nObj_MyChildPeerGroup

	     // Deploy a new jxCOAP-E virtual server within the group 
             // (it provides the service named SJXTA_CoapService0) 
	     SJXTA_Service MyNewService = new SJXTA_Service(); 
	     ErrCode = SJXTA_ServiceManager.DeployNewService(nObj_MyChildPeerGroup, 
                                                             MyNewService,
                                                             "SJXTA_CoapService0", 
                                                             null, 0); 
	     if (ErrCode==0) 
	     { 
	       // Start the new jxCOAP-E virtual server
	       ErrCode = SJXTA_ServiceManager.StartTheService(MyNewService, 
                                                              null, 
                                                              0); 

	       if (ErrCode==0) 
	       { 
                  // Add a new resource to the jxCOAP-E server
		  MyNewService.AddResource  (new FaceAuthentication("faceauth")); 

		  // Wait for the connection of a jxCOAP-E virtual client  
		  System.out.println ("Hit any key to continue..."); 
		  System.in.read(); 

		  ErrCode = SJXTA_ServiceManager.RevokeService(MyNewService, 
                                                               10000, 
                                                               0); 
               } 

               SJXTA_NetObj.StopTheConnection(MyNetObj, 0); 
	       System.exit(0); 
	     } 
	  }
	} 
    } 
  } 
} 
\end{lstlisting}

The client-side program is shown in Listing \ref{Src2}.  The edge client needs to contact a rdvpeer
to start operations. The address of the local rdvpeer can be found through multicast (when ETH or SICS links are
used) or providing the address explicitly. 

\lstset{escapechar=@,style=customc,caption={The source code of the client-side program.},label={Src2},linewidth=15cm}
\begin{lstlisting}[language=C]
package sjxta_main;
 
import sjxta.*; 
import sjxta_classes.*; 

public class Main 
{ 
  private static final String GROUP_PASSW = "0000"; 


  public static void main(String[] args) throws java.io.IOException 
  { 
    SJXTA_NetObj MyNetObj = new SJXTA_NetObj(); 
    SJXTA_NetObj nObj_MyChildPeerGroup = new SJXTA_NetObj(); 
    SJXTA_AuthToken_PSE_String MyAuthToken = null; 

    // Client configured as edge peer

    int ErrCode = SJXTA_NetObj.Init (MyNetObj, "PeerClient", SJXTA_CONST.EDGE); 

    if (ErrCode==0)   // No errors. NetPeerGroup is now associated to the
    {		      // network object named MyNetObj	
	 
      // Set the BT address of the local rdvpeer 
      SJXTA_NetConfig.AddSeedRendezvous(MyNetObj, "btspp://#RAPTOR:1"); 
       
      // Try to join the NetPeerGroup 
      ErrCode = SJXTA_NetObj.StartTheConnection (MyNetObj, 
                                                  SJXTA_CONST.NO_RDV_AUTOSTART); 

      if (ErrCode==0) // No errors. 
      { 	      
	// Create a new authentication token 
	MyAuthToken = new SJXTA_AuthToken_PSE_String (GROUP_PASSW); 

	// Join to the new secure peergroup using the new token for decryption
	ErrCode = SJXTA_PeerGroup.JoinToPeerGroup (MyNetObj, 
                                                   nObj_MyChildPeerGroup, 
                                                   "ChildPeerGroup0",
                                                   MyAuthToken,
                                                   SJXTA_CONST.NO_RDV_AUTOSTART); 

	if (ErrCode==0)  // No errors. The secure peergroup is now associated
	{ 		 // to the network object nObj_MyChildPeerGroup

	   // Connect to the remote service named SJXTA_CoapService0 
 	   // The MSA is automatically searched within the group ChildPeerGroup0 
           // (timeout 30s). There is no need to know the address of the server. 

	   SJXTA_RemoteService MyRemoteService = new SJXTA_RemoteService(); 
	   ErrCode = SJXTA_ServiceManager.
			    ConnectToRemoteService (nObj_MyChildPeerGroup,
                                                    MyRemoteService,
                                                    "SJXTA_CoapService0",
                                                    30000, 0); 


	   // If the MSA is found, EmbJXTAChord recognizes that the 
           // remote JXTA service contains a jxCOAP-E virtual server and 
           // automatically creates (and binds) a new instance of a jxCOAP-E 
           // virtual client.	

	   if (ErrCode==0) 
	   { 
	      // No errors. Now a new instance of jxCOAP-E virtual client is in
	      // the MyRemoteService Java object, bound to the server. 

	      byte[] BioTemplateArray = new byte [4096]; 
	      // (..) Here is the code for grabbing the face image from camera
	
	      // Send a jxCOAP-E request to the remote resource named faceauth. 
              // The array BioTemplateArray is the request payload. 
              // GET parameter n=1. Timeout = 10s. 

	      SJXTA_PNT_TO_JxCoapResponse TheJxCoapResponses = null;
              TheJxCoapResponses = new SJXTA_PNT_TO_JxCoapResponse(); 
	      ErrCode = MyRemoteService.
			 SendCommand_CoapMode (SJXTA_RemoteService.COAP_OPCODE_GET,
                                               "/faceauth?n=1",
                                               CoapMediaTypeRegistry.TEXT_PLAIN,
                                               BioTemplateArray, 10000,
                                               TheJxCoapResponses, 0); 

	      if (ErrCode==0)  // No errors. Read the answer received from server. 
	      { 
	          JxCoapResponse theJxCoapResponse = TheJxCoapResponses.v.get(0); 
		  System.out.println (Response.getPayloadString()); 
	      } 
	    } 
        } 

        SJXTA_NetObj.StopTheConnection(MyNetObj, 0); 
        System.exit(0); 
      } 
    } 
 } 
} 
\end{lstlisting}

In the case shown in Listing \ref{Src2}  EmbJXTAChord exploits the Bluetooth SDP (Service Discovery Protocol) in order to 
translate automatically the address \ttfamily btspp://\#RAPTOR:1 \rmfamily into the real address \ttfamily btspp://00228372FFC0:1. \rmfamily
Once successfully contacted the local rdvpeer, the edge node attempts to authenticate itself thus
joining the custom peergroup. Finally, the edge provides to contact the service \textit{SJXTA\_CoapService0}.
It is important to observe that, differently from UDP CoAP, the client needs to know only the name
of the jxCOAP-E service within the group, not the address of the server providing the service. 
In fact, EmbJXTAChord can automatically determine the provider node looking for the Module Specification
Advertisement of the service  (see Sect. \ref{ServiceInter}) within the custom peergroup. 
Once connection is established, EmbJXTAChord  transparently provides to compress, encrypt and route over 
multiple subnetworks (if it is needed) the jxCOAP-E messages.

\section{Power measures}\label{Sec:PowerMeasures}

The last experiment measured the power consumption and the lifetime of a client
node (\textit{mobile device}) connected to a jxCOAP-E server, when only a battery pack is
available as power source. The test was repeated multiple times, using mobile devices
such as a Raspberry PI, a Raspberry PI-3 and one Android smartphone (an Oukitel U7 Plus
based on a Mediatek MTK6737 CPU with four 1.3 Ghz cores).

For each trial, the PC and the mobile device were configured as the
rendezvous server and the edgepeer client, respectively. Both server and client worked within a
custom peergroup, thus performing AES encryption. For each trial described in this
section, the mobile phone (edgepeer) ran only a single vclient ($v = 1$). For each of
the mobile devices under test, two jxCOAP-E trials were run: \textit{SendDataBlock} and
\textit{SortSquareRoot}. The SendDataBlock trial was devised to simulate operations that are
frequently performed in a smart home.
For instance, in the scenario described in Sect.\ref{Sec:Smart Home}, 
a smartphone that transmits to the AP server the biometric template of a user,
exploiting the face authentication algorithm described in \cite{FACERFID00},
needs to send requests whose payload is 4KB long, next waiting for 
a response. 
During the SendDataBlock trial the client sent to the server consecutive requests containing
a payload of $\eta$ = 4800 bytes (randomly generated). For each request, the server sent
a response whose payload is a 256-chars long string. The SortSquareRoot service was
already described in Sect. \ref{Subsec:Scalability}. For this service, the parameter $\eta$ = 30 was always used
in all the trials described in this section. Each trial was run for $T = 180s$.

\newcommand{\TABl} 
{
	\begin{tabular}{|c|c|c||c||c||c|c||c||c|c||c|c|}%
		\hline 
		Device &CPU  &Link   &Transm.  &jxCOAP-E      &Power  &Power  &Average&Batt.  &Batt.  &Estim. &Estim.\\
                       &     &to     &chipset  &test          &cons.  &cons.  &latency&Voltage&Max    &life   &life\\
                       &     &server &         &              &(mW)   &(mW)   &(ms)   &(V)    &Charge &time   &time\\
                       &     &       &         &              &(before&(during&       &       &(mAh)  &(Min)  &(Max)\\
                       &     &       &         &              &test)  &test)  &       &       &       &       & \\
                       &     &       &         &              &$\mathcal{P}_{b}$ &$\mathcal{P}$ &$\mathcal{L}_{jxcoapE}$ & $V_{batt}$ & $C_{batt}$ & $T^{min}_{life}$ & $T^{max}_{life}$ \\ 
		\hline
		RaspPI &BCM  &ETH    &built-in &SortSqRoot    &1004.0 &1161.5 &355.96 &5.0    &10000  &43h02m &49h48m \\
		       &2835 &       &built-in &SendDataBlock &1004.0 &1159.2 &307.62 &5.0    &10000  &43h07m &49h48m \\
		\cline{3-12}
                       &     &BT     &CSR      &SortSqRoot    &1064.7 &1250.0 &1153.2 &5.0    &10000  &40h00h &46h57m \\
                       &     &v2.1   &BT v2.1  &SendDataBlock &1064.7 &1202.4 &984.1  &5.0    &10000  &41h35h &46h57m \\
		\cline{3-12}
                       &     &SICS   &STM      &SortSqRoot    &1294.8 &1482.0 &2428.5 &5.0    &10000  &33h44m &38h36m \\
                       &     &       &MB950    &SendDataBlock &1294.8 &1432.6 &2956.1 &5.0    &10000  &34h54m &38h36m \\
		\hline
              RaspPI-3 &BCM  &ETH    &built-in &SortSqRoot    &1156.9 &1584.0 &45.09  &5.0    &10000  &31h33m &43h13m \\
                       &2837 &       &built-in &SendDataBlock &1156.9 &1536.0 &38.68  &5.0    &10000  &32h33m &43h13m \\
		\cline{3-12}
                       &     &BT     &built-in &SortSqRoot    &1297.4 &1831.6 &520.4  &5.0    &10000  &27h17m &38h32m \\
                       &     &v4.1   &built-in &SendDataBlock &1297.4 &1783.4 &495.0  &5.0    &10000  &28h02m &38h32m \\
		\cline{3-12}  
                       &     &SICS   &STM      &SortSqRoot    &1482.0 &2101.5 &2414.3 &5.0    &10000  &23h47m &33h44m \\
                       &     &       &MB950    &SendDataBlock &1482.0 &2152.8 &2844.2 &5.0    &10000  &23h13m &33h44m \\
		\cline{3-12}
                       &     &WiFi   &built-in &SortSqRoot    &1200.0 &1968.0 &54.78  &5.0    &10000  &25h24m &41h40m \\
                       &     &       &built-in &SendDataBlock &1200.0 &1920.0 &53.93  &5.0    &10000  &26h02m &41h40m \\
		\hline
             Oukitel&Mediatek&BT     &built-in &SortSqRoot    &773.5  &855.0  &632.1  &4.35   &2500   &12h43m &14h03m \\
	     U7 Plus&6737    &v4.0   &built-in &SendDataBlock &773.5  &810.0  &587.9  &4.35   &2500   &13h25m &14h03m \\
		\cline{3-12}
	            &        &WiFi   &built-in &SortSqRoot    &824.4  &914.0  &282.1  &4.35   &2500   &11h53m &13h11m \\
	            &        &       &built-in &SendDataBlock &824.4  &868.3  &333.1  &4.35   &2500   &12h31m &13h11m \\
		\hline
	\end{tabular}
}
\ifdefined\ACM
\begin{table}
	\tbl{Measures about energy consumption and battery lifetime estimation.}
	{      
		\TABl{}
	}
	\label{PowerConsump_Table}
\end{table}
\else
\begin{table}[t]
	\setlength{\tabcolsep}{.16em}
	\scriptsize
	{
		\caption{\scriptsize {Measures about energy consumption and battery lifetime estimation.} }
		\vspace*{-0.1in}
		\begin{center}
			\TABl{}
		\end{center}\label{PowerConsump_Table}
	}
	\vspace*{-0.2in}
\end{table}
\fi

A digital wattmeter (connected to the USB port of the mobile device) was used to
measure the power consumption during the jxCOAP-E transmission ($\mathcal{P}$). The whole
procedure was repeated using different links (ETH, BT, SICS, Wi-Fi) between the
server and the mobile device. Before starting each trial,
the power consumption $\mathcal{P}_{b}$ of the device was measured when the transmitter is
powered on, but no elaboration or transmission is performed. A comparison between $\mathcal{P}$ and 
$\mathcal{P}_{b}$ allows to estimate the power consumption due to message transmission and cpu
elaboration.

The measures for Wi-Fi and BT referring to RaspPI-3 were performed using the
transmitter modules integrated in the BCM43438 chipset \cite{Broadcom1}. The measures related to the Oukitel smartphone 
were realized using the USB port as the only power source (i.e. after disconnecting 
the phone battery). As the smartphone ran Android 6.0, a dedicated version of
EmbJXTAChord based on OpenJDK 9 \cite{OpenJDK9} was developed for the experiment.

The lifetime of the mobile device was calculated assuming the RaspPI
and RaspPI-3 connected to a ROMOSS battery \cite{ROMOSS01} (voltage $V_{batt}$ = 5 V, maximum
charge $C_{batt}$ = 10000 mAh), and the smartphone connected to the battery provided
by the producer (voltage $V_{batt}$ = 3.8 V, maximum charge $C_{batt}$ = 2500 mAh). 
An estimation of the battery lifetime can be obtained calculating the minimum 
and maximum value through the formulas 
$T^{min}_{life} = (3600 \cdot V_{batt} C_{batt})/\mathcal{P}$ and
$T^{max}_{life} = (3600 \cdot V_{batt} C_{batt})/\mathcal{P}_{b}$. 

Tab. \ref{PowerConsump_Table} shows that RaspPI-3 ensures lower latencies than RaspPI in 
all trials, but at the cost of a higher power consumption. Moreover, the use of SICS determines a high power
consumption, as the Raspberry boards need to be connected to an external MB950 daughter
board. Tab. \ref{PowerConsump_Table} also shows that all the tested mobile devices, in the conditions 
previously described, can remain fully operative for many hours without the need for recharging the
battery.

\section{Conclusions}\label{Sec:Conclusions} 

EmbJXTAChord enables to support IoT applications
allowing any sensor or actuator device to be connected in secure \textit{peergroups of objects}, 
regardless of both the transport protocol (TCP, HTTP, Bluetooth RFCOMM, 6LoWPAN) and the features of the link that is 
actually used for communication. The P2P protocol provides functionalities such as node and resource discovery, 
secure peergroups, routing over subnetworks, reliable and unreliable connections between nodes. 
In order to ensure good performance even on narrowband networks, EmbJXTAChord exploits a compression protocol 
that reduces the size of a JXTA message to 25\% of the original one. 
The protocol is light enough to run on RaspberryPI and RaspberryPI-3  boards, 
thus allowing the implementation of low-cost heterogeneous IoT networks based on a RESTful architecture.

jxCOAP-E allows the creation of a RESTful architecture over the heterogeneous network. 
The support for AES-128 protected peergroups allows to create group of peers that shares services and
resources, regardless of the underlying network topology.  
jxCOAP-E leverages on the underlying P2P architecture in order to provide a distributed and 
fault-tolerant service discovery. 
\
In all the trials performed over Ethernet, Bluetooth and Bluetooth Smart networks, jxCOAP-E provided
latency values that are low enought to allow the use for applications that require
a responsive behaviour (such as home management and automation, assisted living or health monitoring). 
Conversely, the latency values measured over a 6LoWPAN wireless sensor network were higher, but they are still 
acceptable for tasks that do not require a high level of responsiveness (such as smart metering and enviromental 
control). 

Future work will deal with the implementation of new Message Transport Binding  
modules for other protocols, such as G3-PRIME for smart grid \cite{PRIME00}, 
and a plugin for the browser to allow monitoring the JXTA peergroup resource through a REST-ful API. 

\section*{References}

\bibliography{MainPaper}

\end{document}